%% file: zh_llbb_prd.tex
\documentclass[aps,prd,amsmath,amssymb,superscriptaddress,twocolumn,floatfix]{revtex4}

\usepackage{epsfig}
\usepackage{psfrag}
\usepackage{graphicx}% Include figure files
\usepackage{dcolumn}% Align table columns on decimal point
\usepackage{bm}% bold math
\usepackage{epsfig}
\usepackage{multirow}
\usepackage{slashed}
\usepackage{rotating}
\usepackage[caption=false]{subfig}

\def\et {E_{\mathrm{T}}}

\newcommand{\ppbar}{\mbox{$p\overline{p}$}}

\newcommand{\bbbar}{\mbox{$b\overline{b}$}}
\newcommand{\ccbar}{\mbox{$c\overline{c}$}}
\newcommand{\csbar}{\mbox{$c\overline{s}$}}
\newcommand{\ttbar}{\mbox{$t\overline{t}$}}

\newcommand{\zbb}{\mbox{$Z+\bbbar$}}
\newcommand{\zcc}{\mbox{$Z+\ccbar$}}

\newcommand{\dzero}{D0}

\newcommand{\runiia}{Run~$2$a}
\newcommand{\runiib}{Run~$2$b}

\newcommand{\ee}      {\ensuremath{ee}}
\newcommand{\mumu}    {\ensuremath{\mu\mu}}
\newcommand{\eeicr}   {\ensuremath{ee_{\rm ICR}}}
\newcommand{\mumutrk} {\ensuremath{\mu\mu_{\rm trk}}}

\newcommand{\mutrk}   {\ensuremath{\mu_{\rm trk}}}

\newcommand{\gev} {\ensuremath{\mathrm{Ge\kern -0.1em V}}}
\newcommand{\tev} {\ensuremath{\mathrm{Te\kern -0.1em V}}}

\newcommand{\mll}{\mbox{$M_{\ell\ell}$}}

\newcommand{\dr}{\Delta\mathcal{R}}
\newcommand{\pt}{\mbox{$p_{\rm T}$}}
\newcommand{\ptbb}{\mbox{$p_{\rm T}^{bb}$}}
\newcommand{\ptbone}{\mbox{$p_{\rm T}^{b1}$}}
\newcommand{\ptbtwo}{\mbox{$p_{\rm T}^{b2}$}}

\newcommand{\etadet}{\mbox{$\eta_{\mathrm{det}}$}}

\newcommand{\costhst}{\mbox{$\cos\theta^*$}}

\newcommand{\alpgen}{{\sc alpgen}}
\newcommand{\pythia}{{\sc pythia}}

\newcommand{\geant}{{\sc geant3}}

\newcommand{\met}{\mbox{$\slashed{E}_{\rm T}$}}

\def\explim{5.1}%%expected limit at MH=125
\def\obslim{7.1}%%observed limit at MH=125

\def\vzRFobs{3.5}
\def\vzRFerr{2.5}
\def\vzRFobsSF{0.8}
\def\vzRFerrSF{0.6}

\begin{document}

\hspace{5.2in} \mbox{FERMILAB-PUB-13-067-E}

\title{\boldmath{ 
Search for $ZH \rightarrow \ell^+\ell^-b\bar{b}$ production in
$9.7$~fb$^{-1}$ of $p\bar{p}$ collisions with the \dzero\ detector}}
\input author_list.tex

\date{13 March 2013}

\begin{abstract}
  We present a search for the standard model (SM) Higgs
  boson produced in association with a $Z$ boson in 9.7~fb$^{-1}$ of
  $p\bar{p}$ collisions collected with the \dzero\ detector at the
  Fermilab Tevatron Collider at $\sqrt{s}$~=~1.96~\tev.  Selected events
  contain one reconstructed $Z\rightarrow e^+e^-$ or $Z\rightarrow
  \mu^+\mu^-$ candidate and at least two jets, including at least one
  jet likely to contain a $b$ quark.   To validate the search procedure,
  we also measure the cross section for $ZZ$ production, 
  and find that it is consistent with the SM expectation.  
  We set upper limits at the $95\%$ C.L. on the product of
  the $ZH$ production cross section and branching ratio
  $\mathcal{B}(H\to\bbbar)$ for
  Higgs boson masses $90 \le M_H \le 150~\gev$. The observed (expected)
  limit for $M_H = 125~\gev$ is a factor of \obslim~(\explim) larger than the
  SM prediction. 
\end{abstract}
\pacs{13.85.Ni, 13.85.Qk, 13.85.Rm, 14.80.Bn}

\maketitle

\section{Introduction}\label{sec:intro}

In the standard model (SM), the spontaneous breaking of the
electroweak gauge symmetry generates masses for the $W$ and $Z$ bosons
and produces a new scalar elementary particle, the Higgs
boson~\cite{higgs}.  Precision electroweak data, including the latest
$W$ boson mass measurements from the CDF \cite{cdf-wmass} and D0
\cite{d0-wmass} collaborations and the latest Tevatron combination
for the top quark mass~\cite{tev-top}, constrain the mass of the SM
Higgs boson to $M_H<$ 152 GeV \cite{ew-fit} at the 95\% confidence level (C.L.).
Direct searches at the CERN $e^+e^-$ Collider (LEP)~\cite{lep-higgs}, by the
CDF and D0 collaborations at the Fermilab Tevatron $\ppbar$
Collider~\cite{tev-results}, and by the ATLAS and CMS collaborations
at the CERN Large Hadron Collider (LHC)~\cite{atlas-results,
  cms-results} further restrict the allowed range to 
$122.1 < M_{H} < 127.0$ GeV. 
ATLAS and CMS have discovered a new boson with properties
consistent with those of the SM Higgs boson at 
${M_H~\approx~126~\gev}$~\cite{new-atlas-results,new-cms-results},
primarily through its decays into $\gamma\gamma$ and $ZZ$,
while the CDF and D0 collaborations have reported combined 
evidence for a particle consistent with such a boson produced
in association with weak bosons and decaying to $b\bar{b}$~\cite{tev-results2}.

For ${M_H \lesssim 135~\gev}$, the dominant Higgs boson decay is
to the $b\bar{b}$ final state.  At the Tevatron the best sensitivity
to a low mass Higgs boson is obtained from the analysis of its
production in association with a $W$ or $Z$
boson and its subsequent decay into pairs of $b$~quarks. 
Evidence for a signal in this decay mode complements the
ATLAS and CMS observations and provides further indication that the new particle is 
consistent with the SM Higgs boson that also couples directly to fermions.

We present a search for the process $ZH \rightarrow \ell^+\ell^-b\bar{b}$, where
$\ell$ is either a muon or an electron, in $9.7$~fb$^{-1}$ of $p\bar{p}$ 
collisions at $\sqrt{s}$~=~1.96~\tev\ using the D0 detector. 
This Article is a detailed description of a published Letter~\cite{zh_prl}
providing inputs
included in the CDF and D0 combination described in Ref.~\cite{tev-results2}.
The CDF collaboration has performed a search in the same final state \cite{cdf-zhiggs}.
This analysis extends and supersedes the previous D0 result obtained on
4.2~fb$^{-1}$ of integrated luminosity~\cite{pubzh}. 

We select events that contain a $Z$ boson candidate, reconstructed
in one of four independent channels defined by lepton identification
criteria.
Selected events must also contain a Higgs boson candidate,
reconstructed from two jets.  At least one jet must be identified as
likely to originate from a $b$ quark (``$b$~tagged''). 
The backgrounds to this selection include the production of a $Z$ boson
in association with jets, $\ttbar$ production, diboson production, and
multijet events with non-prompt muons or electrons, or with jets misidentified
as electrons.  They are estimated using Monte Carlo (MC)
simulations and control samples in the data.
We employ a kinematic fit to improve the reconstruction of the $H\to\bbbar$
resonance.  Subsequently, we develop a two-stage multivariate analysis
to separate the signal from the backgrounds and extract results from
the shapes of the resulting multivariate discriminants.  
To validate the search procedure,
we also present a measurement of the $ZZ$ production cross section in
the same final state used for the Higgs boson search. 

We describe the \dzero~detector in Section~\ref{sec:detector} and the event
selection in the four analysis channels in Section~\ref{sec:evtsel}. 
Background and signal MC simulations are detailed in Section~\ref{sec:mcsim}
and multijet estimation is described in Section~\ref{sec:mjbkg}. In
Section~\ref{sec:normalization} we discuss the normalization applied to
the background samples. The kinematic fit is described in Section~\ref{sec:kinfit}.
We describe the multivariate analysis strategy in Section~\ref{sec:mva} and
the systematic uncertainties affecting the final results in Section~\ref{sec:syst}.
We present the results for Higgs boson production and diboson production in
Section~\ref{sec:results} and summarize our results in Section~\ref{sec:summary}.

\section{The \dzero~detector}\label{sec:detector}

The \dzero~detector~\cite{d0det,run1det} consists of a central
tracking system in a 2~T superconducting solenoidal magnet,
surrounded by a central preshower (CPS) detector, a liquid--argon
sampling calorimeter, and a muon spectrometer.  The central tracking
system consists of a silicon microstrip tracker (SMT) and a
scintillating fiber tracker (CFT), and provides coverage for charged
particles in the pseudorapidity~\cite{d0coords} range $|\etadet| < 3$,
where $\etadet$ is the pseudorapidity measured with respect
to the center of the detector.  The CPS is located immediately before
the inner layer of the calorimeter, and has about one radiation length
of absorber, followed by three layers of scintillating strips.  The
calorimeter consists of a central cryostat (CC), covering
$|\etadet|<1.1$, and two end cryostats (EC), covering up to $|\etadet|
\approx 4.2$.  In each cryostat the calorimeters are divided into
electromagnetic (EM) layers on the inside and hadronic layers on the
outside.  Plastic scintillator detectors improve the calorimeter
measurement in the inter-cryostat regions (ICRs, $1.1<|\etadet|<1.5$) 
between the CC and the ECs.
The muon spectrometer is located beyond the calorimeter and consists of a
layer of tracking detectors and scintillation trigger counters before
a 1.8 T iron toroidal magnet, followed by two similar layers after the
toroid. It provides coverage up to $|\etadet| \approx 2$.  The instantaneous
luminosity is measured by a system composed of two disks of
scintillators positioned in front of the ECs.
A three-level trigger system selects events for data
logging and subsequent offline analysis.

\section{Event Selection}\label{sec:evtsel}

The search is performed in four independent channels defined by the subdetectors 
used for lepton identification:
the dimuon channel (\mumu), the muon + isolated track channel (\mumutrk),
the dielectron channel (\ee), and the electron + ICR electron channel (\eeicr).
The data for this analysis were collected 
from April 2002 to February 2006 (\runiia), and from June 2006 to September 2011
(\runiib).  Between \runiia\ and \runiib, a new layer of the SMT
was installed and the trigger system was upgraded
\cite{d0upgrade}. \runiia\ corresponds to an integrated luminosity of
1.1~fb$^{-1}$. \runiib\ is further sub-divided into three 
periods that we analyze independently to account for time-dependent
effects in the performance of the detector.  We refer to
them as Runs 2b1 (corresponding to an integrated luminosity of
1.2~fb$^{-1}$), 2b2 (3.0~fb$^{-1}$), and 2b3 (4.4~fb$^{-1}$).  

\subsection{Triggering}\label{sec:trigger}

In the \ee\ and
\eeicr~channels we analyze events acquired predominantly with triggers
that provide real-time identification of electrons and jets.
In the \ee~channel we accept events that satisfy any trigger
requirement, with a measured efficiency consistent with 100\% within 1\%.
In the \eeicr~channel the set of triggers used has an efficiency
of 90--100\% depending on the region of the detector toward which 
the electron points, and we apply
the trigger efficiency, measured in data and parametrized by
electron $\eta$, electron $\phi$, and jet transverse momentum, 
to the MC events as a weight. Specific selection requirements applied 
to the two channels are described in Sec.~\ref{sec:offlevtsel}.

\begin{figure*}[htbp]\centering
\includegraphics[scale=0.43]{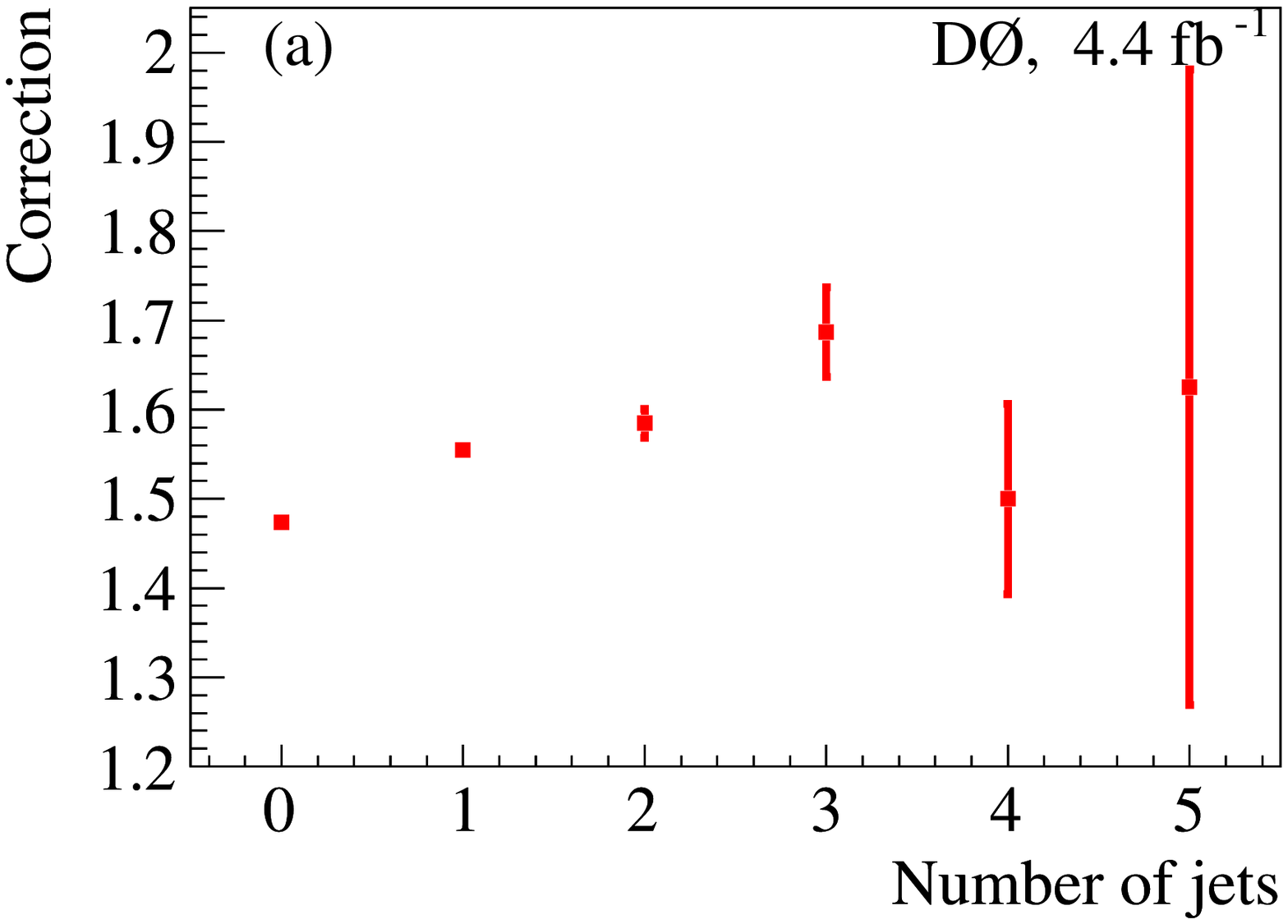}
\includegraphics[scale=0.43]{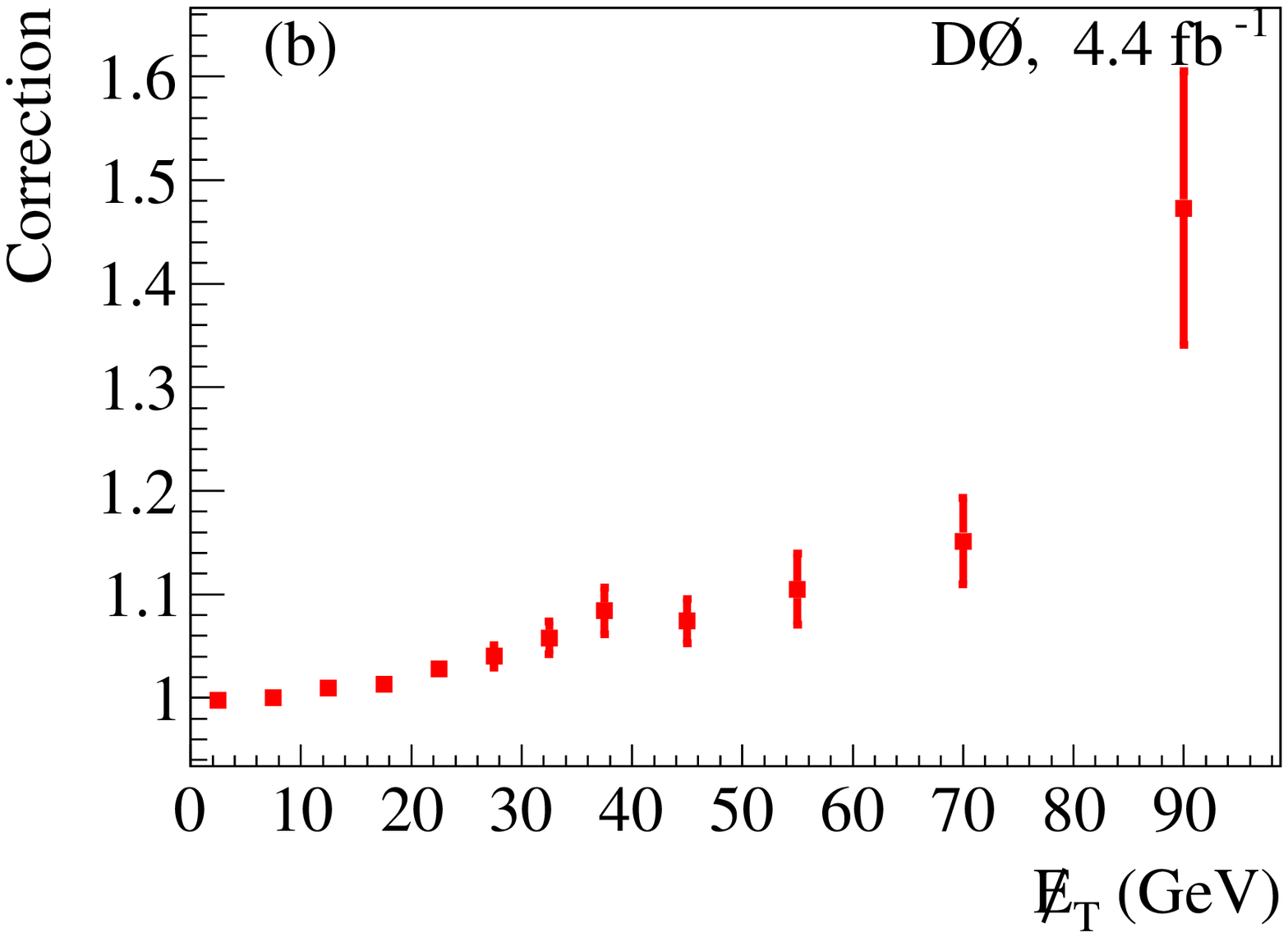}
\includegraphics[scale=0.43]{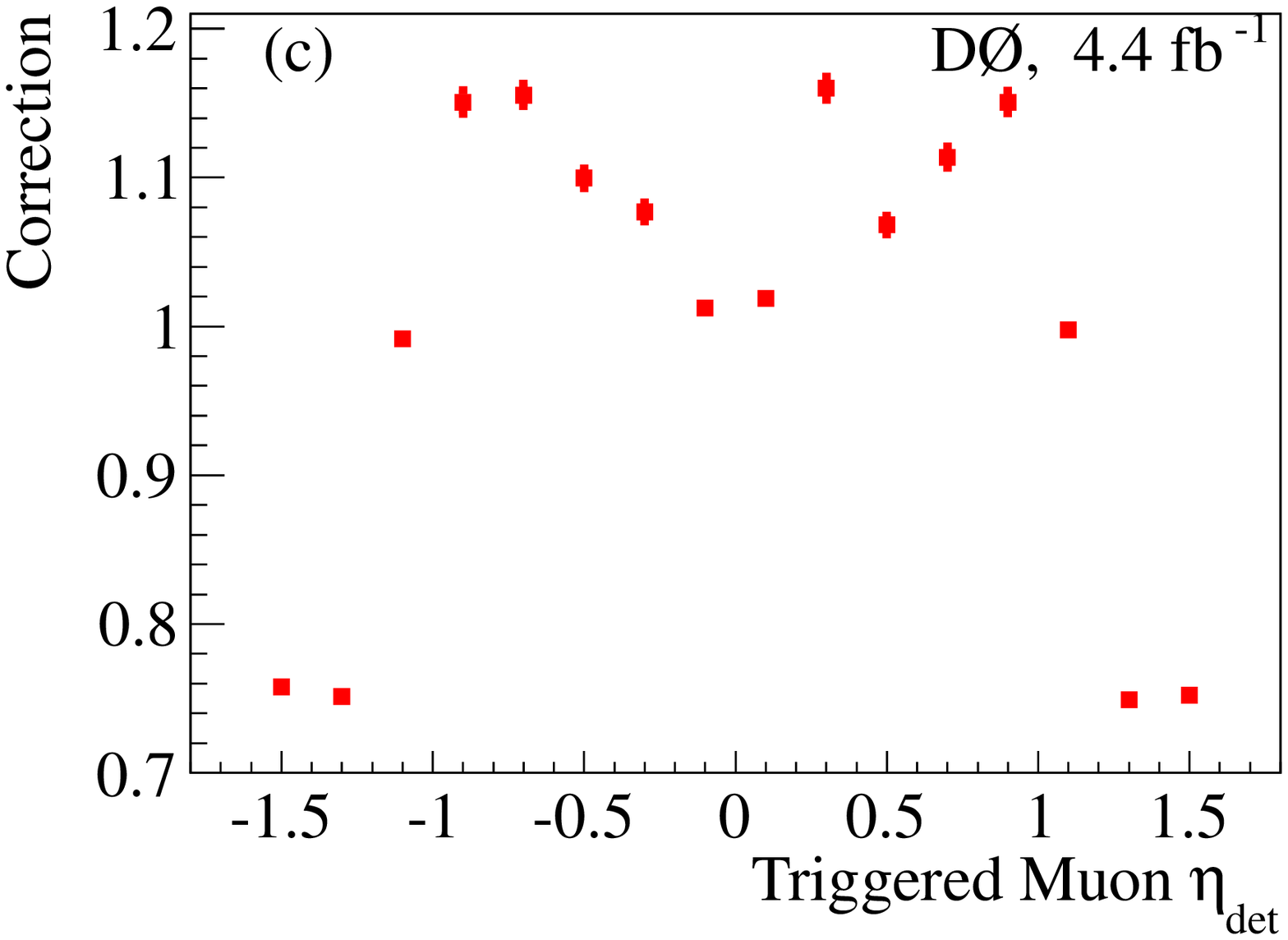}
\includegraphics[scale=0.43]{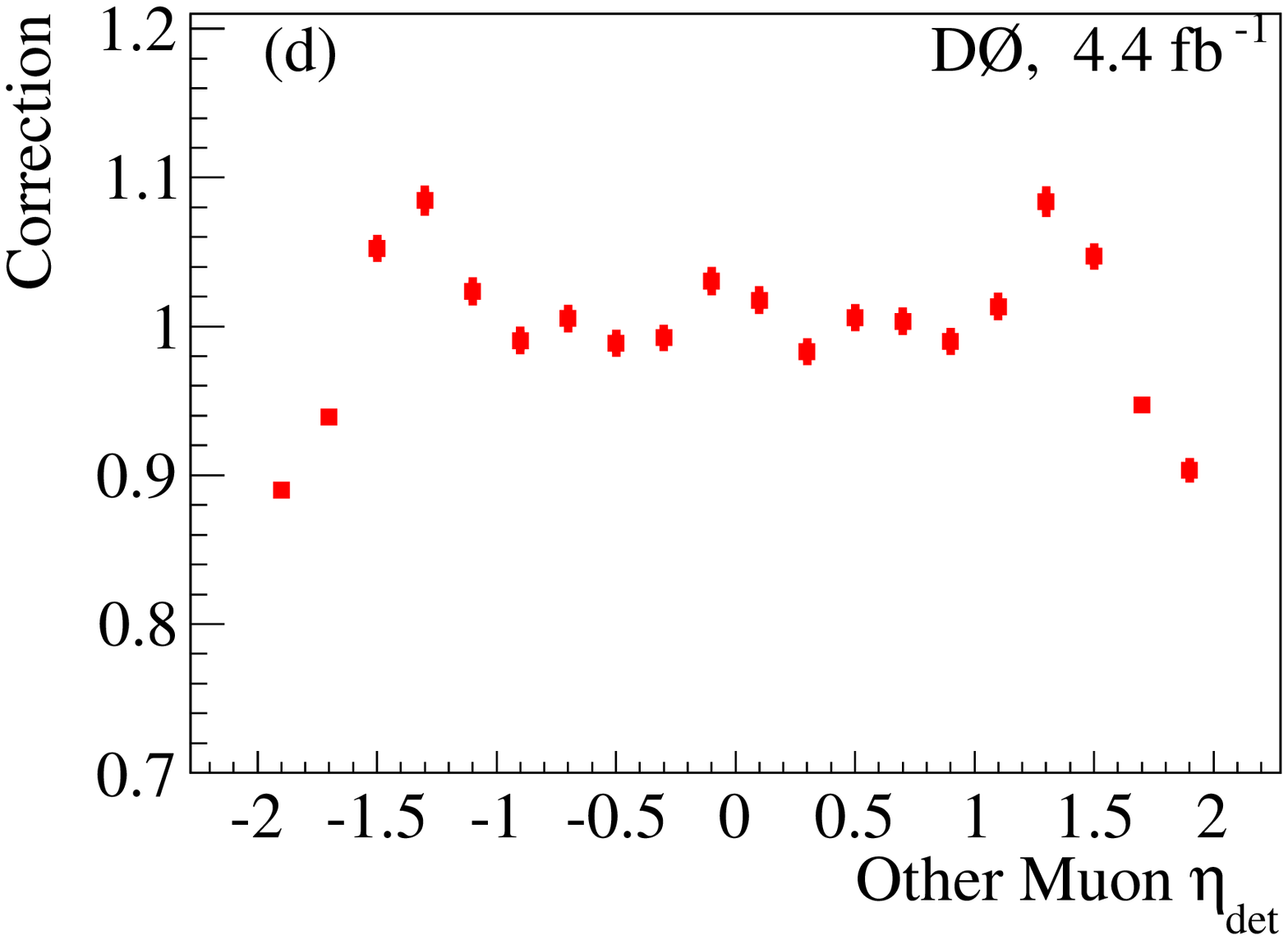}
\caption[Inclusive Trigger Correction (\mumu~Run 2b34)]
  {\label{fig:mutrig_2b34}(color online). Trigger correction factors for the \mumu\ channel in \runiib3
  as a function of (a) jet multiplicity, (b) \met,
  (c) $\etadet$ of the triggered muon, and 
  (d) $\etadet$ of the other muon.
  The correction applied to the single muon trigger is the product
  of all four components.}
\end{figure*}

In the \mumu~and \mumutrk~channels we accept events that 
satisfy any trigger requirement, although most were recorded
using triggers that contain muon selection terms. To
correctly model the efficiency of the inclusive set of triggers
for these events, we develop a
correction based on a reference data sample, for which we demand that the
leading muon with $|\etadet|<1.5$ satisfies one of the triggers that
require a single muon.
We confirm that this reference sample is well
modeled by the MC when we apply the corresponding trigger
efficiencies.  We then derive a normalization correction factor equal to the ratio of the
number of events in the inclusively triggered sample to 
the single-muon trigger sample in bins of the number of jets in the event.
Shape-only correction factors are determined in zero-jet events in bins of 
$\eta$ of each of the two muons and the transverse energy imbalance (\met). 
To account for changes
in the trigger conditions, and hence efficiency, with time, we derive separate corrections
for each of the four data-taking periods.  Figure \ref{fig:mutrig_2b34} 
shows as an example the correction factors for the $\mumu$ channel in \runiib3.

After imposing data quality requirements, the integrated luminosity recorded
by these triggers is 9.7~fb$^{-1}$ in each channel.

\subsection{Offline Event Selection}\label{sec:offlevtsel}

The event selection in all channels requires a $\ppbar$
interaction vertex (PV) that has at least three associated tracks, and
is located within $\pm$60 cm of the center of the detector along the
beam direction. In the dimuon channel ($\mumu$) we select events 
with at least two muons identified in the muon system, matched to 
central tracks with transverse momenta $\pt >10~\gev$ and $|\etadet|~<~2$.
At least one muon must have $|\etadet|~<~1.5$ and $\pt > 15~\gev$.
The two muons must also have opposite charges.
The distance between the PV and each of the muon tracks
along the $z$ axis, $d_{\mathrm{PV}}^z$, must be less than 1 cm.  The distance
of closest approach of each muon track to the PV in the
plane transverse to the beam direction, $d_{\mathrm{PV}}$, must be less than
$0.04~$cm for tracks with at least one hit in the SMT.  Muon tracks
without any SMT hits must have $d_{\mathrm{PV}}~<~0.2$~cm, and the momentum
resolution of these tracks is improved through a constraint to the position 
of the PV in the transverse plane.

At least one muon must be separated from all jets (see below) by
$\dr~=~\sqrt{(\Delta\eta)^2 + (\Delta\phi)^2} > 0.5$,
where the jets must have $\pt~>~20~\gev$ and $|\etadet|~<~2.5$.
If only one muon satisfies this criterion, we also require that the
ratio ($R_{\mathrm{TRK}}$) of the vector sum of the 
transverse momenta of all tracks 
in a cone of $\dr~<~0.5$ around that muon to its $\pt$ satisfy 
$R_{\mathrm{TRK}}~<~0.2$,
and that the ratio ($R_{\mathrm{CAL}}$) of the transverse energy
deposited in the calorimeter in a hollow cone with $0.1~<~\dr~<~0.4$
around that muon to its $\pt$ satisfy $R_{\mathrm{CAL}}~<~0.2$.
If both muons are separated from
jets, then only the leading muon must satisfy the additional 
track and calorimeter isolation requirements described above.  
To reduce contamination from
cosmic rays, the muon tracks must not be back-to-back in
$\eta$ and $\phi$. 

The \mumutrk~channel is designed to recover dimuon events in which one
muon is not identified in the muon system, primarily because of gaps
in the muon system coverage. In this channel we require
the presence of exactly one muon with $|\etadet|~<~1.5$ and
$\pt~>~15~\gev$ that must satisfy the same tracker and calorimeter 
isolation requirements used
for the \mumu~channel. We also require the presence of an isolated
track with $|\etadet| < 2$ and $\pt > 20~\gev$, separated from the
muon by $\dr~>~0.1$.  This track-only muon (\mutrk) must have at least one SMT
hit, $d_{\mathrm{PV}} < 0.02$~cm, and $d_{\mathrm{PV}}^z < 1$~cm.  
It must be separated from all jets 
having $\pt~>15~\gev$ and $|\etadet| < 2.5$ by $\dr~>~0.5$.  
It must also satisfy the same tracker and calorimeter isolation
requirements as the first muon. The muons
must have opposite charges.  To ensure that the
\mumu~and \mumutrk~selections do not overlap, we reject events that
contain any additional muons with $|\etadet|~<~2$ and $\pt~>10~\gev$.
For the small fraction of events (approximately 0.1\%) with more than
one track passing these requirements, the track whose invariant mass
with the muon is closest to the $Z$ boson mass is
chosen.

In the dielectron ($\ee$) channel we select events with at least two
electrons with $\pt~>~15~\gev$ that pass selection requirements based
on the energy deposition and shower shape in the calorimeter and the
CPS.  Electrons are accepted in the CC with $|\etadet| < 1.1$ and in the EC
with $1.5 < |\etadet| < 2.5$, but at least one of the electrons must be 
identified in the CC. Electrons are selected from
EM clusters reconstructed within a cone of radius $\mathcal{R}~=~0.2$ 
and satisfying the following requirements: (i)~at least 90\%
(97\%) of the cluster energy is deposited in the EM calorimeter of the
CC (EC); (ii)~the calorimeter isolation variable $I =
[E_{\mathrm{tot}}^{0.4}-E_{\mathrm{EM}}^{0.2}]/E_{\mathrm{EM}}^{0.2}$ 
is less than 0.09 (0.05) in
the CC (EC), where $E_{\mathrm{tot}}^{0.4}$ is the total energy in a cone of
radius $\mathcal{R}~=~0.4$ and $E_{\mathrm{EM}}^{0.2}$ is the EM energy in
a cone of radius $\mathcal{R}~=~0.2$; (iii)~the scalar sum of
the transverse momenta of all tracks in a hollow cone of
$0.05<\Delta\mathcal{R}<0.4$ around the electron is less than 
4~\gev\ in the CC, and less than or equal to $0$ to $2$~\gev\
in the EC, depending on $\etadet$ of the electron; 
(iv)~the output of an artificial neural
network -- which combines the energy deposition in
the first EM layer, track isolation, and energy deposition in the CPS -- is
consistent with that expected from an electron; (v)~CC electrons must match central
tracks or a set of hits in the tracker consistent with that of an
electron trajectory; and (vi)~for EC electrons the energy-weighted cluster
width in the third EM layer must be consistent with that expected from an EM shower.

In the $\eeicr$ channel, events must contain exactly one electron in
either the CC or EC with $\pt>15~\gev$, and a track pointing toward one of the 
ICRs, where electromagnetic object identification is compromised.
This ICR track must be matched to a 
calorimeter energy deposit with $\et>15~\gev$.  
The ICR electron must satisfy a requirement on
the output of a neural network, designed to separate electrons from jets, that combines
the track quality, the track isolation and the energy deposition in the 
scintillator detectors located in the ICR.
If the electron is found in the EC, we require that the ICR electron 
has the same rapidity sign. In both the $\ee$ and the $\eeicr$ channels,
any tracks matched to electrons must have $d_{\mathrm{PV}}^z < 1$~cm.

We reconstruct jets in the calorimeter using an iterative midpoint
cone algorithm \cite{runiicone} with a cone of $\mathcal{R}~=~0.5$.  The energies
of jets are corrected for detector response, presence of noise
and multiple \ppbar\ interactions, and energy flowing out of (into) the jet cone from 
particles produced inside (outside) the cone~\cite{d0jes}. 
In all lepton channels, jets must have
$\pt > 20~\gev$ and $|\etadet| < 2.5$.  
To reduce the impact from multiple $\ppbar$
interactions at high instantaneous luminosities, jets must contain at
least two associated tracks originating from the PV. We further require
that each of these tracks have at least one hit in the SMT. Jets meeting 
these criteria are considered ``taggable''
by the $b$-tagging algorithm described below.
However, jets separated from
electrons selected in the \ee\ and \eeicr\ channels by $\dr~<~0.5$ are
excluded from the analysis, as they are considered to be reconstructed from 
calorimeter activity generated by the electrons themselves.

We use ``inclusive''
to denote the event sample selected by requiring the presence of two
leptons with an invariant mass $40 < M_{\ell \ell} < 200~\gev$.
We use ``pretag'' to denote the sample that meets the
additional requirements of having at least two taggable jets with $\pt >
20~\gev$ and $|\etadet| < 2.5$, and $70 < M_{\ell \ell} < 110~\gev$. 

To distinguish events containing a $H \to b\bar{b}$ decay from
background processes involving light quarks ($uds$), $c$~quarks, and gluons,
jets are identified as likely to originate from the decay of $b$~quarks
($b$~tagged) if they pass ``loose'' or ``tight'' requirements on the
output of a neural network trained to separate $b$~jets from light
quark or gluon jets.
This discriminant is an improved version of the neural network $b$-tagging
discriminant described in Ref. \cite{bid}, using a larger number
of input variables related to secondary vertex 
information, as well as a more sophisticated multivariate strategy.
The $b$-jet tagging efficiency for taggable jets 
with $|\eta|<1.1$ and $\pt\approx50~\gev$ and the corresponding
misidentification rate of light jets are $72\%$ and
$7\%$ for loose $b$~tags, and $47\%$ and $0.4\%$ for tight $b$~tags.
We classify events with at least one tight and one loose $b$~tag as
double-tagged (DT).  If an event fails the DT requirement, but
contains a single tight~$b$ tag, we classify it as single-tagged (ST).
The $H\rightarrow~b\bar{b}$ candidate is composed of the two
highest-$\pt$ tagged jets in DT events, and the tagged jet plus the
highest-$\pt$ non-tagged jet in ST events.  

\section{Monte Carlo Simulation}\label{sec:mcsim}

The dominant background process for the $ZH$ search is the production 
of a $Z/\gamma^*$ boson (referred to hereafter as a $Z$ boson) in
association with jets, with the $Z$ boson decaying to leptons ($Z+$jets).
The light-flavor component ($Z+$LF) includes jets from only light
quarks or gluons.  The heavy-flavor component ($Z+$HF) includes
$Z+b\bar{b}$ and $Z+c\bar{c}$ production. 
The $Z+$LF, $Z+b\bar{b}$, and $Z+c\bar{c}$
backgrounds are generated separately, and overlaps between them are removed.
The remaining backgrounds are from
$t\bar{t}$, diboson ($WW$, $WZ$, and $ZZ$) and
multijet production with non-prompt muons or electrons, or with jets misidentified
as electrons.

We simulate $ZH$ and diboson production with
\pythia~\cite{pythia}.  In the $ZH$ samples, we consider the
$\ell^+\ell^-b{\bar b}$, $\ell^+\ell^-c{\bar c}$, and
$\ell^+\ell^-\tau^+\tau^-$ final states.  The $\ell^+\ell^-\bbbar$ final state accounts
for 99\% (97\%) of the signal yield in the DT (ST) sample.  The
$Z+$jets and $t\bar{t}$
processes are simulated with \alpgen~\cite{alpgen}.  
The events generated with \alpgen\ use \pythia\ for parton showering and
hadronization.  Because this procedure can generate additional jets,
we use the MLM matching scheme~\cite{mlm} to avoid double counting partons produced
by \alpgen\ and those subsequently added by the showering in
\pythia. All simulated
samples are generated using the {\sc CTEQ6L1}~\cite{cteq6}
leading-order parton distribution functions (PDF). 
To simulate the underlying event, consisting of all particles
not originating from the hard scatter of interest in the $\ppbar$ collision,
we use D0 Tune A~\cite{d0tunea}.

All samples are processed using a detector
simulation program based on \geant~\cite{geant}. Events from randomly
chosen beam crossings with the same instantaneous luminosity distribution
as the data are overlaid on the generated events to model the
effects of multiple $\ppbar$ interactions and detector noise.  Finally,
the simulated events are reconstructed using the same off\-line algorithms
used to process the data.

We take the cross section and branching ratios for the signal from 
Refs.~\cite{zhxsec,hbr}.  For the diboson processes, we use
next-to-leading order (NLO) cross sections from {\sc mcfm}~\cite{mcfm}.
We scale the inclusive $Z$ boson cross sections to 
next-to-NLO \cite{dyxsec}, and apply additional NLO
heavy-flavor correction factors, also calculated from {\sc mcfm}, of 1.52 and 1.67 to the
normalizations of the $Z+b\bar{b}$ and $Z+c\bar{c}$ samples, respectively. For
the $\ttbar$ background, we use the approximate next-to-NLO cross
section~\cite{ttbarxsec}.

\subsection{MC Corrections}\label{sec:modelcorr}

Jet energy calibration and resolution are corrected in
simulated events to match those measured in data, and
we smear the energies of simulated leptons to reproduce the resolution
observed in data. We apply scale factors to MC events to account
for differences in reconstruction efficiency between the data and
simulation for jets and leptons. We also correct the efficiency for 
jets to be taggable and to satisfy $b$-tagging requirements in the simulation 
to reproduce the respective efficiencies in data.

To improve the modeling of the $\pt$ distribution of the $Z$ boson,
we reweight the simulated $Z$ boson events to be consistent with
the observed $Z$ boson $\pt$ spectrum in data \cite{zptrw}.  
In our signal samples, we correct the generator-level $\pt$ of the
$ZH$ system to match the distribution from {\sc resbos} \cite{resbos}.

Additional corrections are applied to improve agreement
between data and background simulation, using two control samples with
negligible expected signal contributions: the inclusive and pretag samples 
discussed in Section \ref{sec:offlevtsel}.
Motivated by a comparison of the {\alpgen} jet angular distributions
with those from data~\cite{zjets}
and the {\sc sherpa} generator~\cite{sherpa},
we reweight the $Z+$jets events to improve the modeling 
of the distributions of the pseudorapidities of the two jets.
The reweighting factors are calculated with the pretag sample as the ratio of the data to
the sum of the simulated $Z+$LF and $Z+$HF backgrounds after having subtracted
all other backgrounds from the data.
Since the energy resolution for jets in the ICR differs from the resolution
for jets in the CC or EC, we exclude jets with $1.0<|\etadet|<1.6$ when determining
these reweighting factors and develop a separate reweighting for jets in the
ICR. These corrections are parametrized in $\eta$ and display
variations of up to 20\%. After applying the corrections, we
renormalize to the yield from {\alpgen}.

\section{Multijet Background}\label{sec:mjbkg}
The multijet backgrounds are estimated from control samples in the data.
The selection criteria in each channel are nearly the same as for the 
inclusive sample, with the differences described below.
For the $\ee$ channel, the electron 
isolation and shower shape requirements are reversed. The multijet sample
in the $\ee$ channel suffers from a bias due to trigger conditions towards tighter
electron identification criteria. The multijet background is therefore
reweighted to correct for this bias, and a systematic uncertainty is
assigned to account for the uncertainty in the fit that calculates the correction.
For the $\eeicr$ channel, the electron in the ICR must fail the neural
network output requirement described in Section~\ref{sec:offlevtsel}.
In the \mumu\ channel, a multijet event must contain a $Z$ boson candidate
that fails any of the isolation requirements.
The two muons forming the $Z$ boson candidate must have the same charge.
In the \mumutrk\ channel, the multijet sample must pass all selection
criteria, except that the two muons should have the same charge.
These samples are used to define templates that are normalized by the procedure
described in Section~\ref{sec:normalization}. The multijet
background comprises approximately 7\% of both the ST and DT samples
after normalization.

\section{Normalization Procedure}
\label{sec:normalization}

We adjust the normalization of the multijet background and all
simulated background and signal samples using a simultaneous template fit of the dilepton
mass ($\mll$) distributions in each channel, data-taking period, and
jet multiplicity bin ($n_{\mathrm{jet}}=$ 0, 1, or $\ge$ 2).  This improves
the accuracy of the background model and reduces the impact of some
systematic uncertainties. The inclusive event sample is used
so that we fit to the inclusive $Z$ boson cross section,
which is known with much greater accuracy than the $Z$ + 2 jets cross section.
The fit minimizes the $\chi^2$:
\begin{equation}
\chi^2 =
\sum_{i,j,m}
\frac{\left(D_m^{ij}-\alpha^{ij} \cdot Q_m^{ij}
-
k_{\epsilon}^{i} \cdot \left(
k_{Z}^{j} \cdot Z_m^{ij} + O_m^{ij}\right)\right)^2}{D_m^{ij}},
\label{eq:Norm_chi2}
\end{equation}
where $m$ runs over the bins of $\mll$, $j$ runs over $n_{\mathrm{jet}}$,
and $i$ indicates the channel.  In the
normalization fit we divide the \ee~channel into two sub-channels:
CC-CC, in which both electrons are in the CC, and CC-EC in which one
electron is in the CC and one electron is in the EC. We
also divide each channel into the four data-taking periods (\runiia,
\runiib1, \runiib2, and \runiib3).

The number of data events are $D_m^{ij}$, and the fit adjusts the normalization of
$Q_m^{ij}$, the multijet sample, $Z_m^{ij}$, the simulated $Z$ boson
(including $Z+\bbbar$ and $Z+\ccbar$) sample, and $O_m^{ij}$,
all other simulated samples.  The fit parameters are 
the multijet scale factors $\alpha^{ij}$ that apply to $Q_m^{ij}$,
the combined luminosity and efficiency scale factors
$k_{\epsilon}^{i}$ for channel $i$ that are applied to
$Z_m^{ij}$ and $O_m^{ij}$, and the $Z$ boson cross section
scale factors $k_{Z}^{j}$ that apply to $Z_m^{ij}$.  The parameters $\alpha^{ij}$
are fixed to unity for the $\mumutrk$ channel, as the only criterion
in this channel for multijet selection is that the two muons
fail the opposite-charge requirement, and a jet
is equally likely to fake a $\mu^{+}$ or a $\mu^{-}$.  We also fix 
$k_Z^0 = 1$, approximately equivalent to assuming that the inclusive $Z$ boson
cross section is known exactly. In the assessment of 
the systematic uncertainty from the background fit, $k_Z^0$ is varied
within the uncertainty on the inclusive $Z$ boson cross section~\cite{zhxsec}.

The $k_{Z}^{j}$ parameters are expected to be independent of
data-taking periods, since these are the cross section scale factors
for $Z+$jets production and any time-dependent detector effects
should be absorbed by $k_{\epsilon}^{i}$. However, we observe
a discrepancy in $k_{Z}^{j}$ between the \runiia\ and
\runiib~data, which we attribute to differences in jet reconstruction and 
identification algorithms between the two epochs.
For this reason, we perform two
separate fits for the $k_{Z}^{j}$: 1) using the \runiia\
period only, and 2) using the
\runiib\ period only, but keeping the separation between \runiib1,
\runiib2, and \runiib3 for the other parameters.  We assign a systematic uncertainty
on the \runiia\ normalization to account for this discrepancy.
Tables~\ref{tbl:cmbnorm_2a} and~\ref{tbl:cmbnorm_2b} show the results
of the fits for \runiia\ and \runiib, respectively.  In Section
\ref{sec:syst} we discuss the uncertainties arising from the normalization
procedure.

\begin{table}[!hdpt]
\caption{Parameters from the combined normalization fit for \runiia.  
         Statistical uncertainties are less than 1\%, and systematic
         uncertainties are on the order of 5\%. There are no
         uncertainties for $\alpha^{ij}$ for the \mumutrk~channel 
         or for $k^0_Z$ since they are fixed.}
\label{tbl:cmbnorm_2a}
\begin{center}
\begin{tabular}{ccccc}
\hline
\hline
Channel & $k_{\epsilon}^i$ & $\alpha^{i0}$ & $\alpha^{i1}$ & $\alpha^{i2}$\\
\hline
\multicolumn{5}{c}{\runiia} \\
\hline
\ee CC-CC & 1.03 & 0.34 & 0.29 & 0.14\\
\ee CC-EC & 1.01 & 0.33 & 0.27 & 0.29\\
\eeicr & 1.02 & 0.12 & 0.07 & 0.01\\
\mumu & 0.93 & 1.4 & 0.46 & 0.44\\
\mumutrk & 0.91 & 1 & 1 & 1\\
\hline
\hline
\end{tabular}
\vskip 1cm
\begin{tabular}{ccc}
\hline
\hline
$k_{Z}^{0}$ &
$k_{Z}^{1}$ &
$k_{Z}^{2}$ \\
\hline
1 &
0.97 &
1.06\\
\hline
\hline
\end{tabular}
\end{center}
\end{table}

\begin{table}[!tbp]
\caption{Parameters from the combined normalization fit for \runiib.  
         Statistical uncertainties are less than 1\%, and systematic
         uncertainties are on the order of 5\%. There are no
         uncertainties for $\alpha^{ij}$ for the \mumutrk~channels 
         or for $k^0_Z$ since they are fixed.}
\label{tbl:cmbnorm_2b}
\begin{tabular}{ccccc}
\hline
\hline
Channel & $k_{\epsilon}^i$ & $\alpha^{i0}$ & $\alpha^{i1}$ & $\alpha^{i2}$\\
\hline
\multicolumn{5}{c}{\runiib1} \\
\hline
\ee CC-CC & 0.99 & 0.18 & 0.13 & 0.14\\
\ee CC-EC & 0.97 & 0.17 & 0.15 & 0.15\\
\eeicr & 0.97 & 0.11 & 0.08 & 0.10\\
\mumu & 0.97 & 1.4 & 0.44 & 0.31\\
\mumutrk & 1.04 & 1 & 1 & 1\\
\hline
\multicolumn{5}{c}{\runiib2} \\
\hline
\ee CC-CC & 1.02 & 0.10 & 0.11 & 0.14\\
\ee CC-EC & 1.01 & 0.099 & 0.11 & 0.14\\
\eeicr & 0.92 & 0.077 & 0.065 & 0.061\\
\mumu & 0.98 & 1.5 & 0.41 & 0.41\\
\mumutrk & 1.03 & 1 & 1 & 1\\
\hline
\multicolumn{5}{c}{\runiib3} \\
\hline
\ee CC-CC & 1.04 & 0.13 & 0.12 & 0.13\\
\ee CC-EC & 1.04 & 0.12 & 0.11 & 0.11\\
\eeicr & 1.01 & 0.080 & 0.071 & 0.061\\
\mumu & 0.99 & 1.2 & 0.44 & 0.35\\
\mumutrk & 1.01 & 1 & 1 & 1\\
\hline
\hline
\end{tabular}
\vskip 1cm
\begin{center}
\begin{tabular}{ccc}
\hline
\hline
$k_{Z}^{0}$ &
$k_{Z}^{1}$ &
$k_{Z}^{2}$ \\
\hline
 1 & 0.90 & 0.94\\
\hline
\hline
\end{tabular}
\end{center}
\end{table}

As a cross-check, we repeat the fit for each channel independently,
and find the results to be consistent with the simultaneous fit.
We assign the RMS of the observed deviations from the combined fit as a
systematic uncertainty. 

Table~\ref{tbl:evtall}
gives the number of events observed in the inclusive, pretag, ST and
DT samples, and the expected number of events for the different
background components and the signal (assuming $M_H~=~125$~GeV),
following all MC corrections and the normalization
fit.

\begin{table*}[hbpt]
\caption{
Expected and observed event yields for all lepton channels combined
after requiring two leptons (inclusive), after also requiring at least two taggable jets
and $70 < M_{\ell\ell} < 110~\gev$ (pretag), and after requiring exactly 
one (ST) or at least two (DT) $b$ tags. 
The $ZH$ yields are given for $M_H=125$ GeV. Expected yields are
obtained following the background normalization procedure described in
Section~\ref{sec:normalization}. The uncertainties quoted on the total background 
and signal include all systematic uncertainties and uncertainties from limited MC 
statistics.}
\begin{tabular}{lcccccccrcl}
\hline\hline
& Data
& Total Background
& MJ
& $Z+$LF
& $Z+$HF
& Diboson
& $\ttbar$
& \multicolumn{3}{c}{$ZH$} \\
\hline
 Inclusive  &$1845610$ & $1841683$        & $160746$  & $1630391$  & $46462$ & $2914$ & $1170$ & $17.3$&$\pm$&$1.1$  \\
 Pretag     &    $25849$ & $25658$       &   $1284$   &    $19253$   &   $4305$ &   $530$ &   $285$ & $9.2$&$\pm$&$0.6$ \\
 ST         &       $886$ &       $824\pm102$ &      $54$   &        $60$   &     $600$ &    $33$ &    $77$ & $2.5$&$\pm$&$0.2$  \\
 DT         &       $373$ &       $366\pm39\phantom{0}$  &      $25.7$ &         $3.5$ &     $219$ &    $19$ &    $99$ & $2.9$&$\pm$&$0.2$  \\
\hline\hline
\end{tabular}
\label{tbl:evtall}
\end{table*}

\begin{figure}[htbp]
\centering
\includegraphics[height=0.24\textheight]{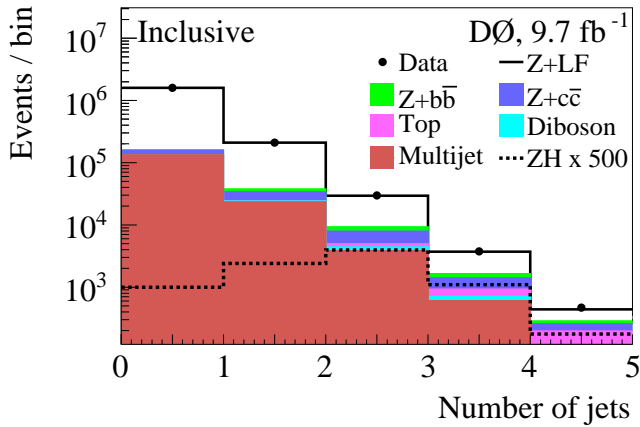}
\caption{\label{fig:njets} 
  (color online). Jet multiplicity distribution in the inclusive sample, summed over all
  lepton channels, along with the background expectation. The signal distribution for $M_H=$ 125 GeV is scaled by a factor of 500.}
\end{figure}

\begin{figure*}[p]
\centering
\begin{tabular}{cc}
\includegraphics[height=0.24\textheight]{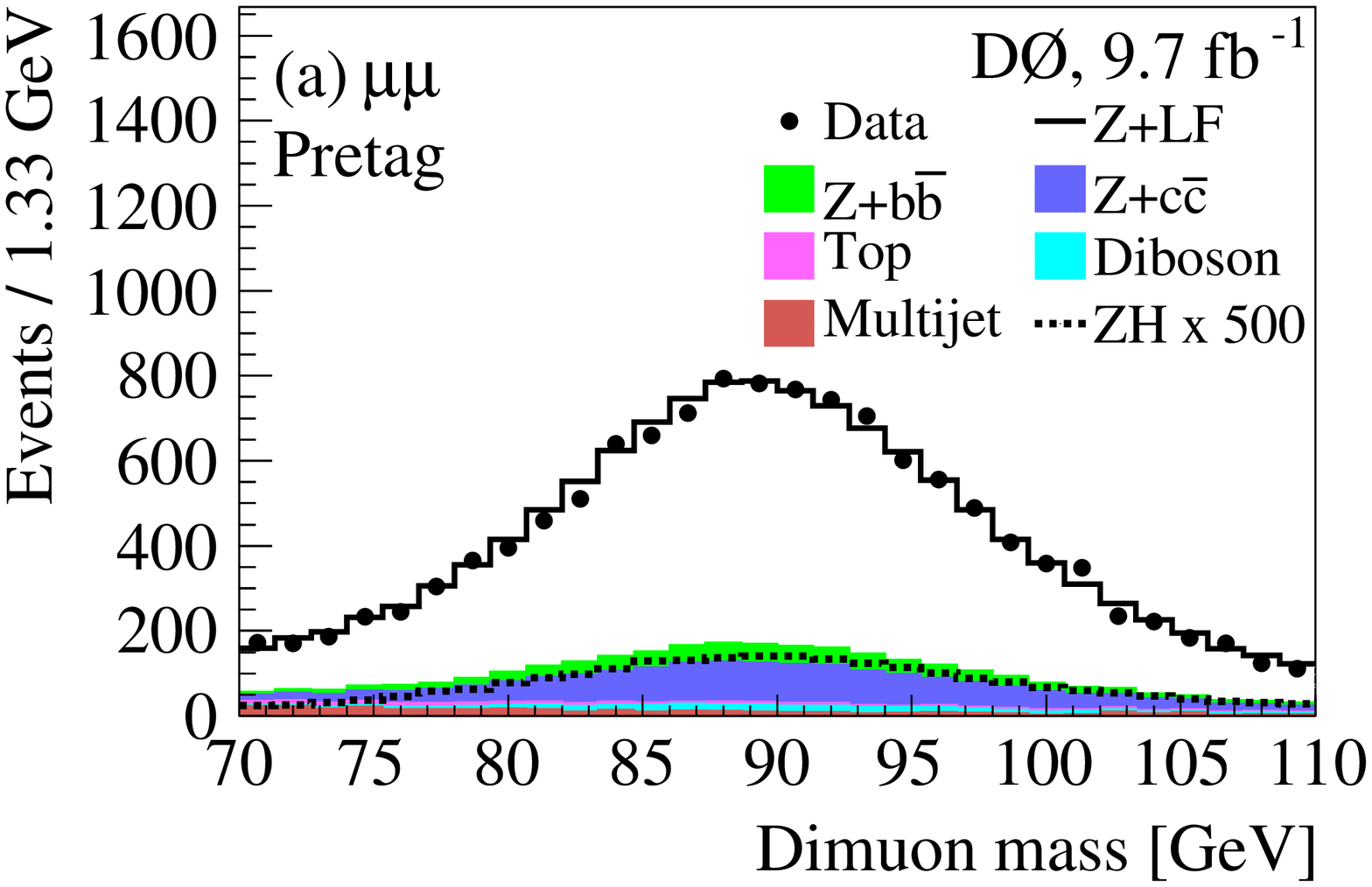} & 
\includegraphics[height=0.24\textheight]{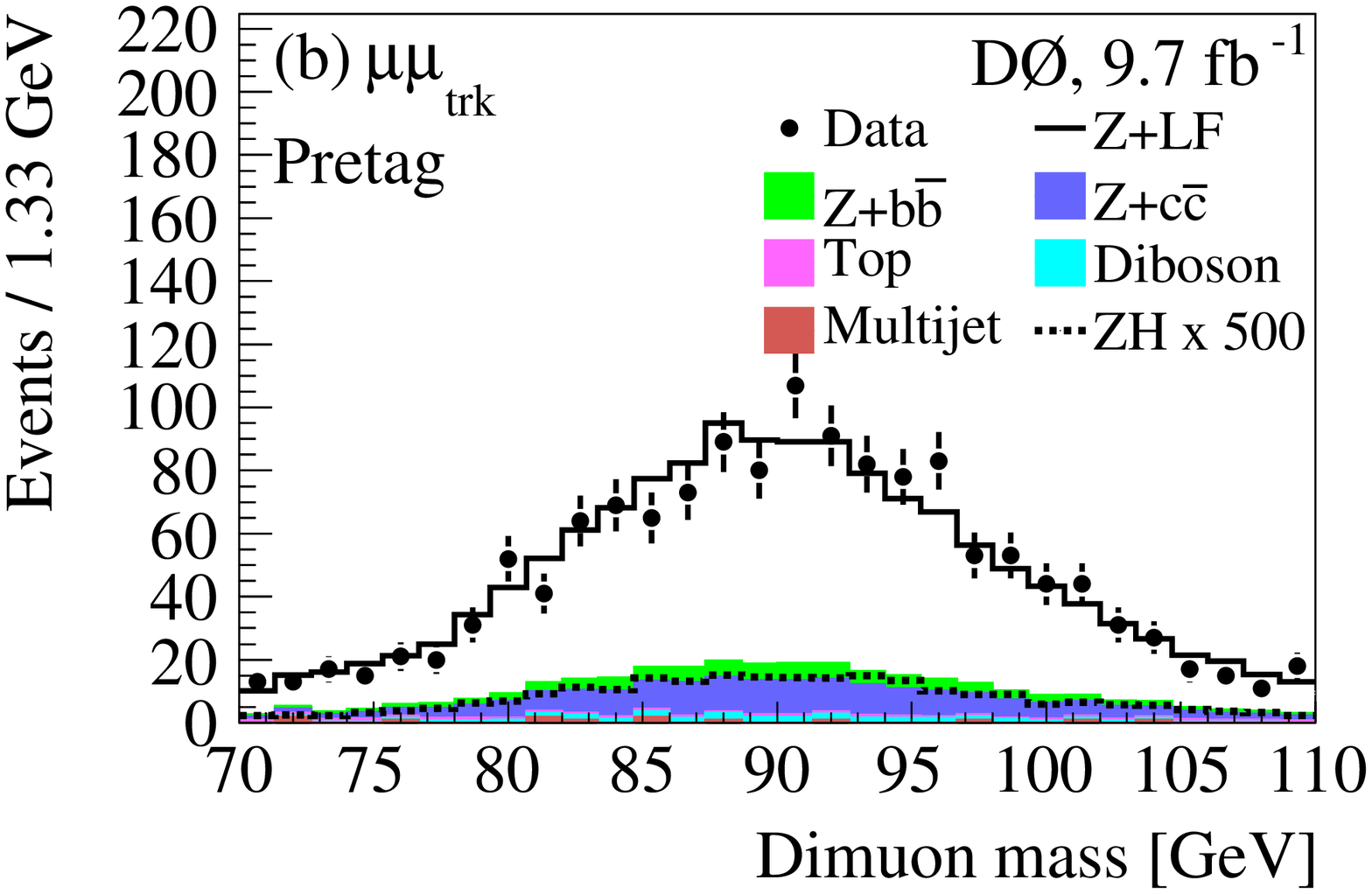} \\ 
\includegraphics[height=0.24\textheight]{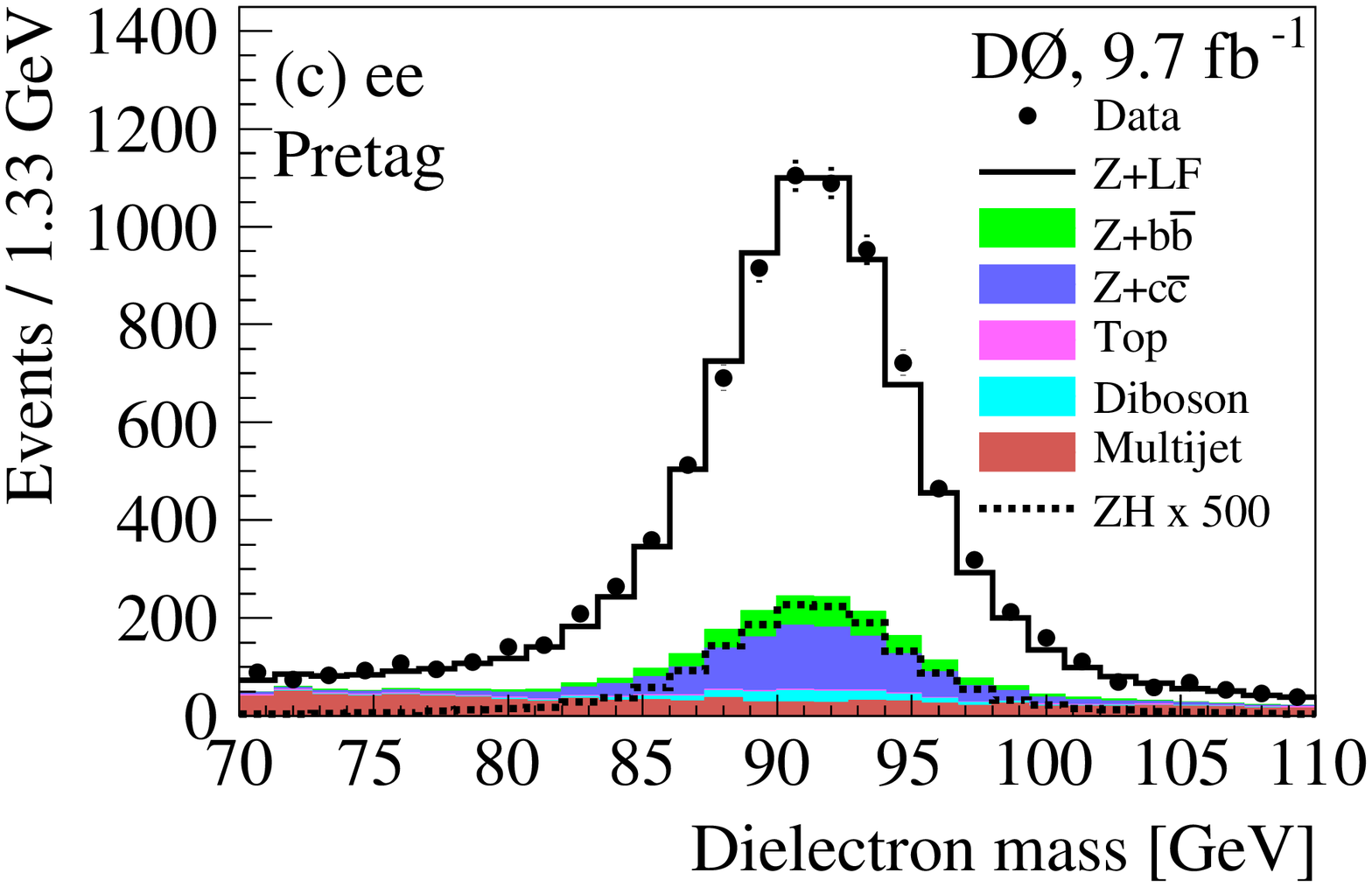} &
\includegraphics[height=0.24\textheight]{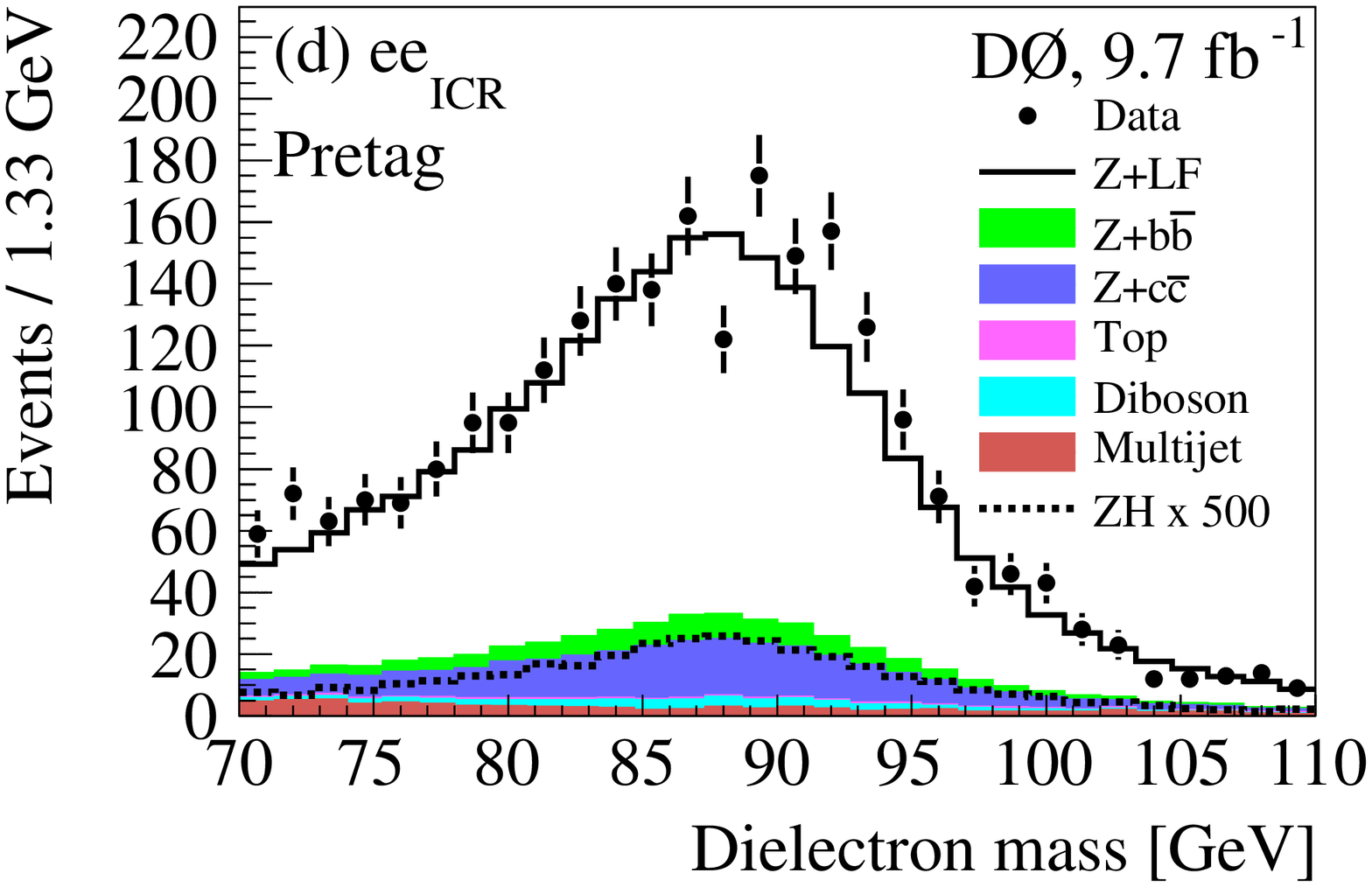} \\
\end{tabular} 
\caption{\label{fig:mll} 
(color online). The dilepton mass spectra, along with the background expectation, for the (a) \mumu, (b) \mumutrk, (c) \ee~and 
(d) \eeicr~ channels in the pretag sample.
The signal distributions for $M_H=$ 125 GeV are scaled by a factor of 500.
}
\end{figure*}

\begin{figure}[h]
\centering
\includegraphics[scale=0.43]{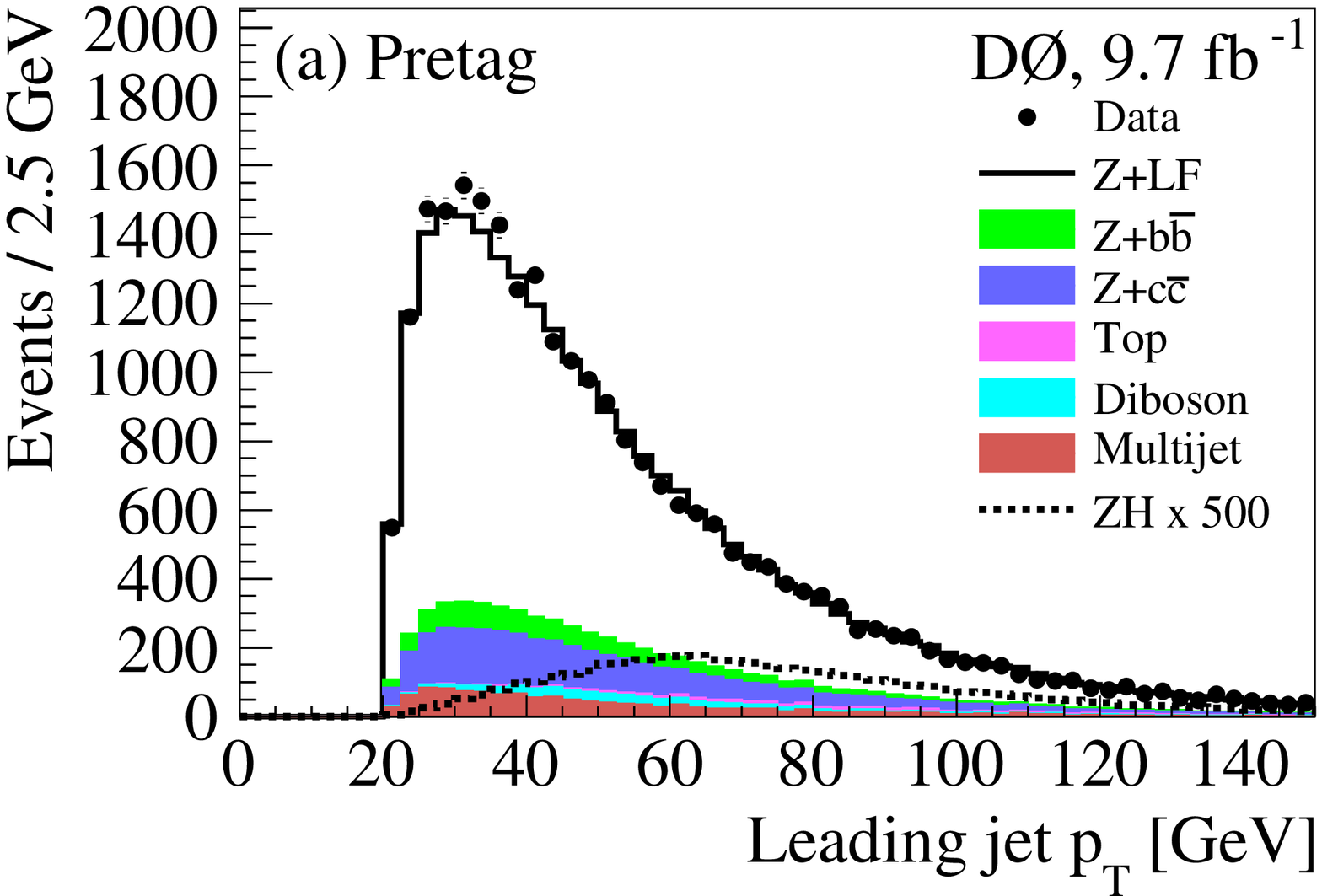}\\
\includegraphics[scale=0.43]{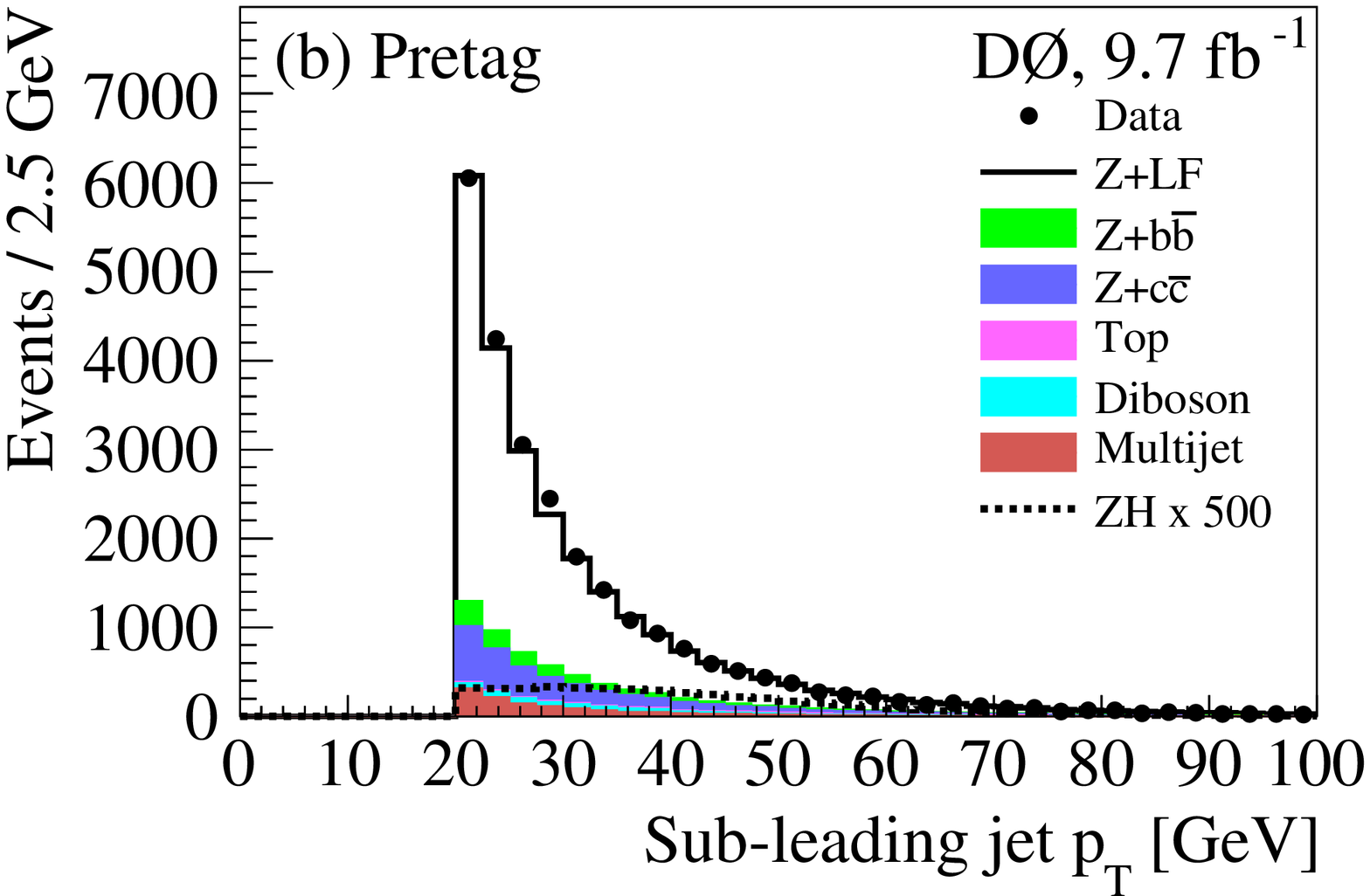}
\caption{\label{fig:jetpt_pretag} 
(color online). The $p_T$ spectra of the  (a) leading and (b) sub-leading jets, 
along with the background expectations, summed over
all lepton channels in the pretag
sample. The signal distributions, for $M_H=$ 125 GeV, are scaled by a factor of 500.}
\end{figure}

\begin{figure}[h]
\centering
\includegraphics[height=0.24\textheight]{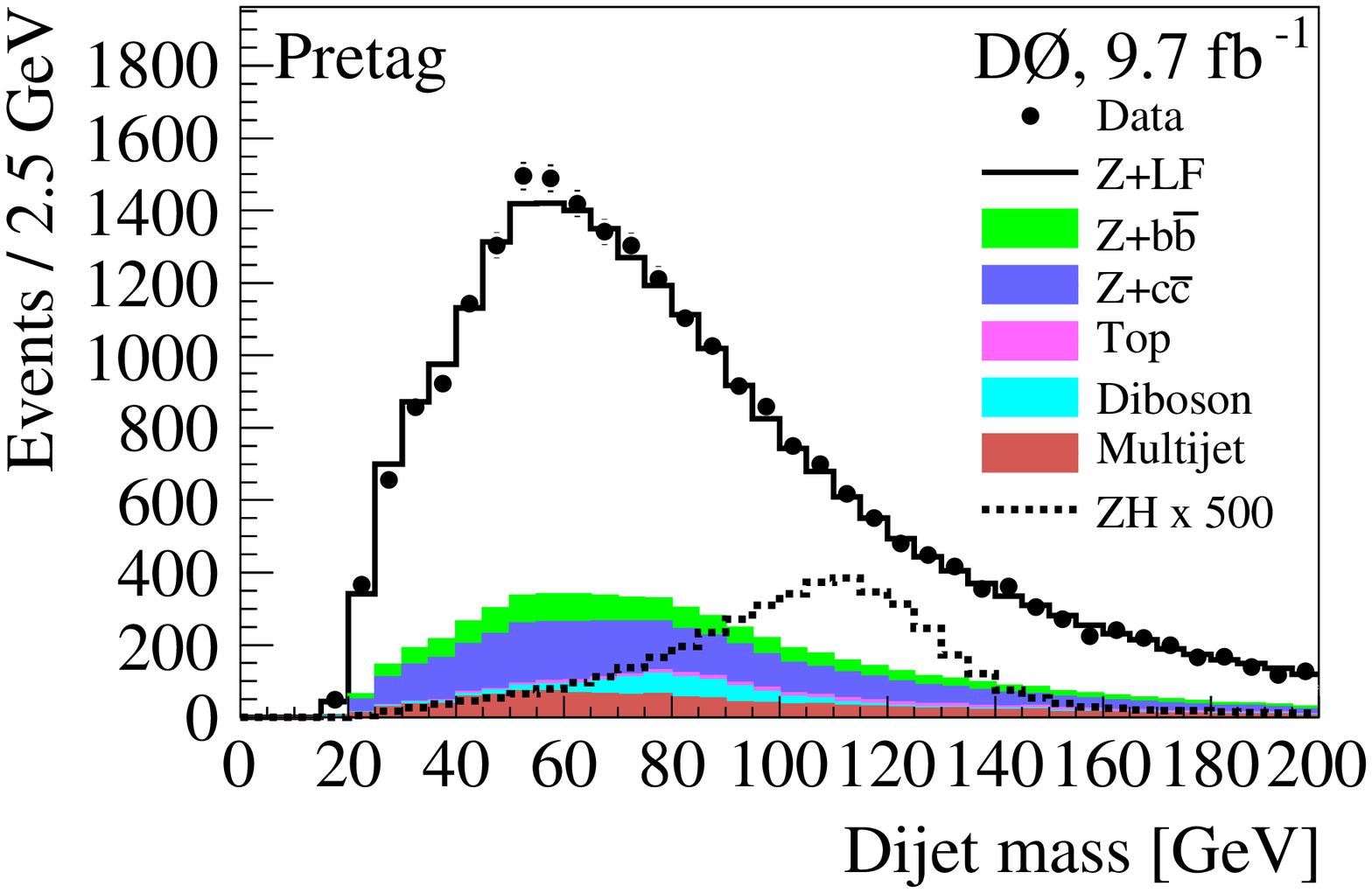}
\caption{\label{fig:mbb_pretag} 
  (color online). Distribution of
  the dijet invariant mass, along with the background expectation, summed over all lepton channels in the pretag sample. 
  The signal distribution, for $M_H=$ 125 GeV, is scaled by a factor of 500.}
\end{figure}

Figure~\ref{fig:njets} shows the jet
multiplicity distribution in the inclusive sample for the combination 
of all channels.
The dimuon and dielectron mass spectra in the pretag sample
are shown in Fig. \ref{fig:mll}.  
In Figs.~\ref{fig:jetpt_pretag} and \ref{fig:mbb_pretag},
we show distributions of the transverse momenta of the two jets with
the highest $\pt$ and the invariant mass of the dijet system constructed
from those two jets. In all plots, data points are shown with error bars
that reflect statistical uncertainty only, and discrepancies in data-MC 
agreement are within the systematic uncertainties described in 
Sec.~\ref{sec:syst}.

\section{Kinematic Fit}\label{sec:kinfit}

We use a kinematic fit to improve the resolution of the dijet
invariant mass.  The fit varies the energies and angles of the two
leptons from the $Z$ boson candidate, and of the two jets that form
the Higgs boson candidate (and of a third jet, if present) within
their experimental resolutions, subject to three constraints: the
reconstructed dilepton mass must be consistent with the $Z$ boson mass and the
$x$ and $y$ components of the vector sum of the transverse momenta of the leptons
and jets must be consistent with zero.

The fit minimizes a negative log likelihood function:
\begin{eqnarray}
-\ln L_{\mathrm{fit}}=-\sum_{i} \ln f_i(y_i^{\mathrm{obs}}, y_i^{\mathrm{pred}}) - \sum_{j} \ln C_j,
\end{eqnarray}
where $C_j$ ($j=$ 1,2,3) are the probability densities for kinematic
constraints, and $f_i$ is the probability density (transfer function) for 
observable $y_i^{\mathrm{obs}}$ whose predicted value is $y_i^{\mathrm{pred}}$.  The fit contains
twelve independent observables for events with two jets: four particles $\times$
three variables ($E$ or $1/\pt$, $\eta$ and $\phi$).  For events with three
jets, there are fifteen observables.

The probability density for the $Z$ boson mass constraint is a Breit-Wigner function
using the values for the mass and width of the $Z$ boson from
Ref.~\cite{pdg}.
The constraints on the total transverse momentum components are
Gaussian distributions with a mean of zero and a width of 7~GeV, as determined
from the simulated $ZH$ samples.

We use Gaussian transfer functions for all observables except the
energies of the jets.  In this case we use three sets of transfer
functions, derived from MC studies for: (i)~jets that
originate from a $b$ quark and do not contain a muon, (ii)~jets that originate from a
$b$ quark and contain a muon, and (iii)~jets that originate from
a light quark or gluon.  For the jets that form the Higgs boson candidate
we use one of the $b$ quark transfer functions, depending on whether they
contain a reconstructed muon.  For the third jet, if present, we use the 
light-quark transfer function.

\begin{figure}[!htbp]\begin{center}
 \includegraphics[height=0.24\textheight]{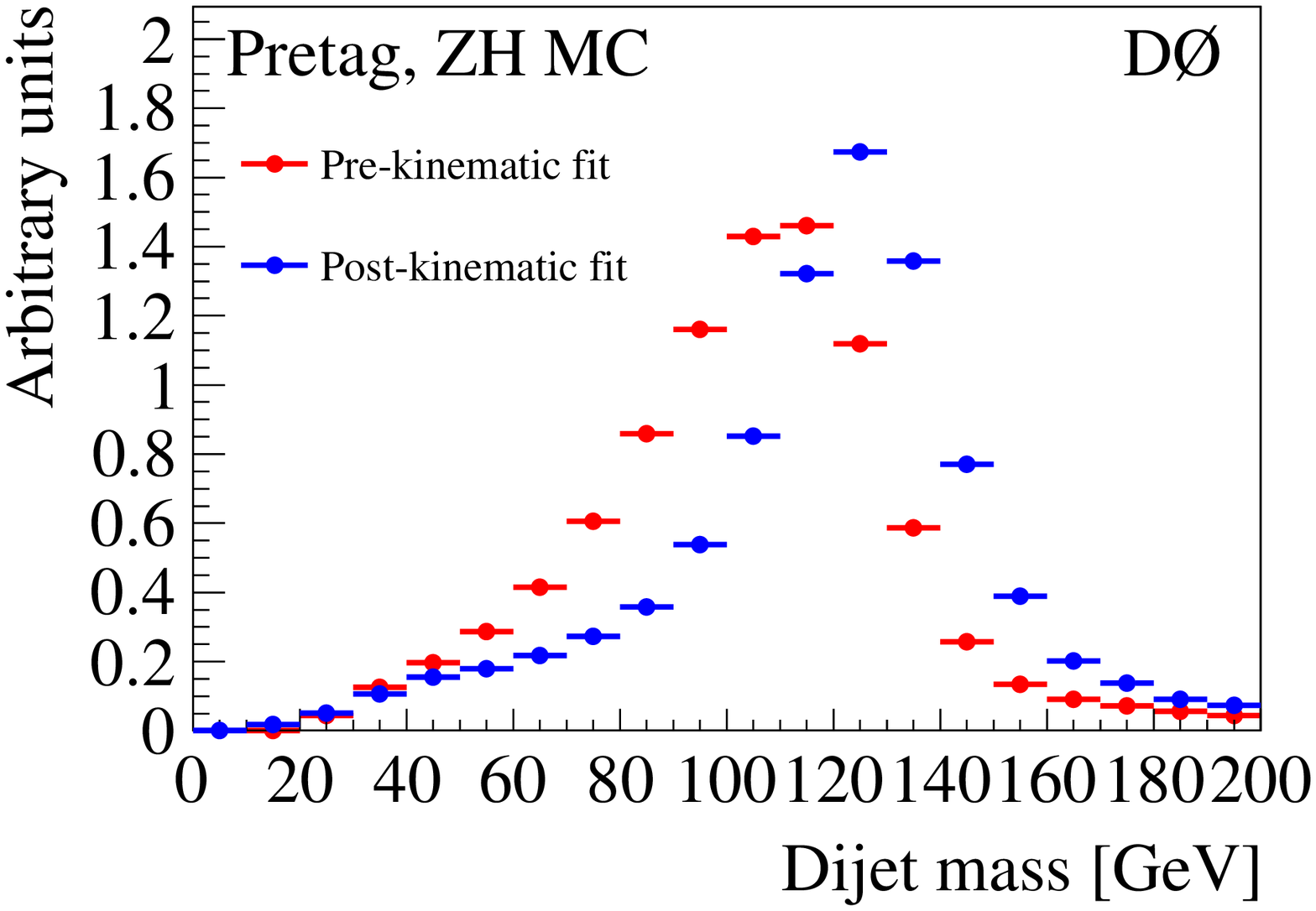}
\caption{(color online). The dijet invariant mass for the simulated $ZH$ signal,
at $M_H=$~125~GeV, summed over all lepton channels in the pretag sample,
shown before and after the kinematic fit.}\label{fig:mbb_fit}
\end{center}
\end{figure}

The kinematic fit improves the dijet mass resolution
by 10$-$15\%, depending on $M_H$. The resolution for
$M_H$~=~125 GeV is approximately 15~GeV (i.e.\ 12\%) after the fit.
Distributions of the dijet invariant mass spectra,
before and after adjustment by the kinematic fit, are shown in
Fig.~\ref{fig:mbb_fit}.

\section{Multivariate Analysis}\label{sec:mva}

We use a two-step multivariate analysis strategy based on random
forest discriminants (RF), an ensemble classifier that consists of many decision
trees~\cite{dtree}, as implemented in the {\sc tmva}
software package~\cite{tmva}, to improve the discrimination of signal from
background.  In a first step, we train a dedicated RF ($\ttbar$~RF)
that considers $\ttbar$ as the only background and $ZH$ as the signal.
This approach takes advantage of the distinctive signature of the
$\ttbar$ background, for instance the presence of large \met.
In a second step, we use the $t\bar{t}$~RF output to
define two independent regions: a $t\bar{t}$-enriched region
 and a $t\bar{t}$-depleted region.
In each region, we train a global RF to separate the $ZH$
signal from all backgrounds.  In both steps we consider ST and DT
events separately and train the discriminants for each value
of the tested Higgs boson mass in the range $90 < M_H < 150$~GeV 
in steps of 5~GeV. Compared to the result
described in Ref.~\cite{pubzh}, this two-step strategy improves sensitivity
to the signal by 5--10\%, depending on $M_H$.

\begin{figure*}[tb]\centering
\begin{tabular}{cc}
\includegraphics[height=0.24\textheight]{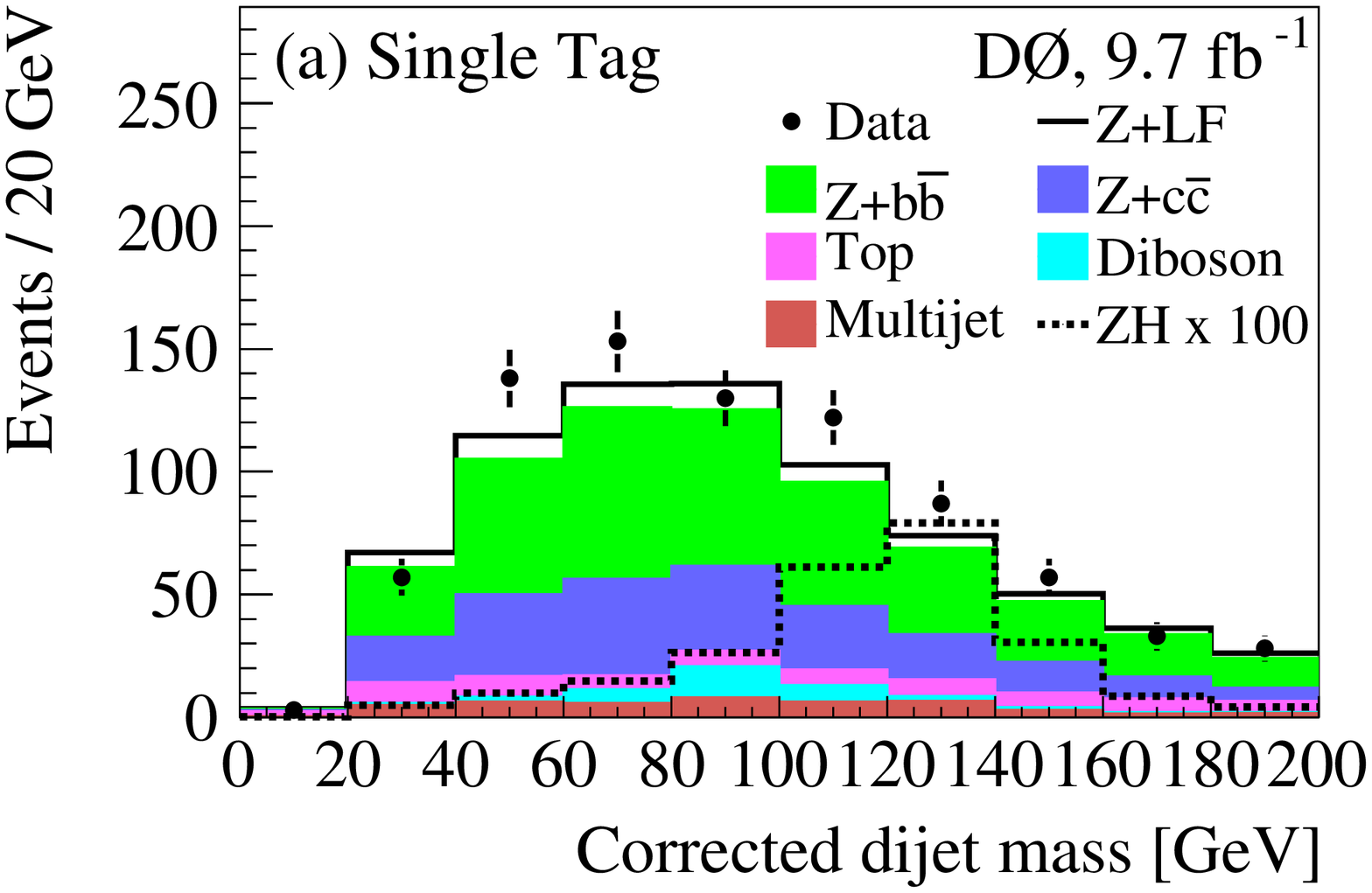}&
\includegraphics[height=0.24\textheight]{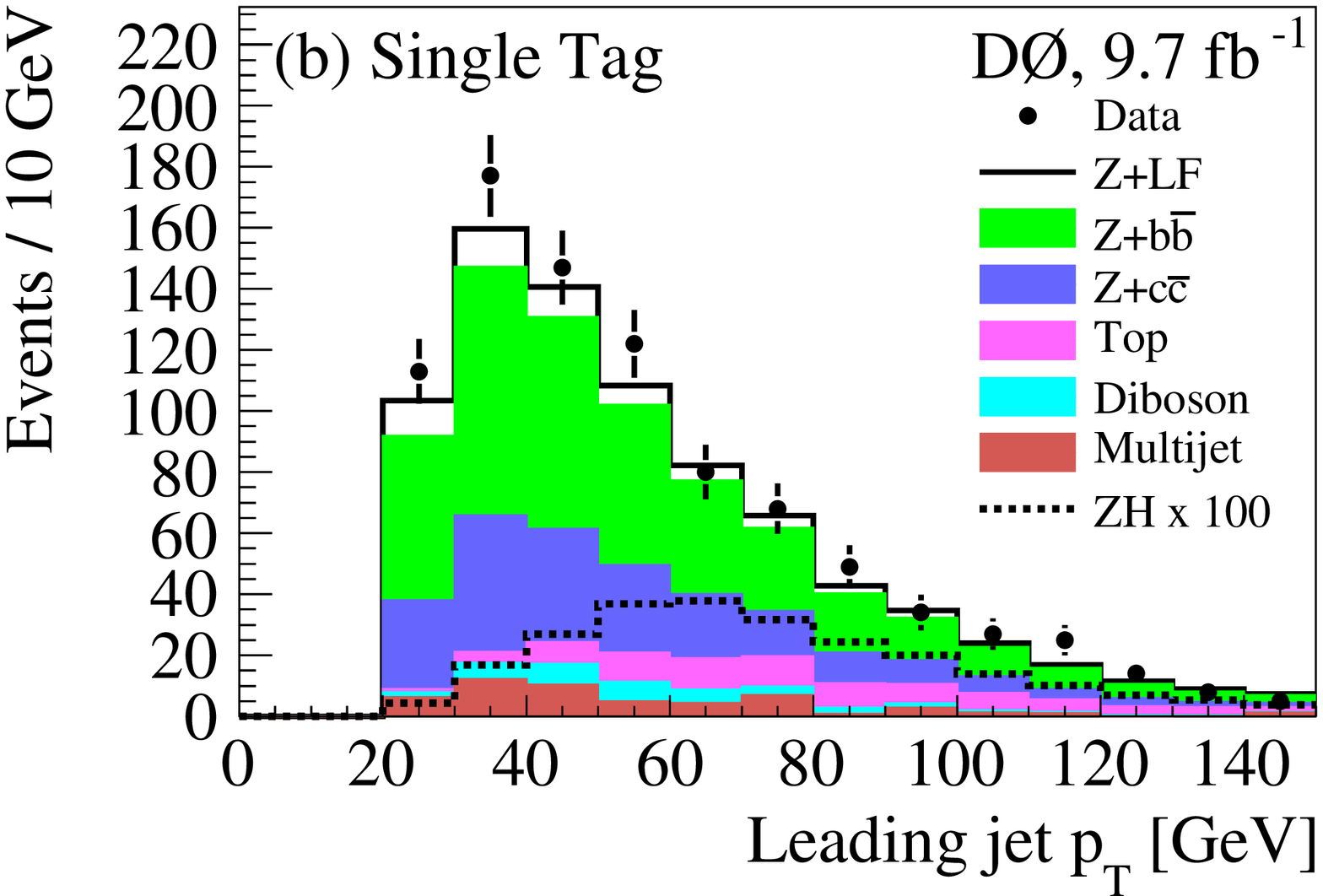}\\
\includegraphics[height=0.24\textheight]{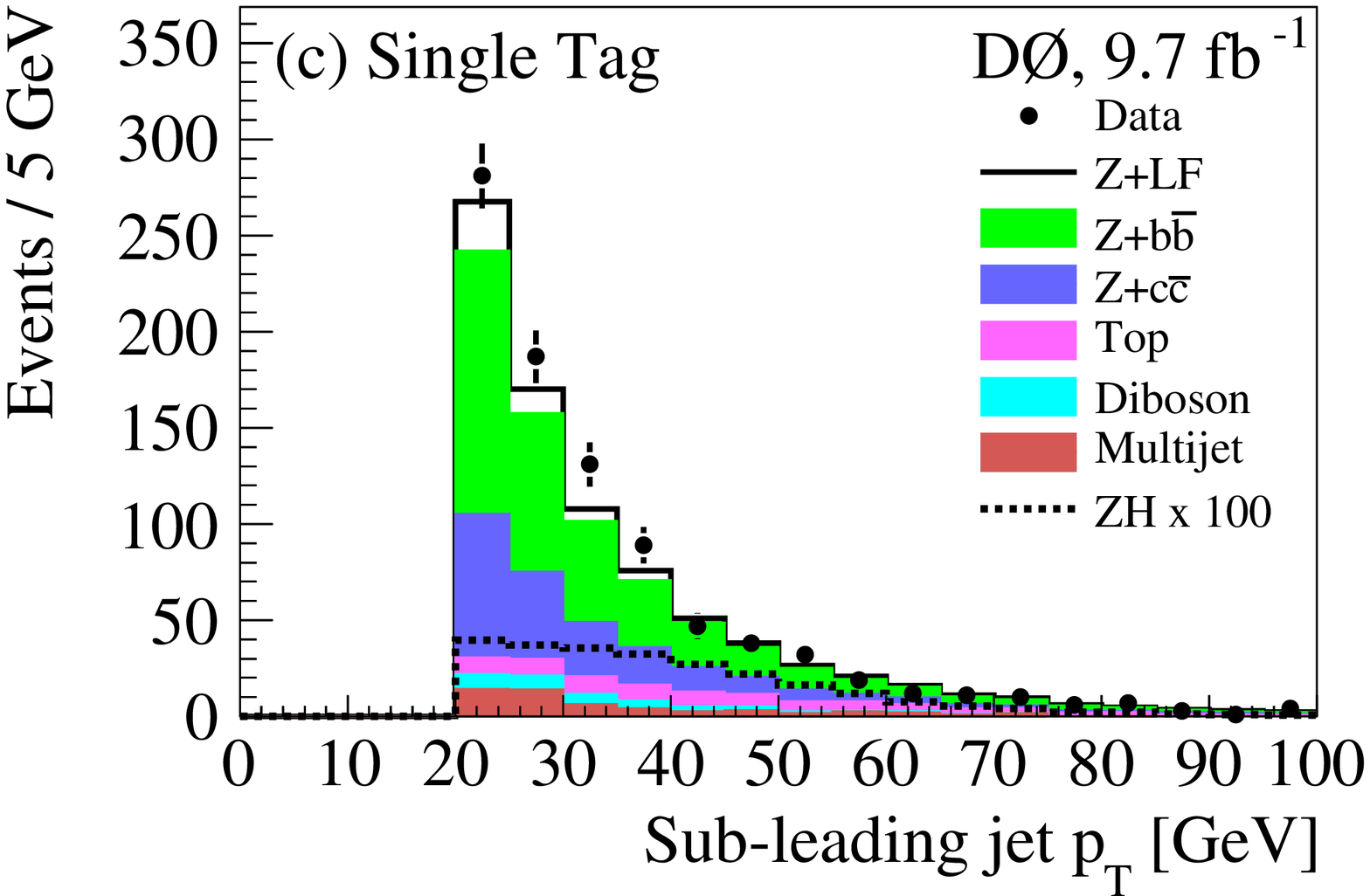}&
\includegraphics[height=0.24\textheight]{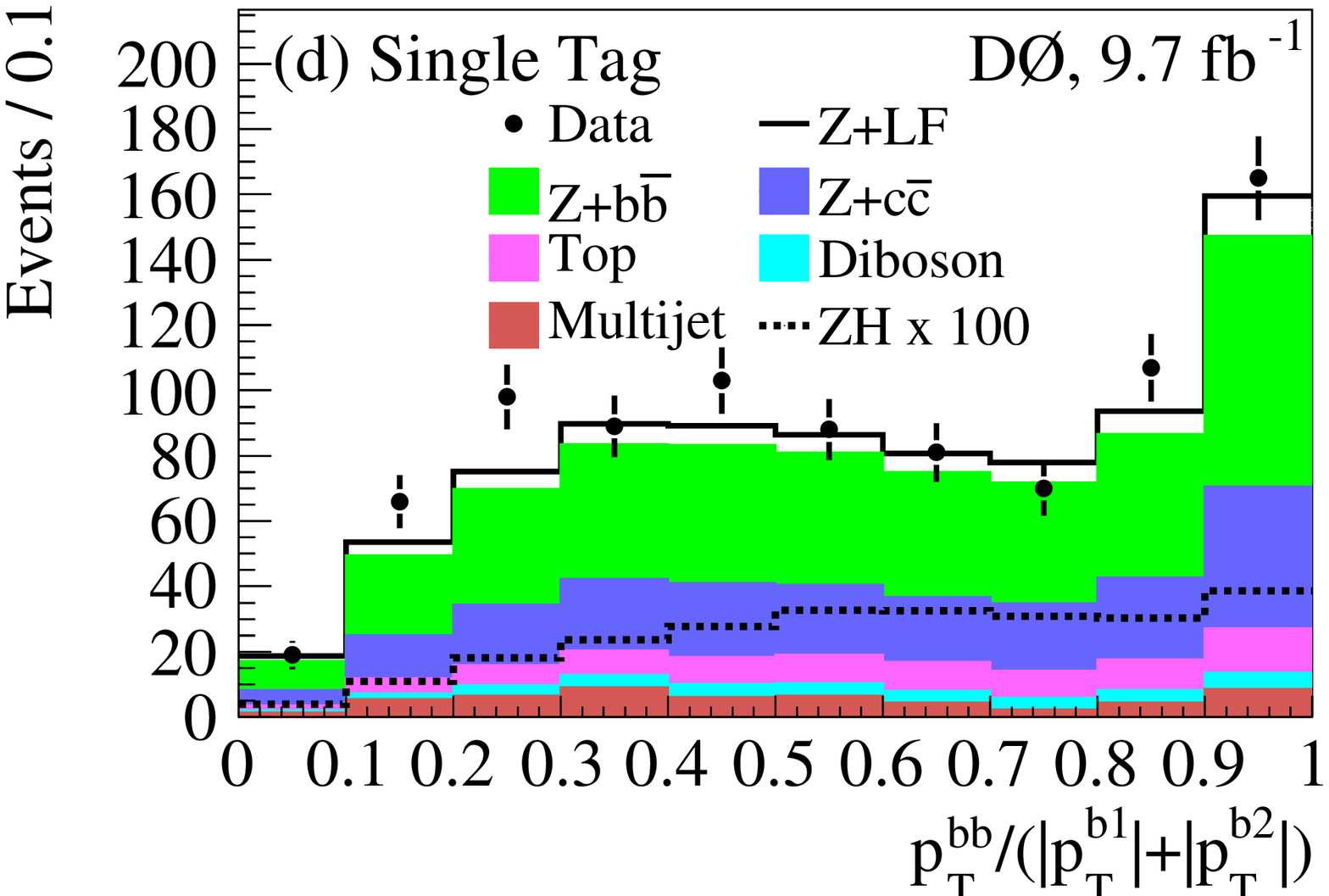}\\
\includegraphics[height=0.24\textheight]{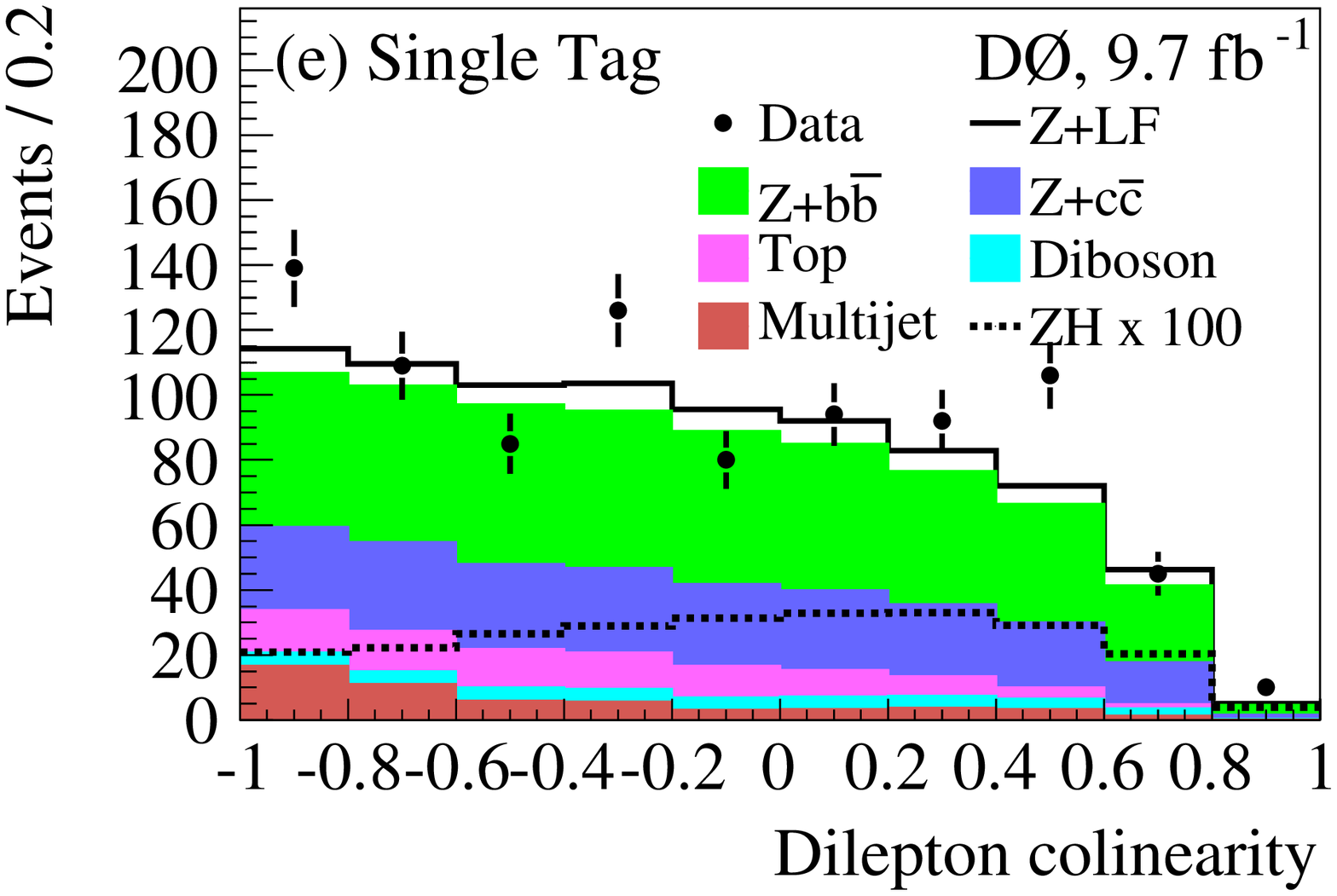}&
\includegraphics[height=0.24\textheight]{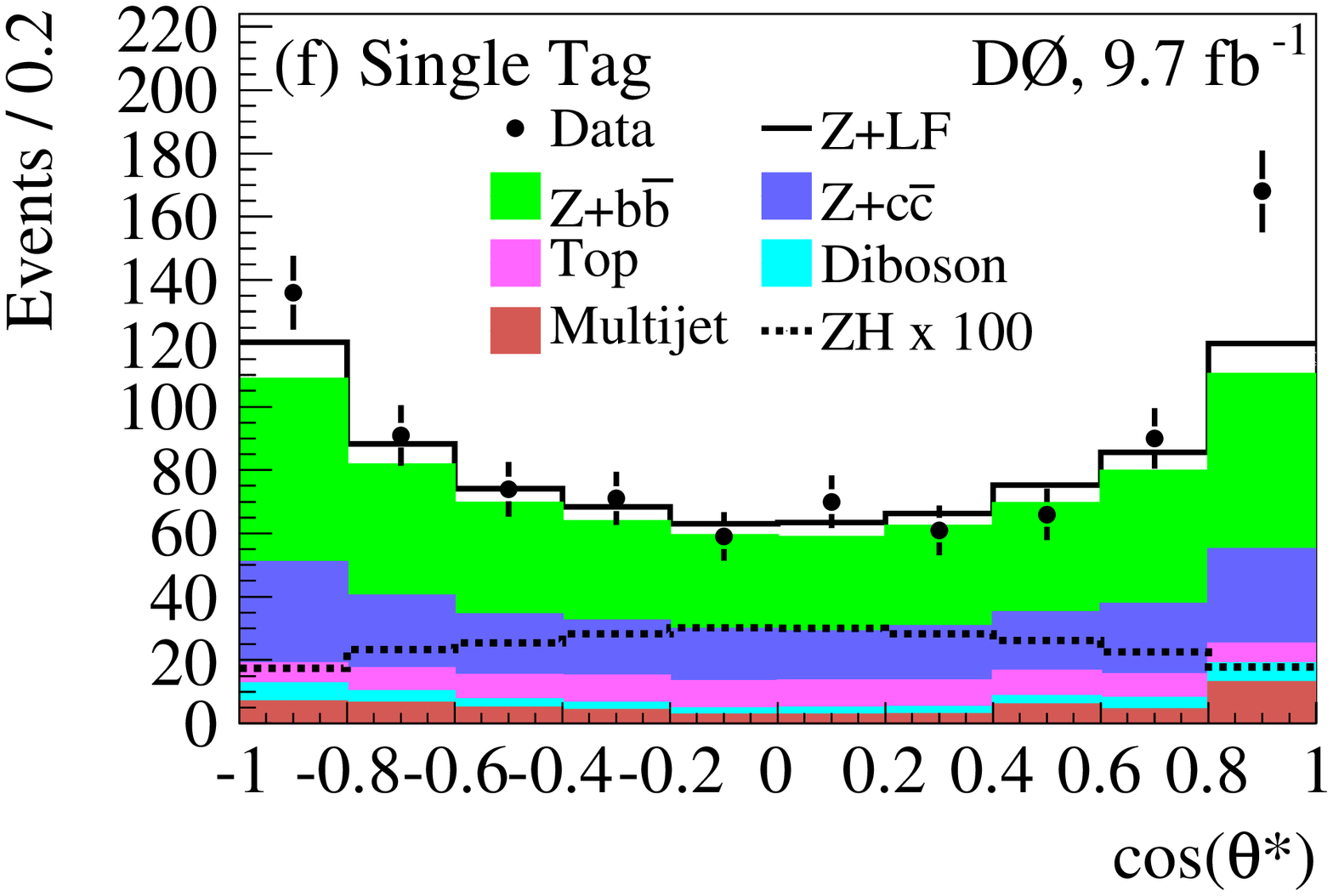}\\
\end{tabular}
\caption{(color online). Distributions in ST events of 
(a) the dijet invariant mass corrected by the kinematic fit,
(b) the $p_T$ of the leading jet from the Higgs boson candidate,
(c) the $p_T$ of the sub-leading jet from the Higgs boson candidate,
(d) the $p_T$ of the dijet system divided by the scalar sum of the transverse momenta of the two jets,
(e) the colinearity of the two leptons, and
(f) \costhst\ \cite{Parke:1999}.
The signal distributions for $M_H=$ 125 GeV are scaled by a factor of 100.
\label{fig:st_rfinputs_1}}
\end{figure*}

\begin{figure*}[htb]\centering
\begin{tabular}{cc}
\includegraphics[height=0.24\textheight]{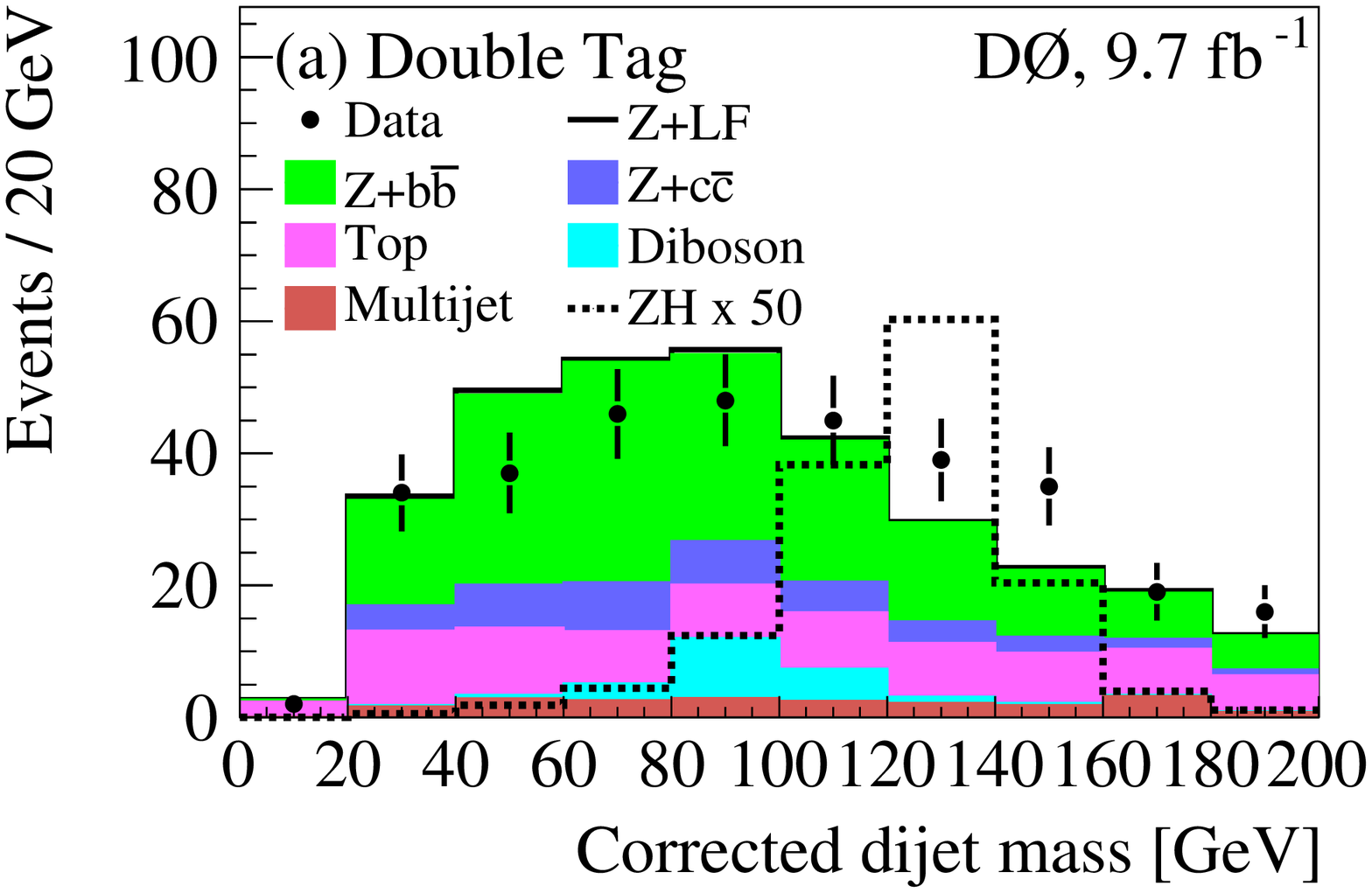}&
\includegraphics[height=0.24\textheight]{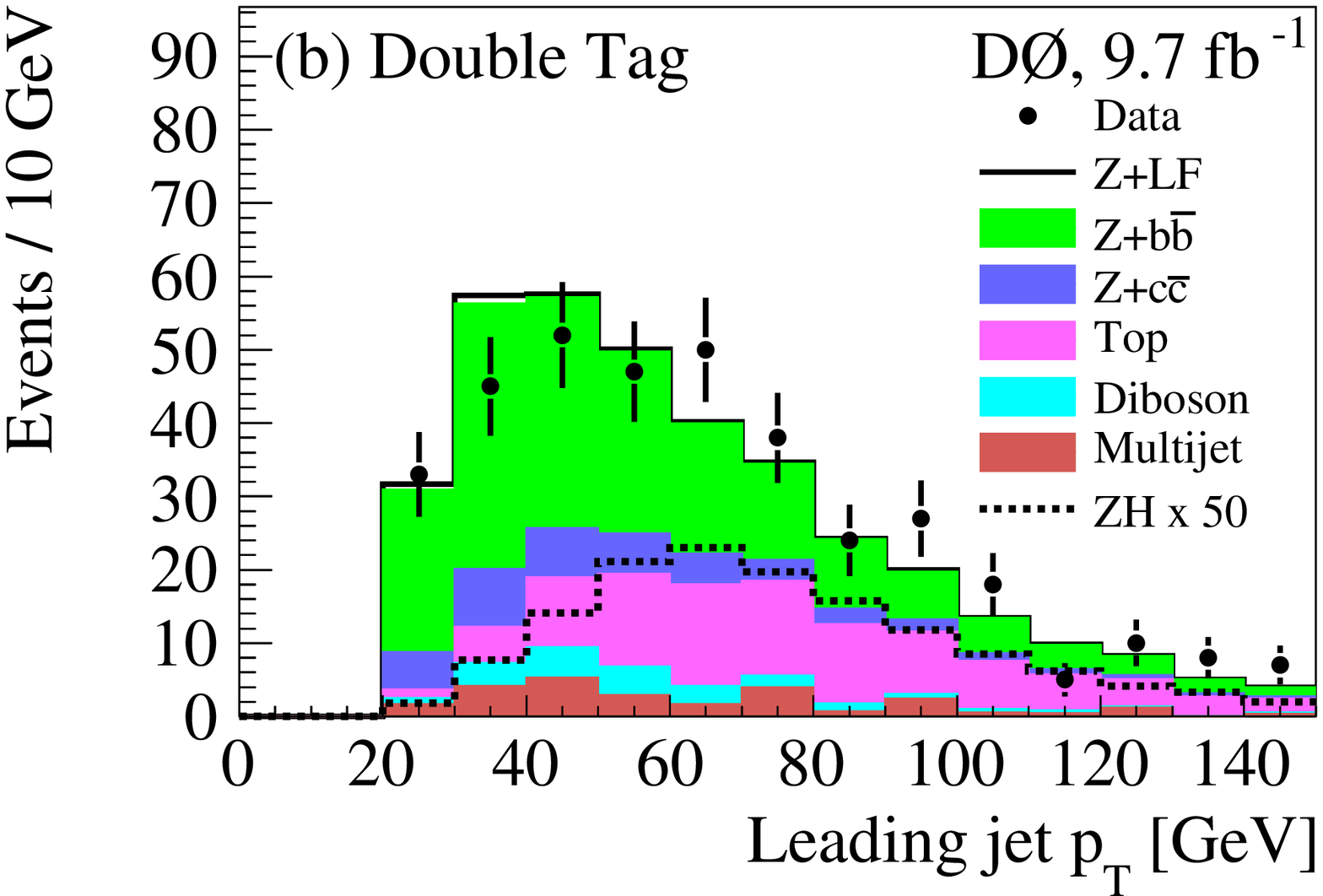}\\
\includegraphics[height=0.24\textheight]{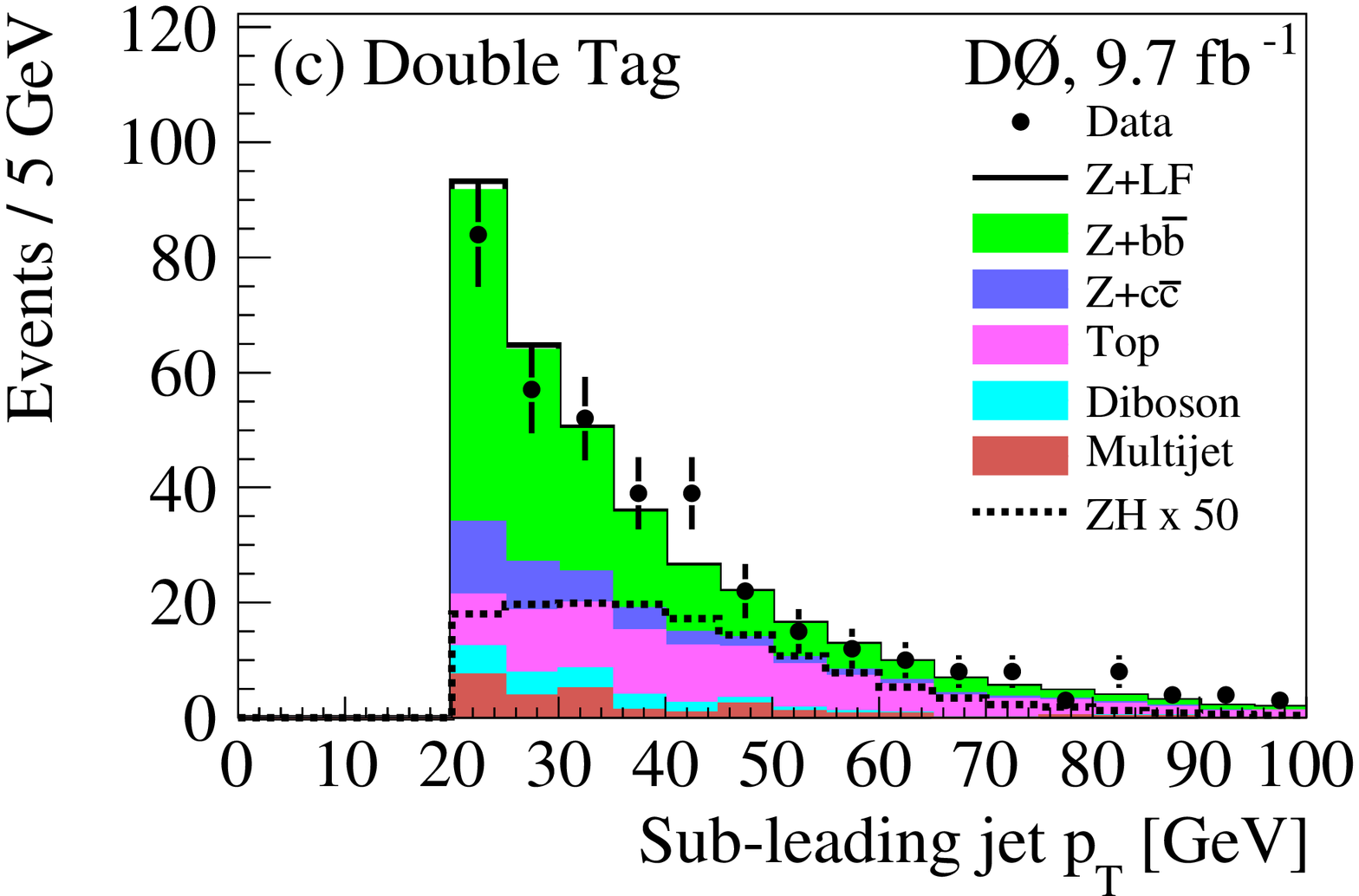}&
\includegraphics[height=0.24\textheight]{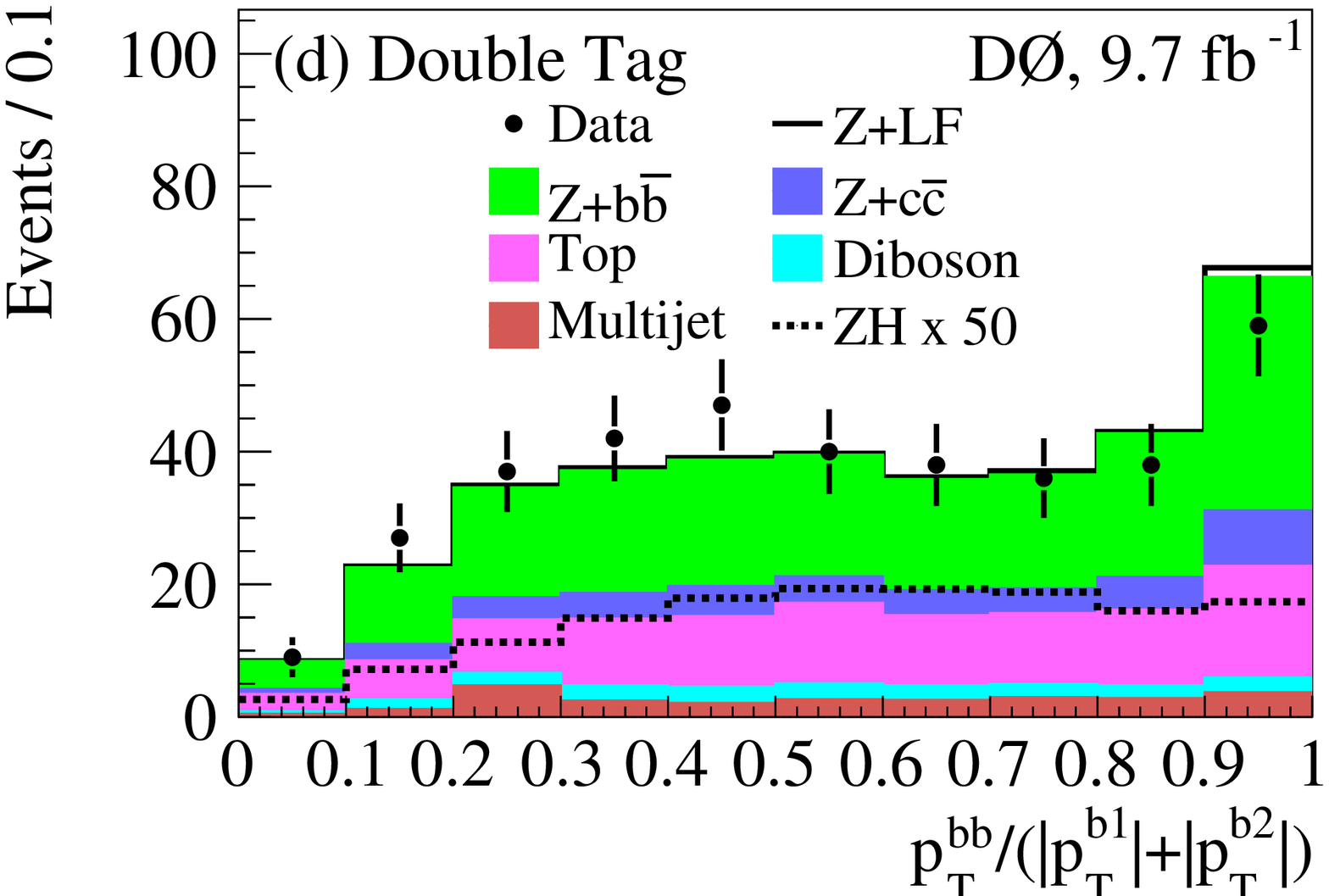} \\
\includegraphics[height=0.24\textheight]{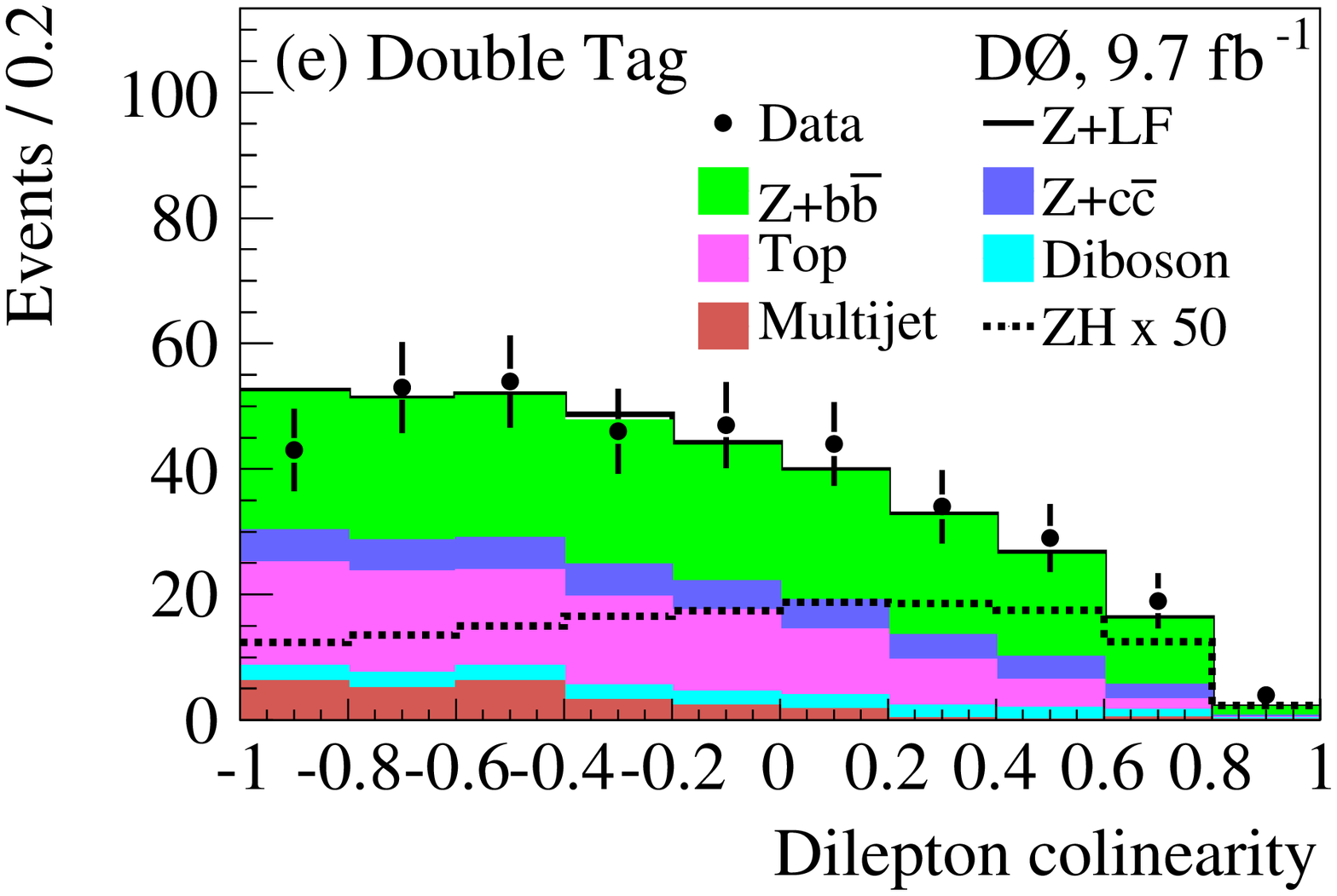}&
\includegraphics[height=0.24\textheight]{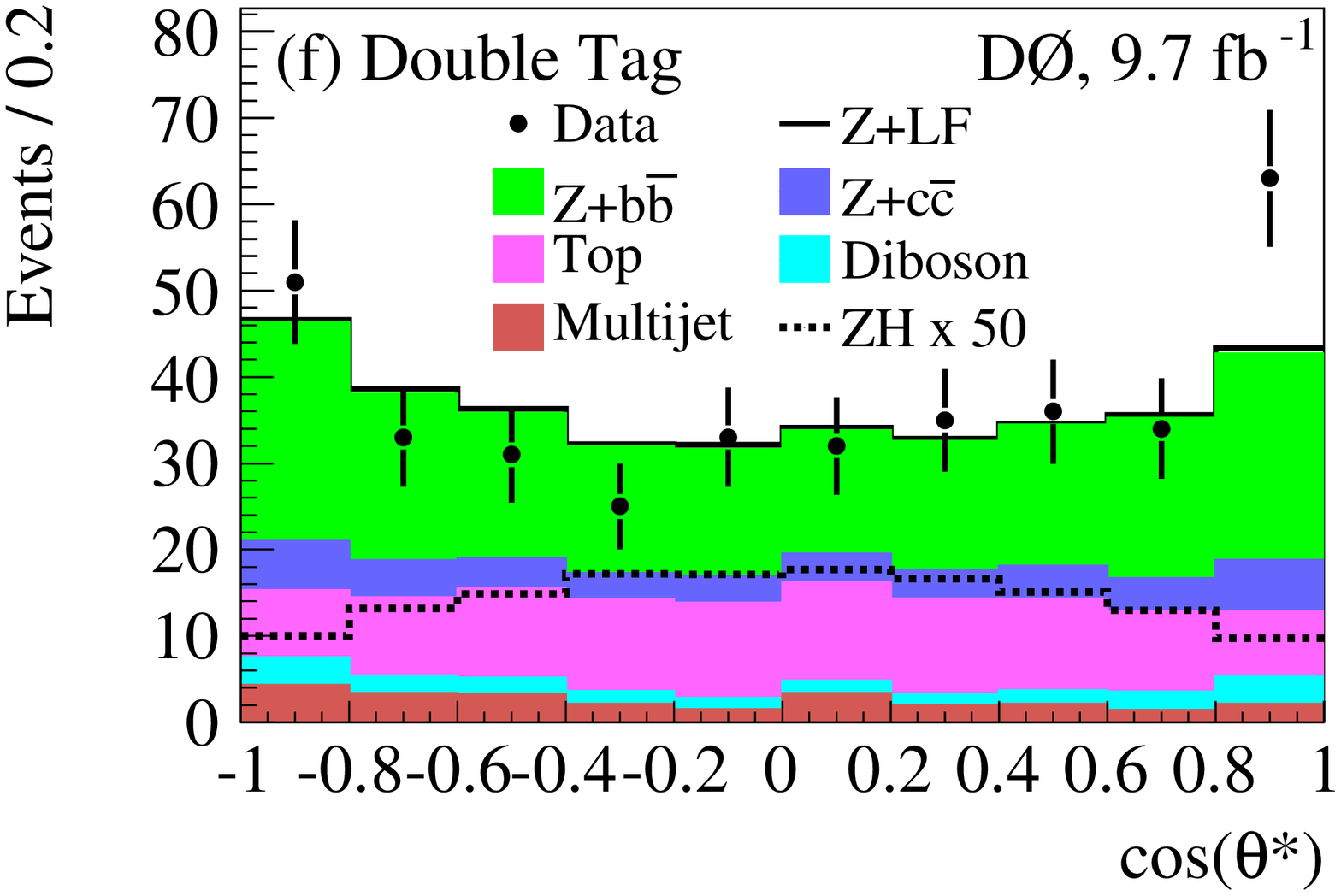}\\
\end{tabular}
\caption{(color online). Distributions in DT events of 
(a) the dijet invariant mass corrected by the kinematic fit,
(b) the $p_T$ of the leading jet from the Higgs boson candidate,
(c) the $p_T$ of the sub-leading jet from the Higgs boson candidate,
(d) the $p_T$ of the dijet system divided by the scalar sum of the transverse momenta of the two jets,
(e) the colinearity of the two leptons, and
(f) \costhst.
The signal distributions for $M_H=$ 125 GeV are scaled by a factor of 50.
\label{fig:dt_rfinputs_1}}
\end{figure*}

The input variables used for the multivariate analysis include the
transverse momenta of the two $b$-jet candidates and the dijet mass,
before and after the jet energies are adjusted by the kinematic fit,
angular differences between the jets, between the leptons, and between the 
dijet and dilepton systems,
the opening angle between the proton beam and the $Z$ boson candidate in the
rest frame of the $Z$ boson, $\costhst$~\cite{Parke:1999}, and composite kinematic
variables, such as the $\pt$ of the dijet system and the scalar sum of
the transverse momenta of the leptons and jets.
Table~\ref{table:RFvariables} provides a complete list of input
variables.  We show selected distributions of the input variables in
Figs.~\ref{fig:st_rfinputs_1} and \ref{fig:dt_rfinputs_1} for ST and DT
events, respectively. 

\begin{table*}[htb] \centering
\caption{Variables used for the $\ttbar$ and global RF training. The 
jets that form the Higgs boson candidate are referred to as $b1$ 
and $b2$, ordered in $\pt$.}
\label{table:RFvariables}
\begin{tabular}{lcc}
\hline
\hline
variables&$t\bar{t}$ RF&global RF\\
\hline
 Invariant mass of the dijet system before (after) the kinematic fit & $\surd$ &$\surd$ \\
 Transverse momentum of the first jet before (after) kinematic fit& $\surd$&$\surd$ \\
 Transverse momentum of the second jet before (after) kinematic fit &$\surd$ &$\surd$\\
 Transverse momentum of the dijet system before the kinematic fit & $\surd$ & $\surd$\\
 $\Delta\phi$ between the two jets in the dijet system &$-$&$\surd$\\
 $\Delta\eta$ between the two jets in the dijet system &$-$&$\surd$\\
 Invariant mass of all jets in the event &$\surd$ & $\surd$\\
 Transverse momentum of all jets in the event & $\surd$& $\surd$\\
 Scalar sum of the transverse momenta of all jets in the event &$\surd$ &$-$\\
 Ratio of dijet system $p_{T}$ over the scalar sum of the $p_{T}$ of the two jets ({\tt $\ptbb/(|\ptbone|+|\ptbtwo|)$}) &$\surd$ & $-$\\
 Invariant mass of the dilepton system &$\surd$ &$-$\\
 Transverse momentum of the dilepton system&$\surd$ &$\surd$ \\
 $\Delta\phi$ between the two leptons &$\surd$ &$\surd$\\
 cosine of the angle between the two leptons (colinearity) &$\surd$ &$\surd$\\
 $\Delta\phi$ between the dilepton and dijet systems &$\surd$ &$\surd$\\
 cosine of the angle between the incoming proton and the $Z$ in the zero momentum frame ({\tt $\costhst$}) \cite{Parke:1999} &$-$&$\surd$\\
 Invariant mass of dilepton and dijet system &$-$&$\surd$\\
 Scalar sum of the transverse momenta of the leptons and jets&$-$&$\surd$\\
 Missing transverse energy of the event &$\surd$ &$-$\\
  \met\ significance \cite{met_sig} &$\surd$ &$\surd$ \\
  Negative log likelihood from the kinematic fit (Eq.~\ref{eq:Norm_chi2})&$\surd$ &$\surd$\\
 $\ttbar$ RF output & $-$&$\surd$\\
\hline
\hline
\end{tabular} 
\end{table*}

To avoid biases in the training procedure, we divide the MC
samples into three independent sub-samples: 25\% of the events are
used to train the RFs (for both the $\ttbar$~RF and the global RF);
25\% of the events are used to test the RF discrimination performance 
and check for overtraining (for both the
$\ttbar$~RF and the global RF), and the remaining 50\% of the events
(the evaluation sub-sample) are used for the statistical analysis to
obtain Higgs boson cross section limits.  

\begin{figure*}[htbp]\centering
\begin{tabular}{ll}
\includegraphics[height=0.24\textheight]{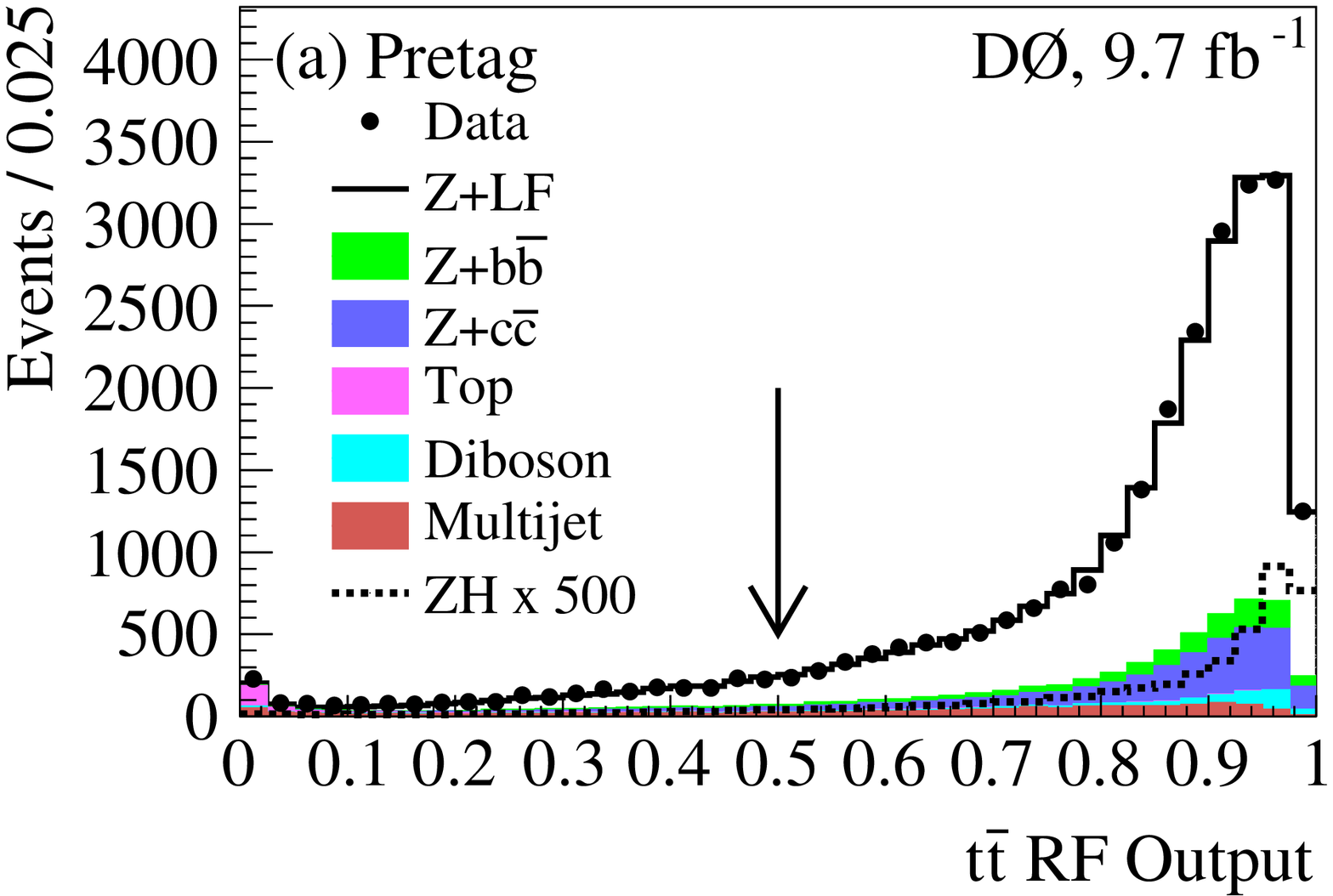} &
\includegraphics[height=0.24\textheight]{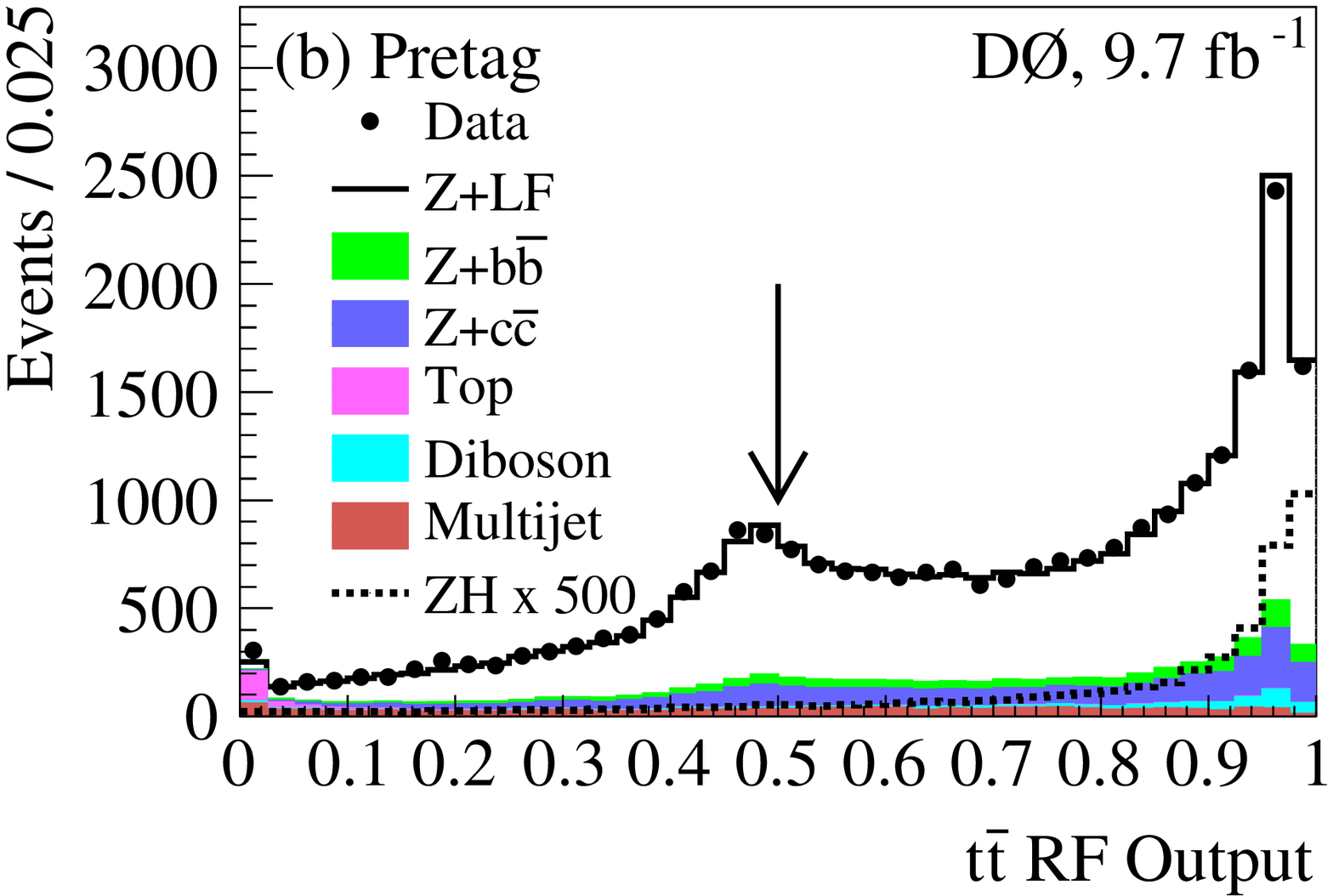}\\
\end{tabular}
\caption{(color online). The
$\ttbar$ RF output ($M_H=125~\gev$) for all lepton channels combined in the pretag sample
(a) trained for ST events and (b) trained for DT events.
The arrows indicate the $\ttbar$~RF selection requirement used to define the  
$\ttbar$-enriched and depleted sub-samples.
The signal distributions for $M_H=$ 125 GeV are scaled by a factor of 500.
\label{fig:ttbar_rf_pretag}}

\begin{tabular}{cc}
\includegraphics[height=0.24\textheight]{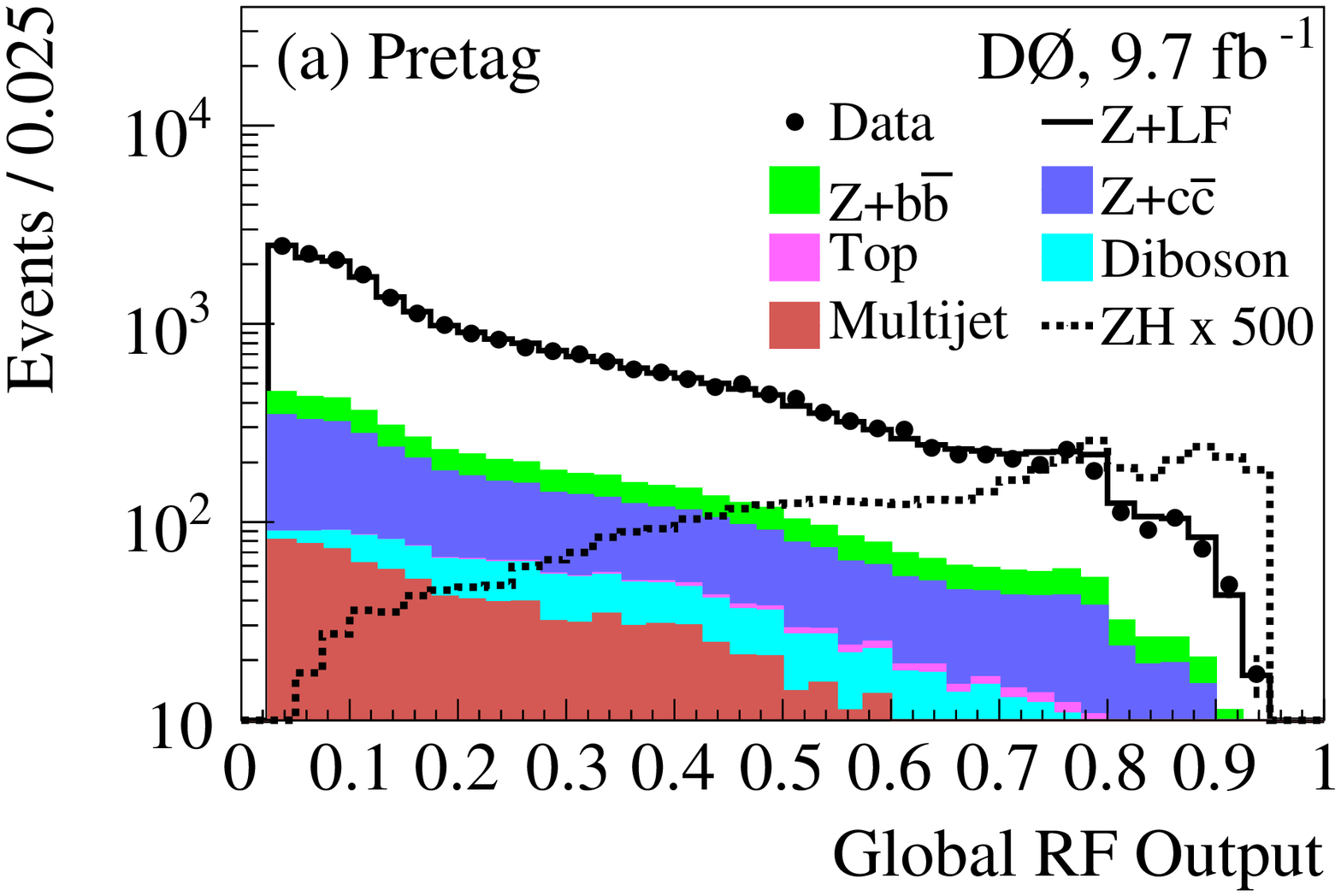}  &
\includegraphics[height=0.24\textheight]{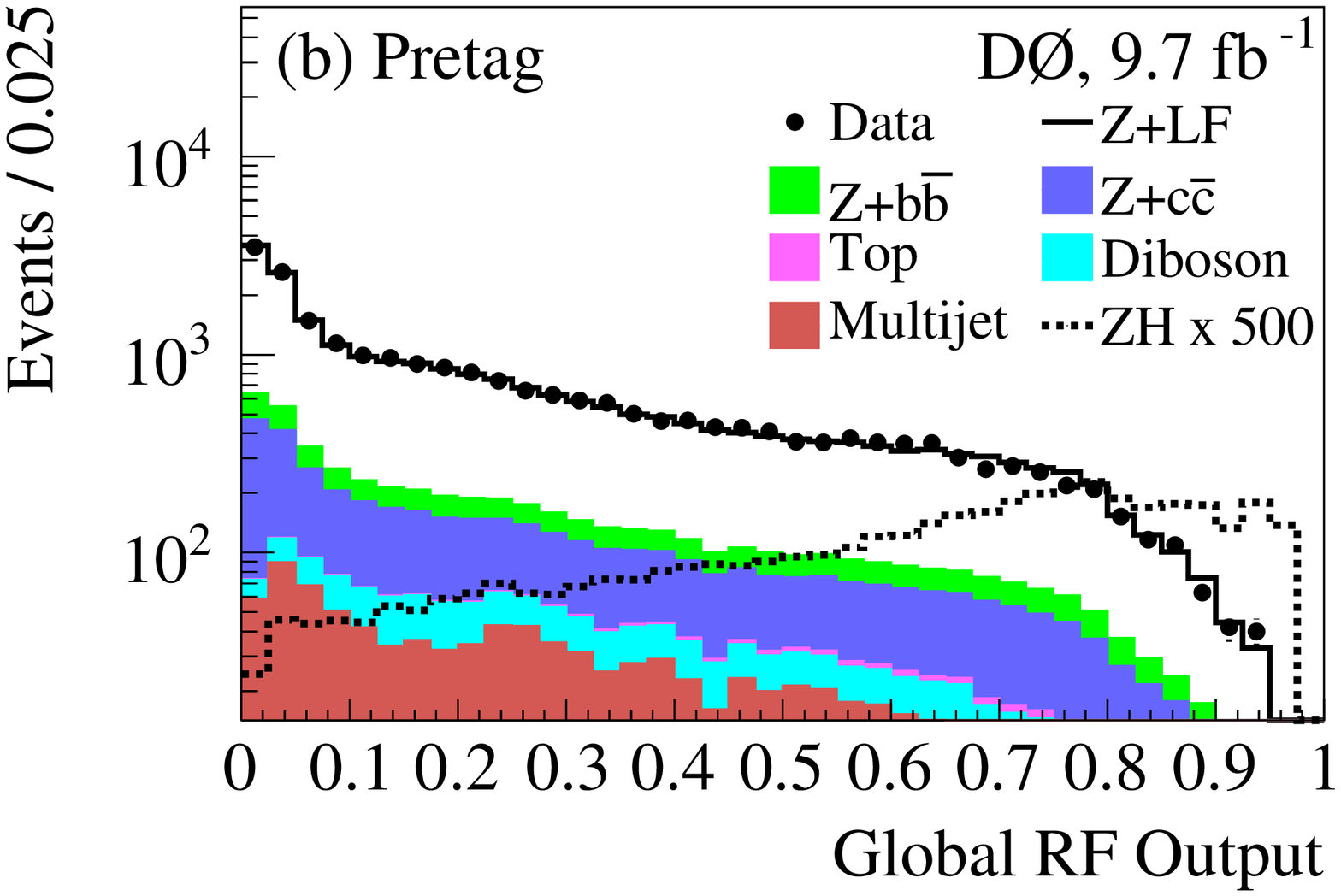} \\
\end{tabular} 
\caption{\label{fig:global_rf_pretag} (color online). Global RF output 
  ($M_H=125~\gev$) for all lepton channels combined 
  for (a) pretag events evaluated with the ST-trained RF and (b) pretag
  events evaluated with the DT-trained RF.
  The signal distributions for $M_H=$ 125 GeV are scaled by a factor of 500.}
\end{figure*}

\begin{figure*}[htdp]\centering
\begin{tabular}{cc}
\includegraphics[height=0.24\textheight]{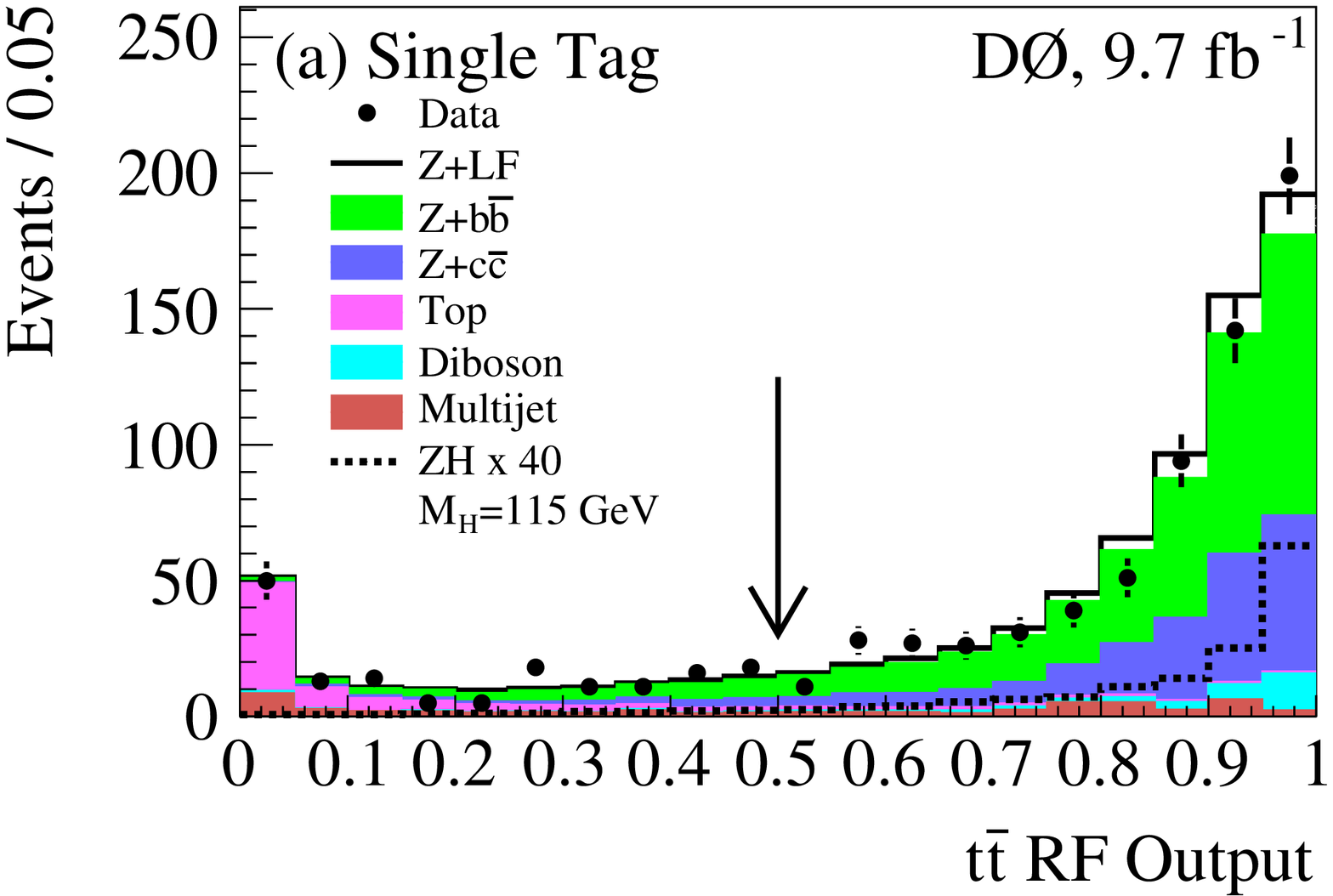}&
\includegraphics[height=0.24\textheight]{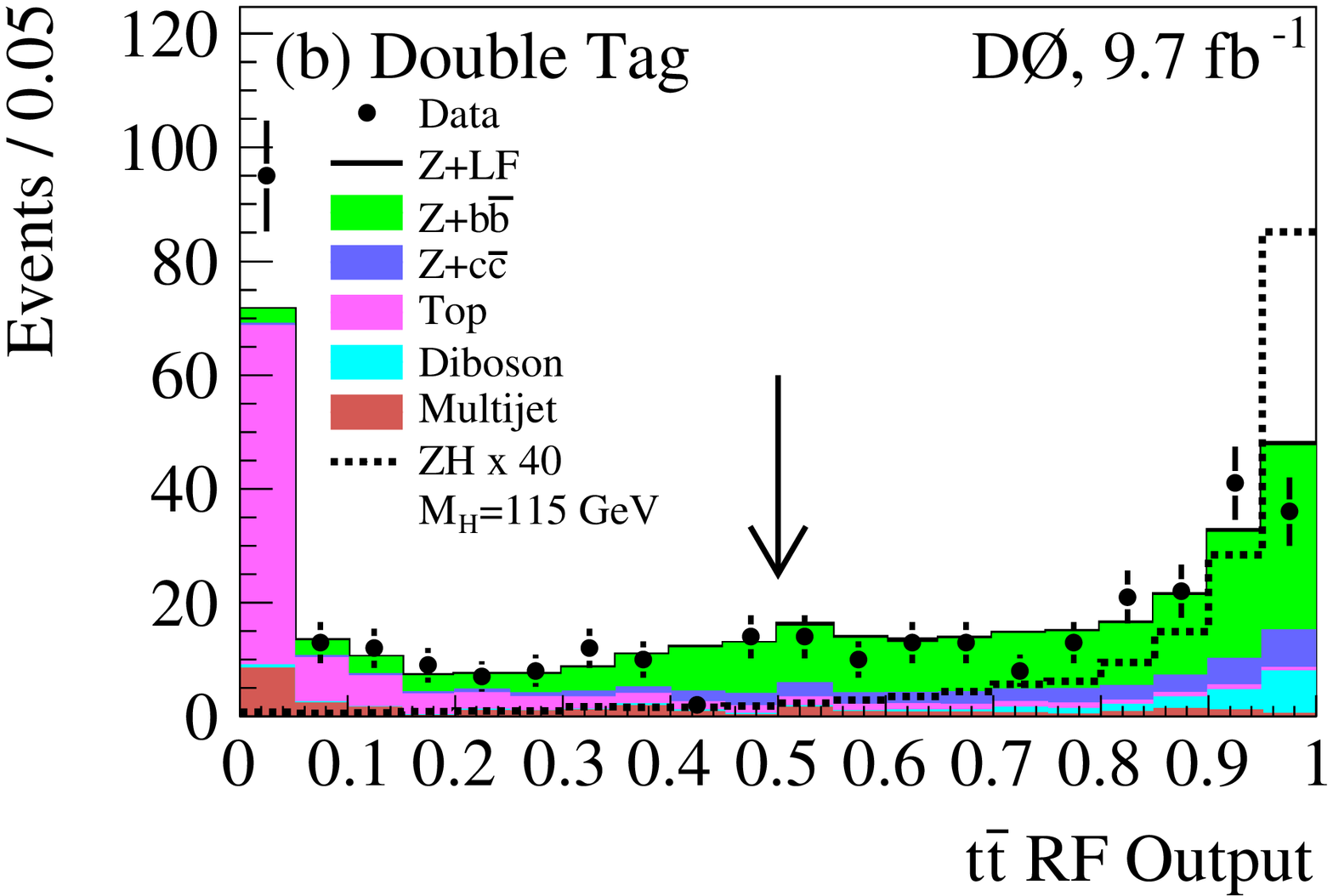}\\
\includegraphics[height=0.24\textheight]{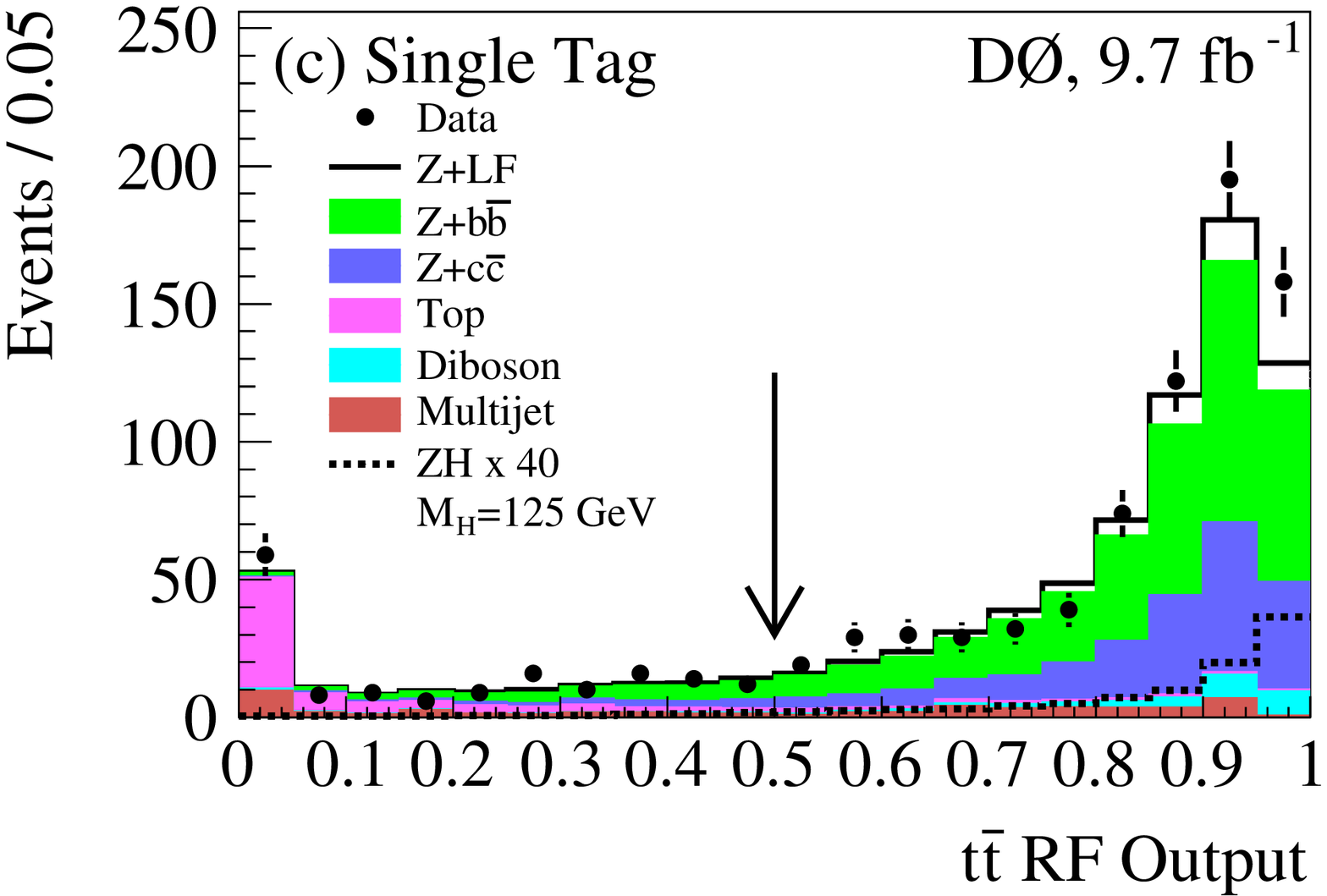}&
\includegraphics[height=0.24\textheight]{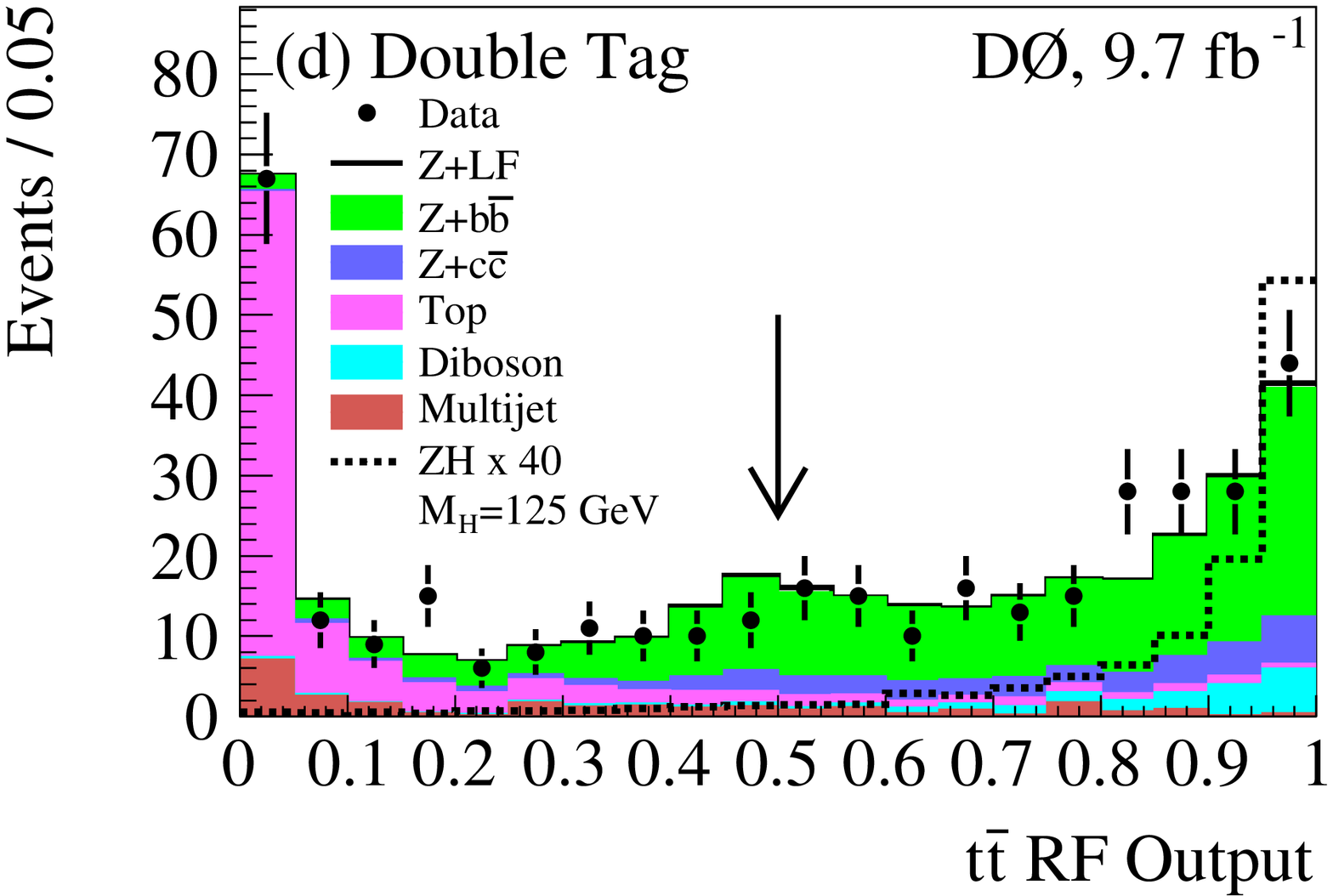}\\
\includegraphics[height=0.24\textheight]{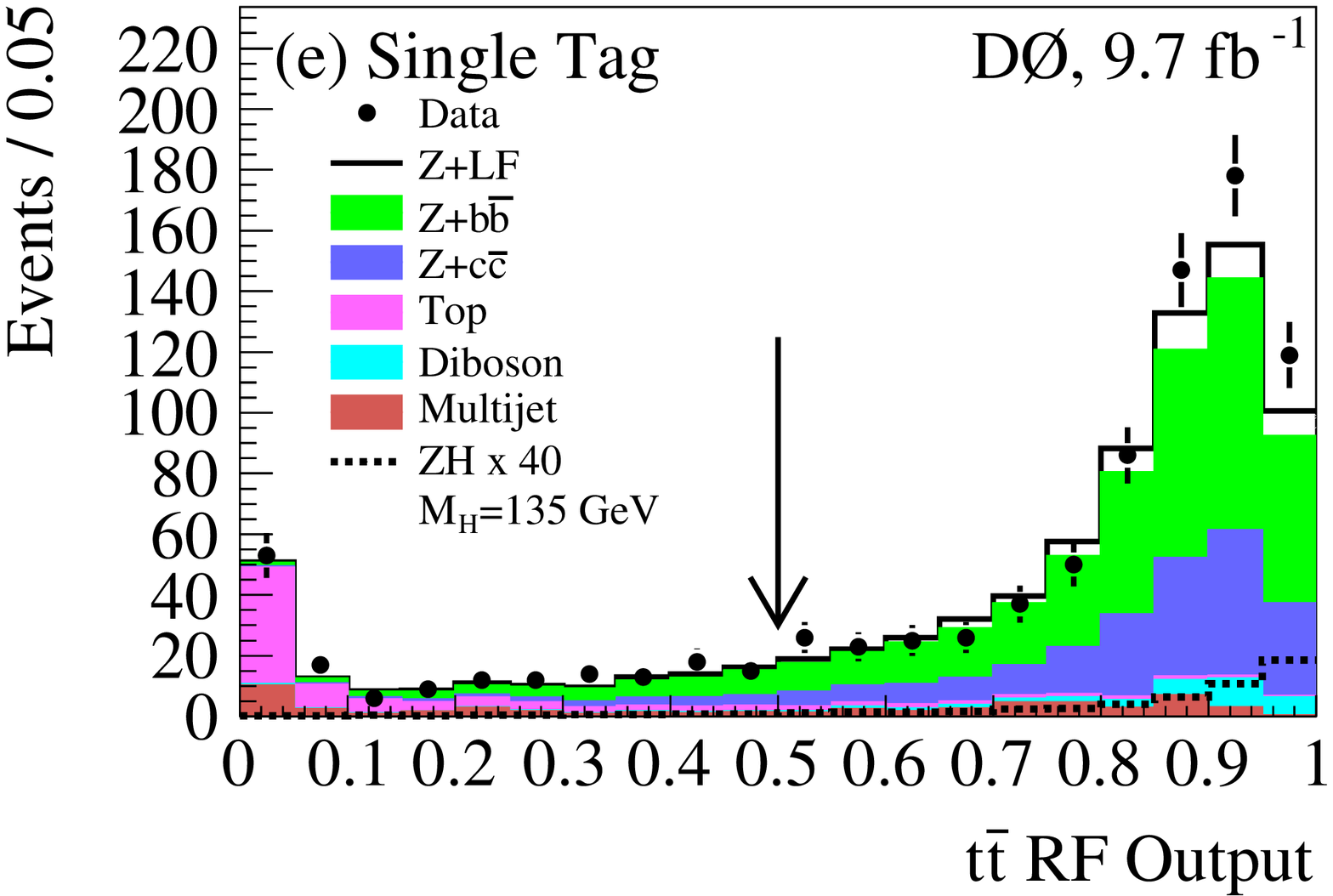}&
\includegraphics[height=0.24\textheight]{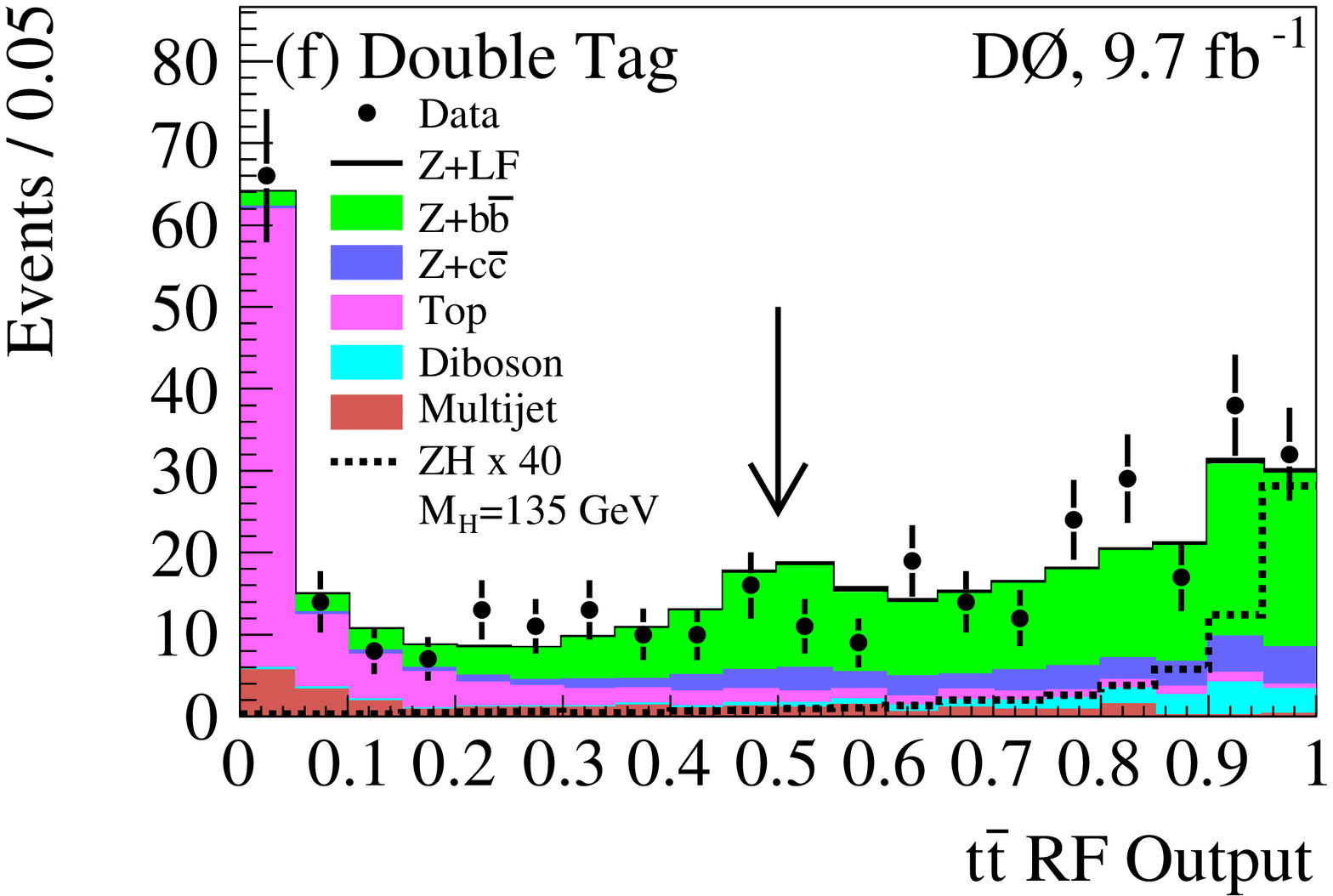}\\
\end{tabular}
\caption{(color online). The $\ttbar$ RF output for all lepton channels combined
in ST and DT events 
for $M_H$ = 115 GeV (a, b),
for $M_H$ = 125 GeV (c, d), and
for $M_H$ = 135 GeV (e, f).
The signal distributions correspond to the $M_H$ used for the RF training and
are scaled by a factor of 40. The arrows
indicate the $\ttbar$~RF selection requirement used to define the 
$\ttbar$-enriched and depleted sub-samples.
\label{fig:ttbar_rf_post_tag}}
\end{figure*}

\begin{figure*}[htbp]\centering
\begin{tabular}{cc}
\includegraphics[height=0.24\textheight]{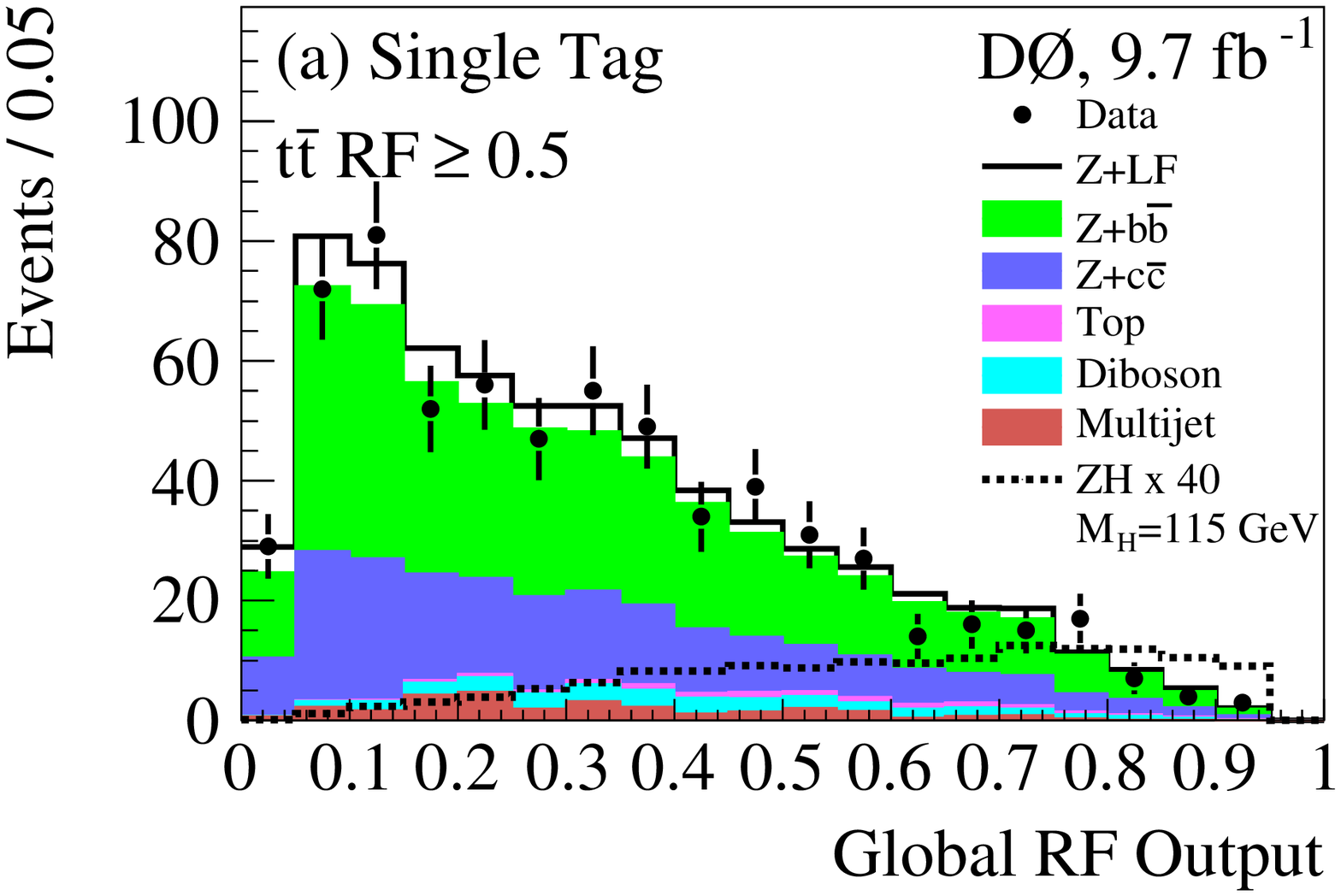} &
\includegraphics[height=0.24\textheight]{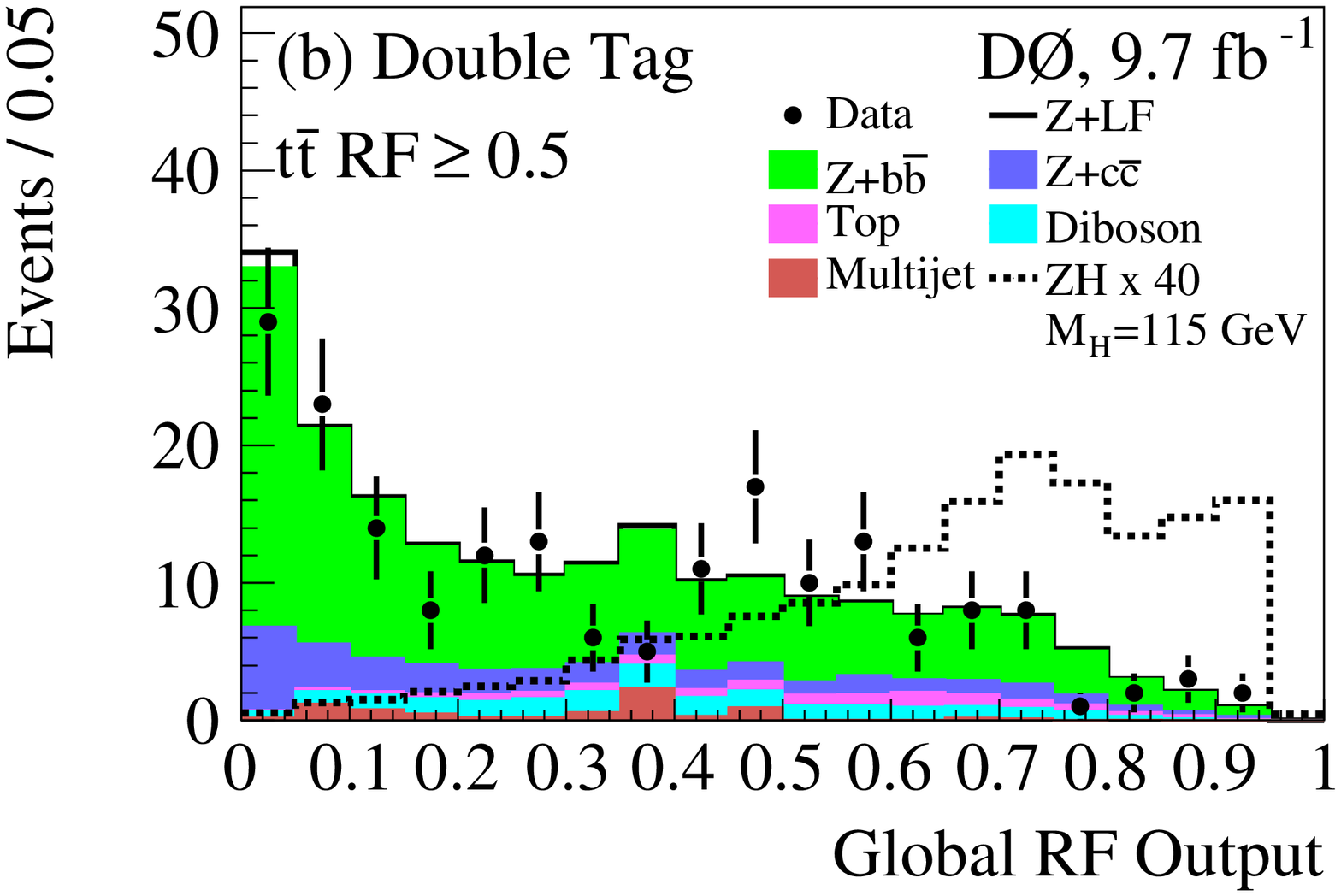} \\
\includegraphics[height=0.24\textheight]{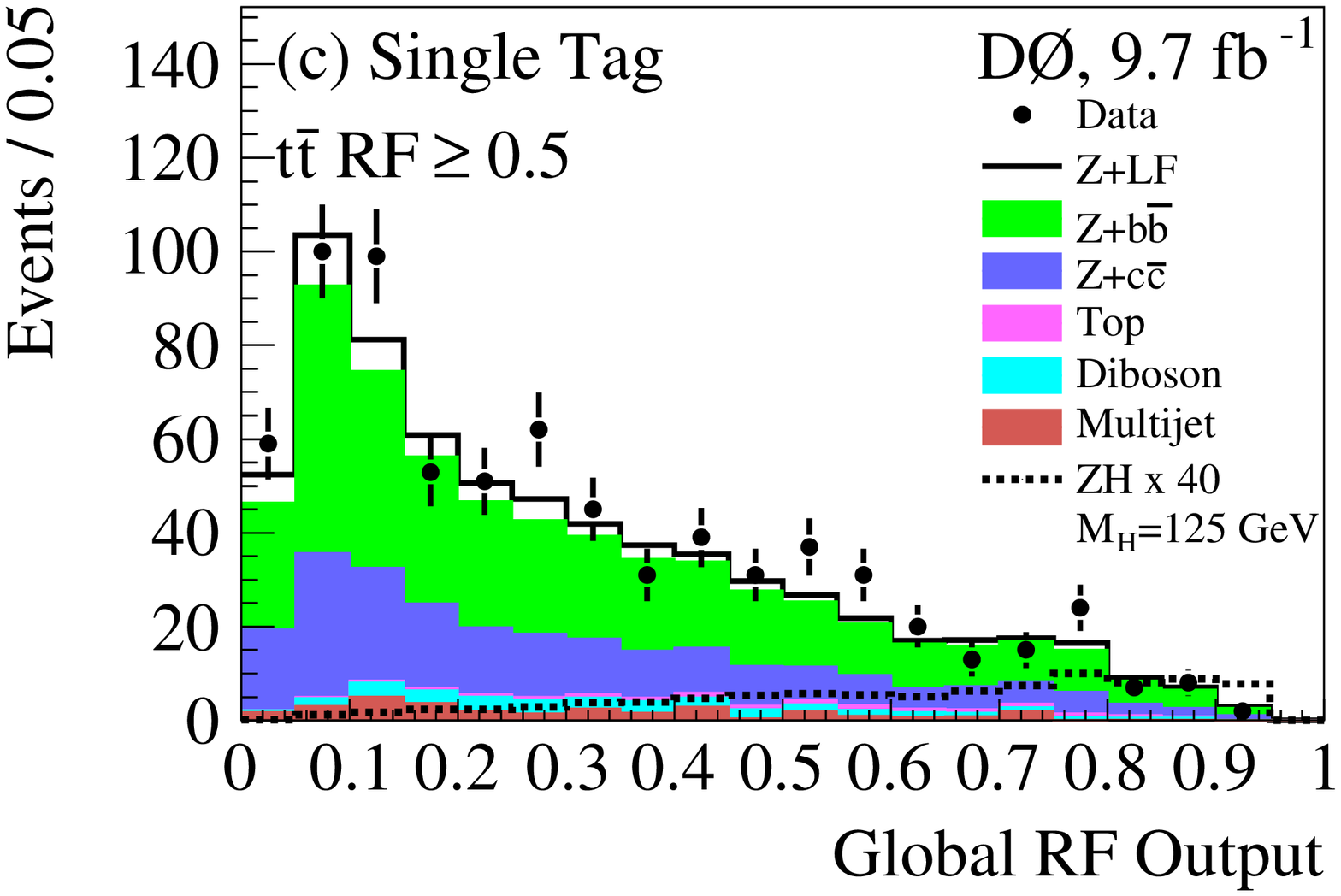} &
\includegraphics[height=0.24\textheight]{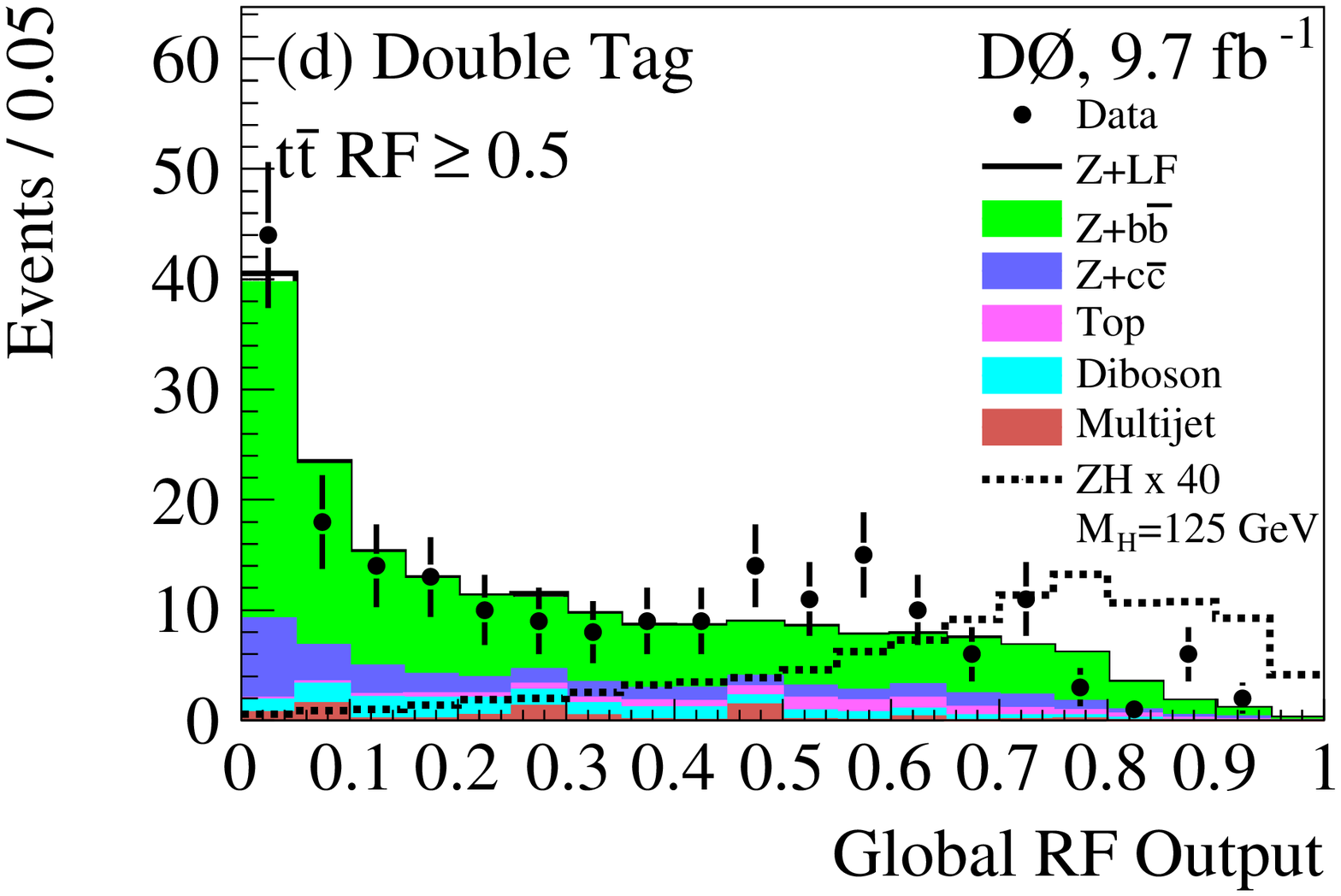} \\
\includegraphics[height=0.24\textheight]{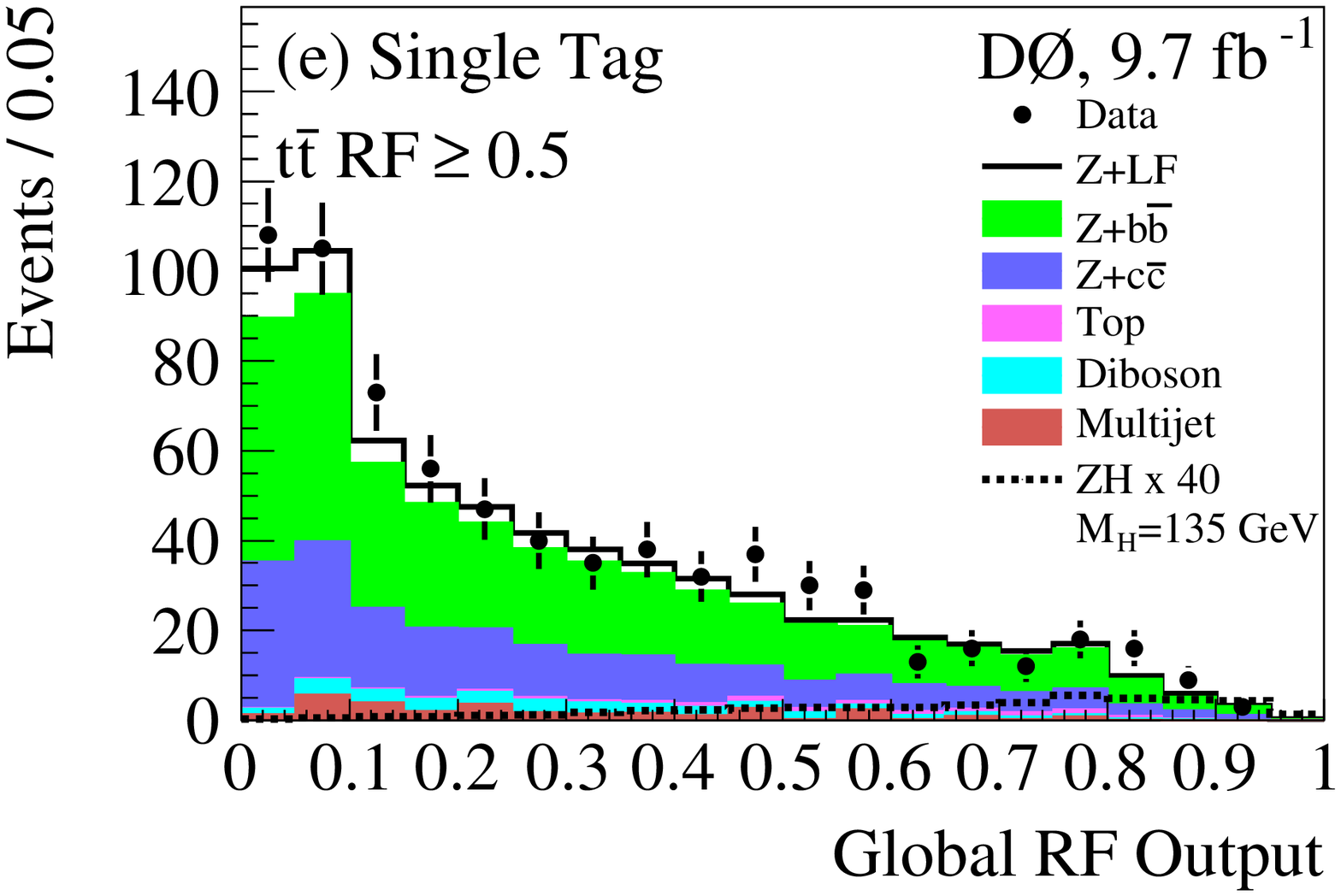} &
\includegraphics[height=0.24\textheight]{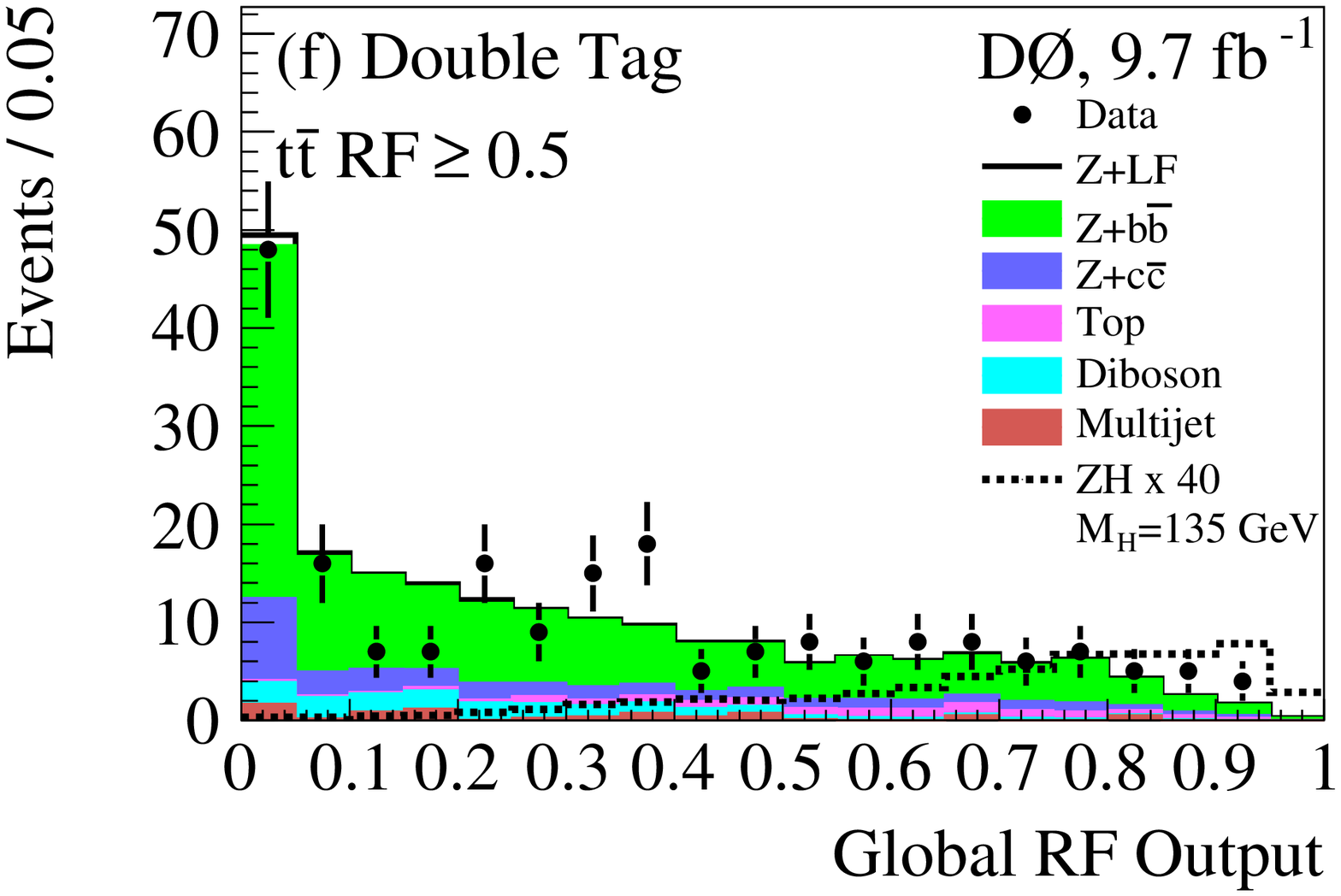} \\
\end{tabular}
\caption{(color online). Global RF distributions for
ST and DT events in the $\ttbar$-depleted region 
for $M_H$ = 115 GeV (a, b),
for $M_H$ = 125 GeV (c, d), and
for $M_H$ = 135 GeV (e, f). The signal distributions correspond to the $M_H$ 
used for the RF training and are scaled by a factor of 40.
\label{fig:global_rf_ttpoor_post_tag}}
\end{figure*}

\begin{figure*}[htbp]\centering
\begin{tabular}{cc}
\includegraphics[height=0.24\textheight]{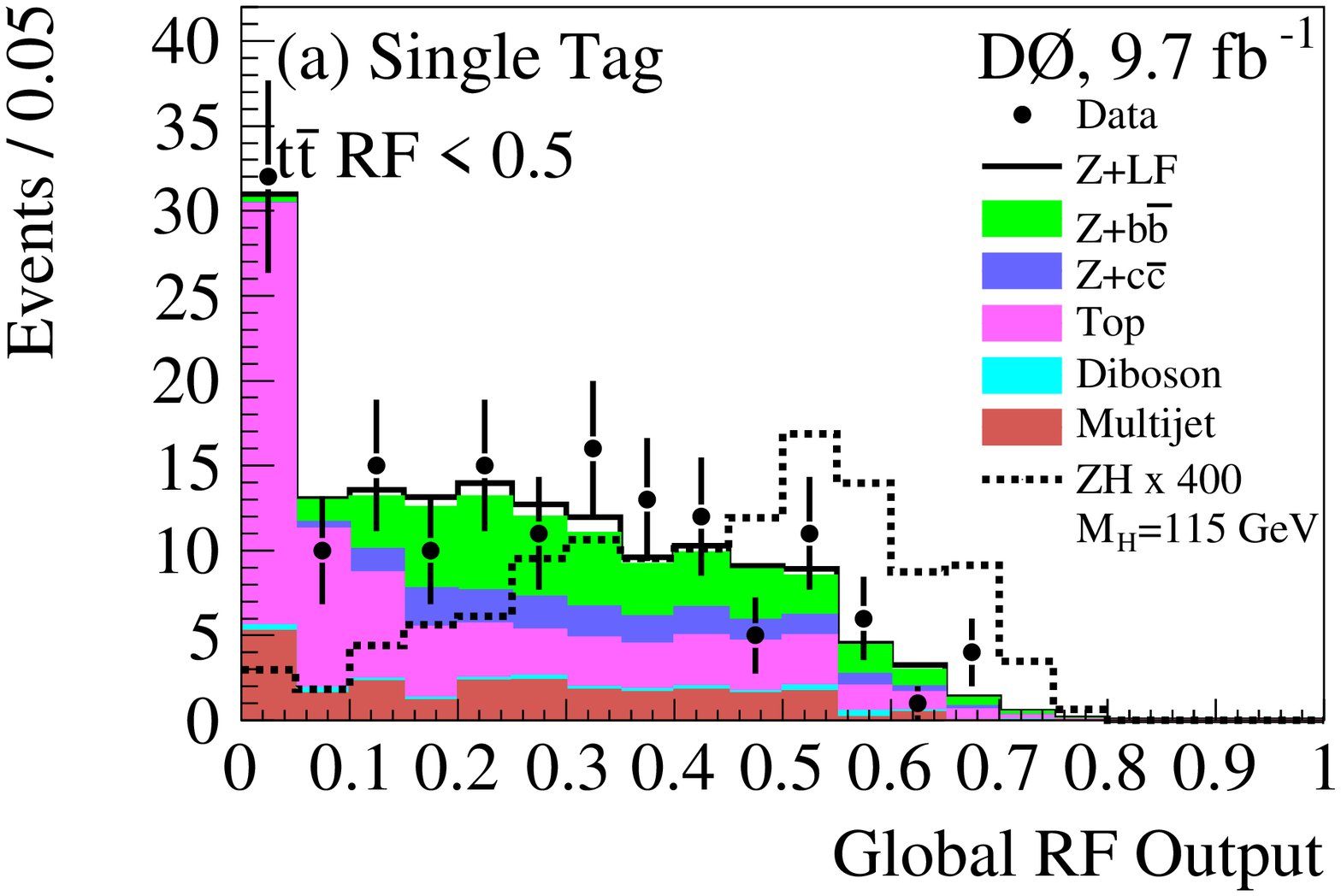} &
\includegraphics[height=0.24\textheight]{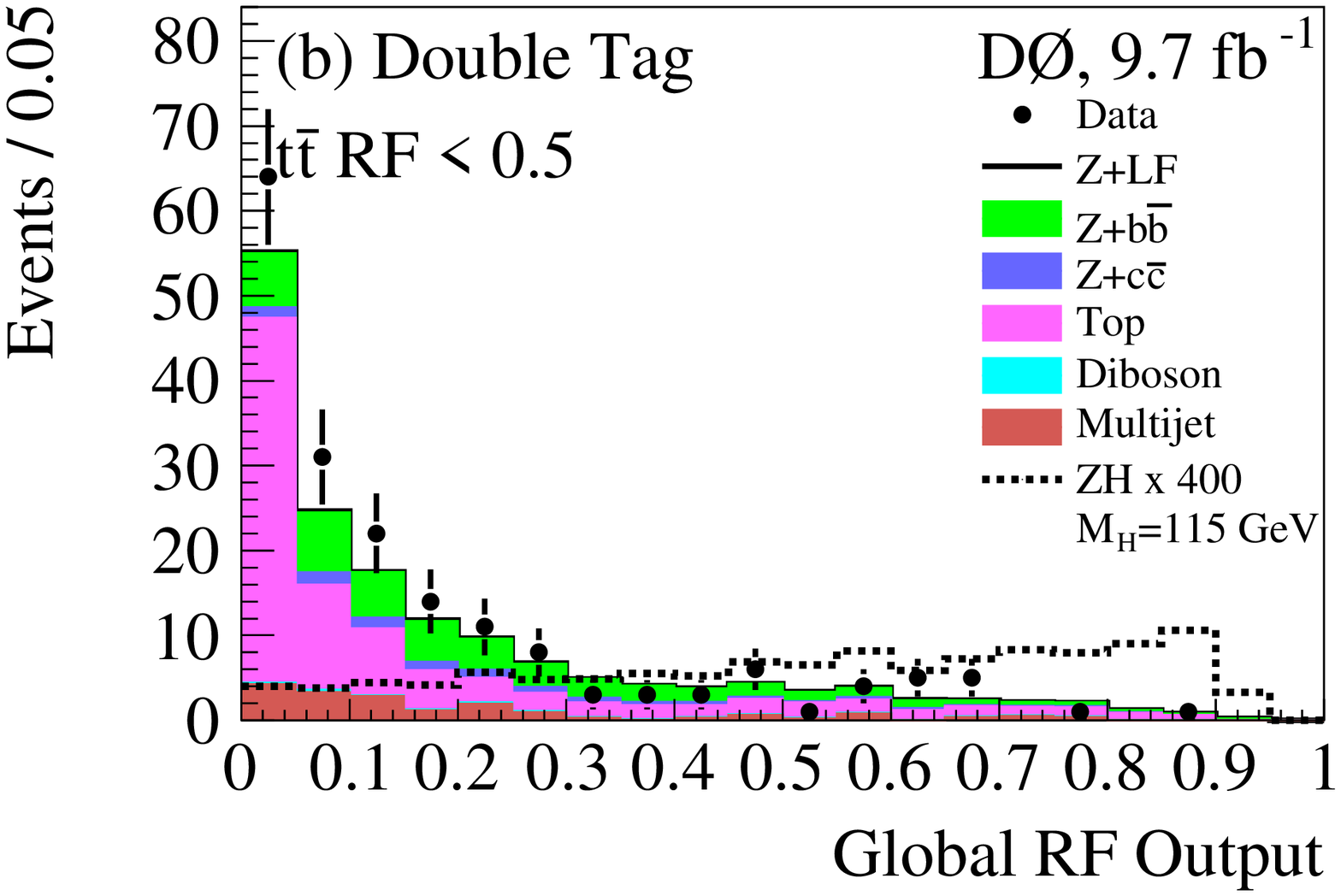} \\
\includegraphics[height=0.24\textheight]{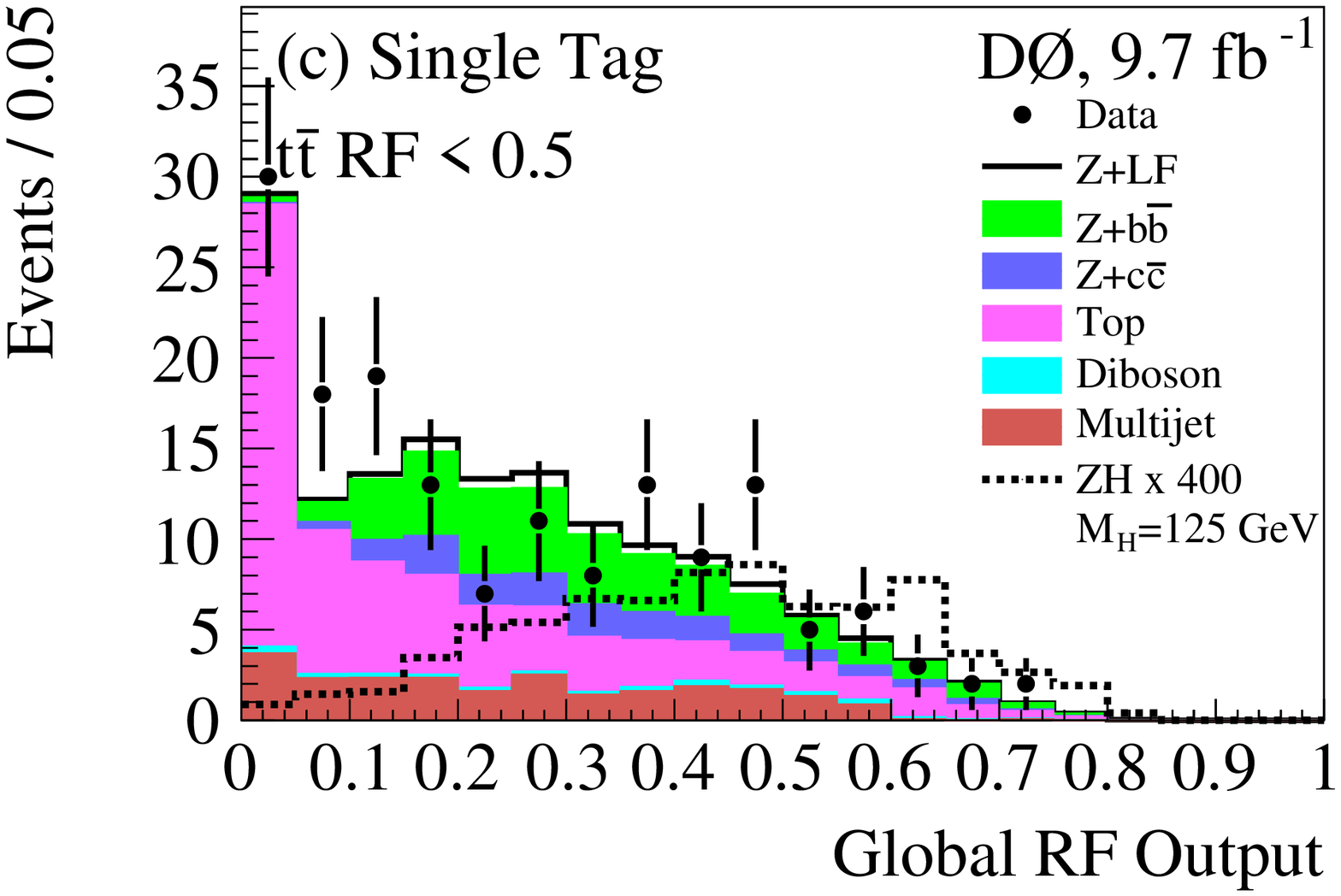} &
\includegraphics[height=0.24\textheight]{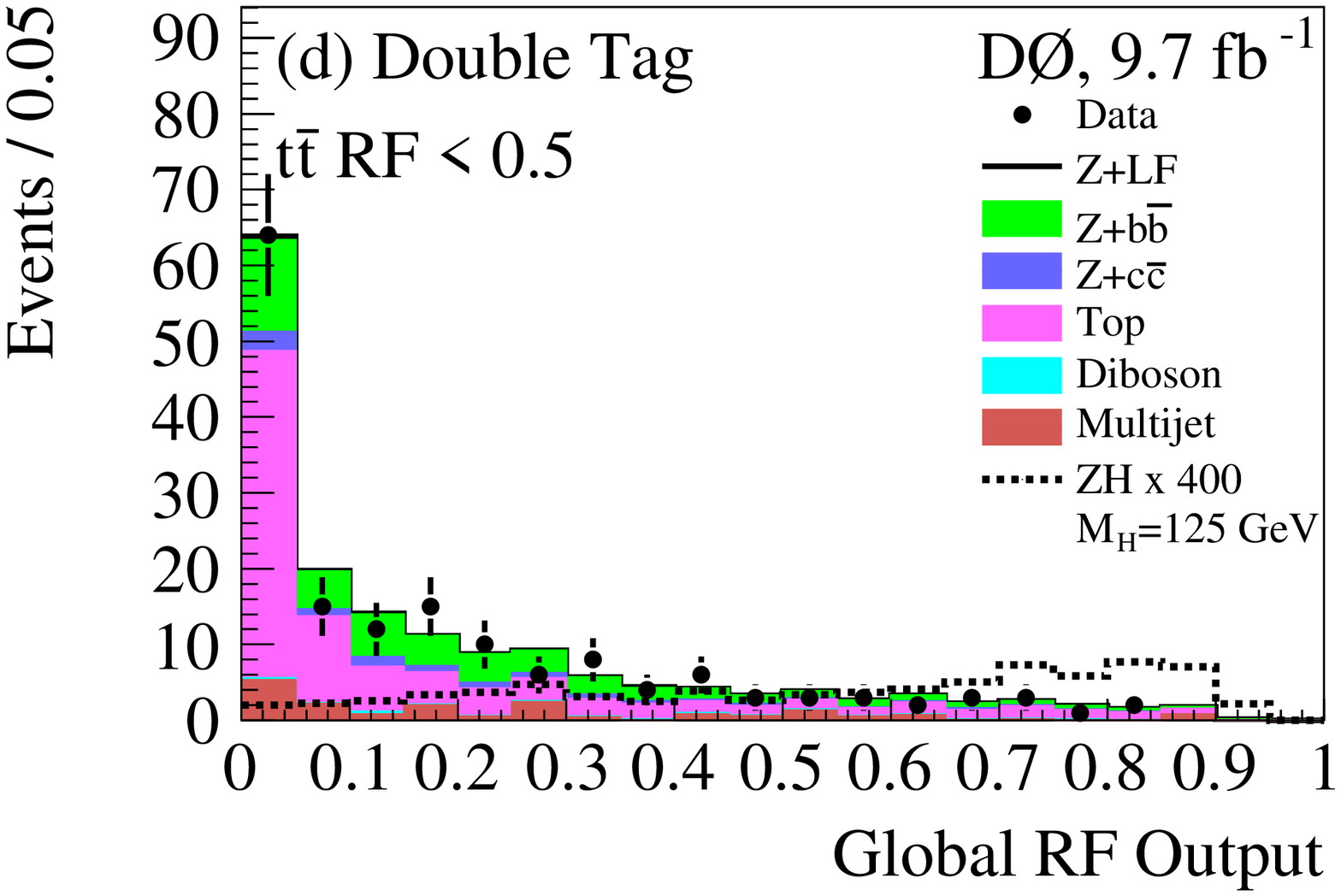} \\
\includegraphics[height=0.24\textheight]{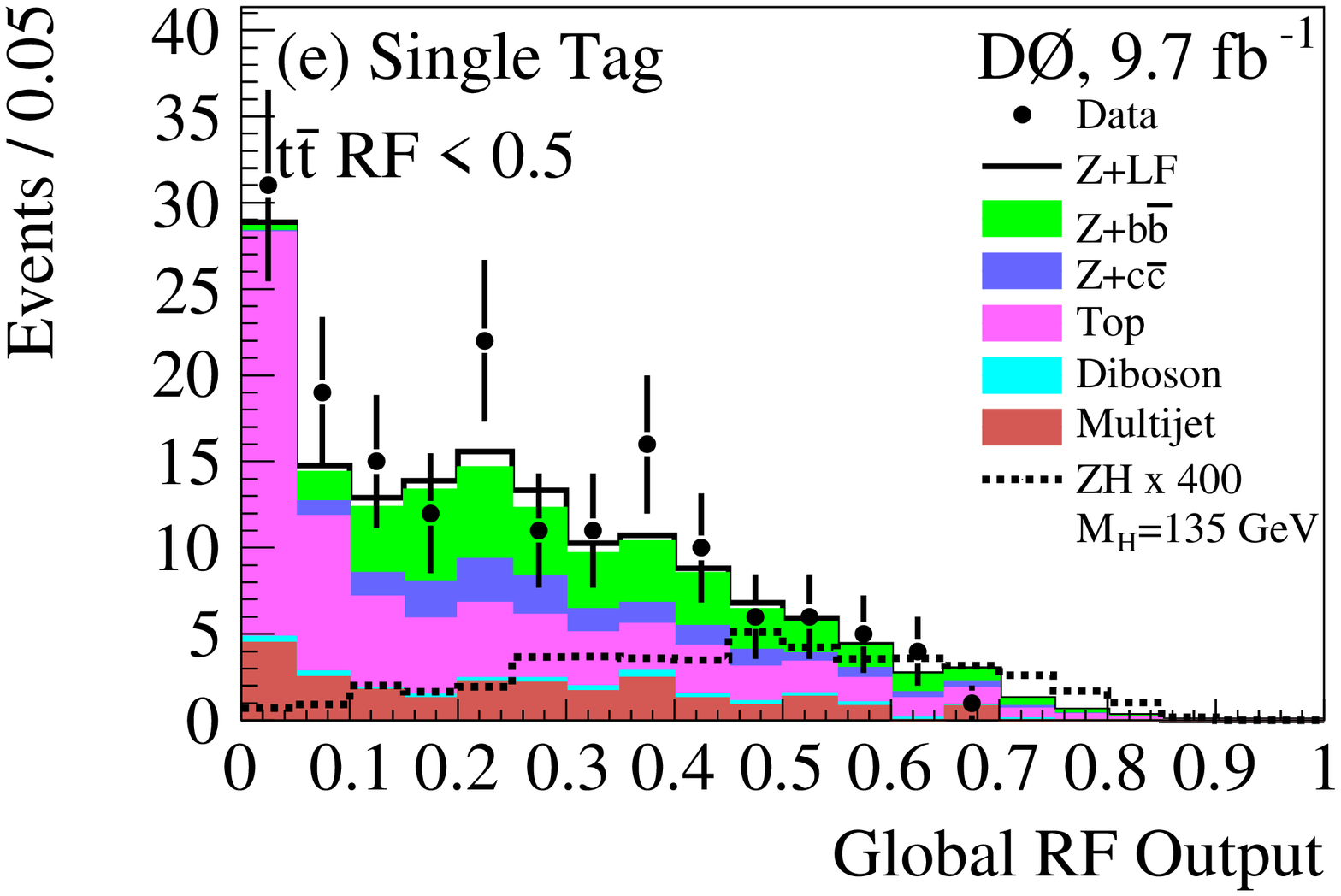} &
\includegraphics[height=0.24\textheight]{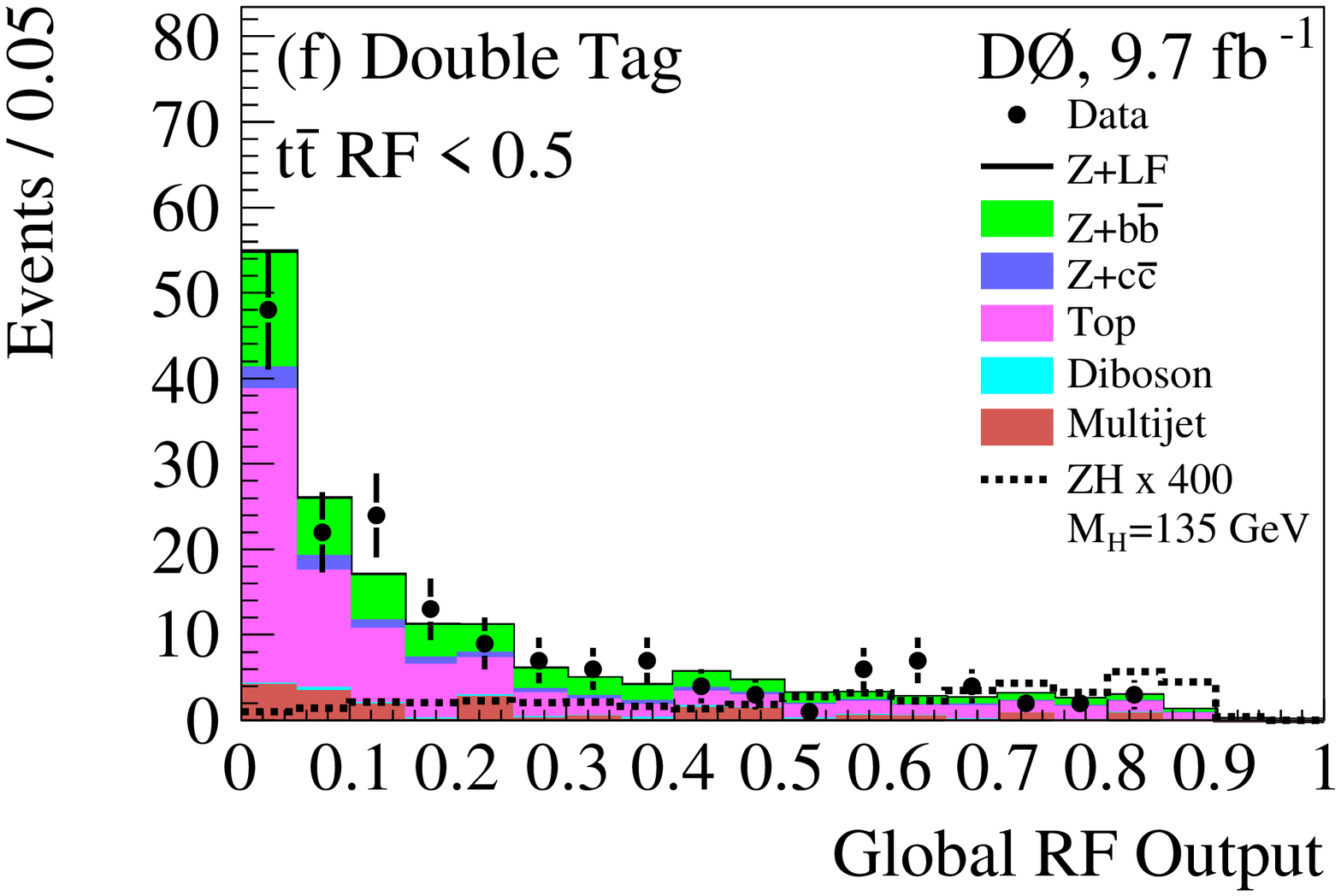} \\
\end{tabular}
\caption{(color online). Global RF distributions for
ST and DT events in the $\ttbar$-enriched region 
for $M_H$ = 115 GeV (a, b),
for $M_H$ = 125 GeV (c, d), and
for $M_H$ = 135 GeV (e, f). The signal distributions correspond to the $M_H$
used for the RF training and are scaled by a factor of 400.
\label{fig:global_rf_ttrich_post_tag}}
\end{figure*}

Figures~\ref{fig:ttbar_rf_pretag} and \ref{fig:global_rf_pretag} show
the pretag distributions of the $\ttbar$ RF and the global RF outputs,
respectively, trained for $M_H$~=~125 GeV.
Figures~\ref{fig:ttbar_rf_post_tag}-\ref{fig:global_rf_ttrich_post_tag}
show the corresponding distributions after applying the $b$-tagging
requirements for several different values of $M_H$. The requirement
that separates the $\ttbar$-depleted region ($\ttbar$ RF $<0.5$)
and the $\ttbar$-enriched region ($\ttbar$ RF $>0.5$) is shown
in Figs.~\ref{fig:ttbar_rf_pretag} and \ref{fig:ttbar_rf_post_tag}.

\section{Systematic Uncertainties}\label{sec:syst}

We assess the impact of systematic uncertainties on both the
normalization and shape of the predicted global RF distributions for
the signal and for each background source.  We summarize the magnitude of
these uncertainties in Tables \ref{tab:all_systs}~--~\ref{tab:dt_systs},
and provide additional details below.  Unless otherwise stated, we
consider each source of systematic uncertainty to be 100\% correlated for
each process across all samples.

The uncertainties on the integrated luminosity and the lepton identification 
efficiencies are absorbed by the uncertainties on the normalization procedure described in
Section~\ref{sec:normalization}. The
uncertainties on the normalization of the multijet background are
determined from the statistical uncertainties on the fit, typically
around 10\%.  These are uncorrelated across channels but are
correlated within a channel (i.e., between the different $b$-tag samples, 
and between the $\ttbar$-depleted and enriched regions).  We
compare the value of $k_Z^2$ from the combined normalization to the
values obtained from independent fits in each channel 
We assess an uncertainty for each channel that is equal to the RMS
(3--5\%) of the observed deviations.  This uncertainty is taken to be
uncorrelated across channels. The normalization of the $Z+$jets background
to the pretag data constrains that sample within the statistical
uncertainty (1--2\%) of the pretag data. Since this sample is
dominated by the $Z$+LF background, the normalization of the $\ttbar$,
diboson, and $ZH$ samples acquires a sensitivity to the inclusive $Z$
boson cross section, for which we assess a 6\% uncertainty~\cite{dyxsec}. We
assign this uncertainty to these samples as a common uncertainty.  We
apply a 9\% uncertainty to the \runiia~prediction of $Z+$LF production to account for the
different values of $k_Z^2$ obtained for \runiia~and \runiib.  For
$Z+$HF production, we evaluate a cross section uncertainty of 20\% based on
Ref.~\cite{mcfm}.  For the diboson and $\ttbar$ backgrounds, we
take the uncertainties on the cross sections to be 7\%
\cite{mcfm} and 10\% \cite{ttbarxsec}, respectively. 
The cross section uncertainty for the signal is 6\%~\cite{zhxsec}.

Sources of systematic uncertainty affecting the shapes of the 
final discriminant distributions are
the jet energy scale, jet energy resolution, 
jet identification efficiency,
and $b$-tagging efficiency.
Shape uncertainties are assessed by repeating the full analysis with 
each source of uncertainty varied by $\pm 1$ s.d.
Other sources include trigger efficiency, multijet modeling
in the \ee\ channel, PDF uncertainties~\cite{pdf},
data-determined corrections to the model for $Z+$jets,
modeling of the underlying event,
the MLM matching applied to {\sc alpgen} $Z$+LF events~\cite{mlm},
and from varying the factorization and
renormalization scales for the {\sc alpgen} $Z+$jets simulation.

\begin{table*}[h]
\begin{center}
\caption{\label{tab:all_systs}Systematic uncertainties that are common
across all sub-samples. Systematic uncertainties for $ZH$ production shown 
in this Table are obtained for $M_H=125$~GeV.
Relative uncertainties are given in percent. When two numbers are given, the first is for
\runiib\ and the second is for \runiia.}
\vskip 0.5cm
{\centerline{Relative uncertainties (\%)}}
\vskip 0.099cm
\begin{ruledtabular}
\begin{tabular}{  l  c  c  c  c  c  c  c  c }   
Contribution              & $ZH$  & Multijet& $Z$+LF    &  \zbb  & \zcc & Dibosons & \ttbar\\ \hline
Multijet Normalization    &   --  &   10    &   --      &   --   &   --  &   --   &   --  \\ %
$k^0_Z$ Uncertainty             & 1.6 / 6.9     &   --    &  --        &  --     &  --    &  1.6 / 6.9    &  1.6 / 6.9    \\ %
$k^2_Z$ Uncertainty      & --     &   --    &  0.7 / 1.8        &  0.7 / 1.8     &  0.7 / 1.8    &  --     &  --    \\ %
$k^2_Z$ RMS          & 5.1 / 3    &   --    &  5.1 / 3      &  5.1 / 3     &  5.1 / 3    &  5.1 / 3     &  5.1 / 3    \\ %
\runiia\ Normalization       & -- / 9    &   --    &  --       &  --      &  --    &  -- / 9     &  -- / 9    \\ %
Theoretical Cross Sections            & 6     &   --    &   --      &  20    &  20   &  7     &  10   \\ %
PDFs                      & 0.6   &   --    &  1.0      &  2.4   &  1.1  &  0.7   &  5.9  \\ %
\end{tabular}
\end{ruledtabular}
\end{center}
\end{table*}

\begin{table*}[h]
\begin{center}
\caption{\label{tab:st_systs}Systematic uncertainties on ST events in the
$\ttbar$-depleted and enriched regions.
Systematic uncertainties for $ZH$ production shown in this Table are obtained for $M_H=125$~GeV.
Relative uncertainties are given in percent.
As these uncertainties change the shape of the global RF distributions,
the numbers refer to average per-bin changes. 
When a range is given, the uncertainty varies by $Z$ boson decay channel.}
\vskip 0.5cm
{\centerline{Relative uncertainties (\%) in the $t\bar t$-depleted region for ST events}}
\vskip 0.099cm
\begin{ruledtabular}
\begin{tabular}{  l  c  c  c  c  c  c  c  c }   
Contribution              & $ZH$  & Multijet& $Z$+LF    &  \zbb  & \zcc & Dibosons & \ttbar\\ \hline
Jet Energy Scale       &  0.6  &   --    &  3.1      &  2.3   &  2.3  &  4.8   &  0.3  \\ %
Jet Energy Resolution  &  0.7  &   --    &  2.7      &  1.3   &  1.6  &  1.0   &  1.1  \\ %
Jet Identification     &  0.6  &   --    &  1.5      &  0.0   &  0.5  &  0.7   &  0.7  \\ %
Jet Taggability            &  2.0  &   --    &  1.9      &  1.7   &  1.7  &  1.8   &  2.2  \\ %
Heavy Flavor Tagging Efficiency  &  0.5  &   --    &   --      &  1.6   &  3.9  &   --   &  0.7  \\ %
Light Flavor Tagging Efficiency  &   --  &   --    &   68      &   --   &   --  &  2.9   &   --  \\ %
Trigger                & 0.4--2 &   --    &  0.1--2   &  0.2--2 &  0.2--2&  0.2--2&  0.5--2 \\ %
$Z$ boson $p_T$ Model          &   --  &   --    &  1.6      &  1.7   &  1.5  &   --   &   --  \\ %
$Z$+jets Jet Angles    &   --  &   --    &  1.7      &  1.7   &  1.7  &   --   &   --  \\ %
\alpgen\ MLM             &   --  &   --    &  0.2      &   --   &   --  &   --   &   --  \\ %
\alpgen\ Scale           &   --  &   --    &  0.3      &  0.5   &  0.5  &   --   &   --  \\ %
Multijet Shape for $ee$ channel    &   --  &   45    &   --      &   --   &   --  &   --   &   --  \\ %
Underlying Event       &   --  &   --    &  0.4      &  0.4   &  0.4  &   --   &   --  \\ %
\end{tabular}
\end{ruledtabular}

\vskip 0.5cm
{\centerline{Relative uncertainties (\%) in the $t\bar t$-enriched region for ST events}}
\vskip 0.099cm
\begin{ruledtabular}
\begin{tabular}{  l  c  c  c  c  c  c  c  c }   
Contribution              & $ZH$  & Multijet& $Z$+LF  &  \zbb & \zcc & Dibosons & \ttbar\\ \hline
Jet Energy Scale       &  7.5  &   --    &  4.6      &  1.7   &  3.9   &  11   &  2.5  \\ %
Jet Energy Resolution  &  0.2  &   --    &  4.5      &  0.7   &  3.1  &  3.9   &  0.7  \\ %
Jet Identification                 &  1.2  &   --    &  2.1      &  1.0   &  1.2  &  0.9   &  0.7  \\ %
Jet Taggability            &  2.1  &   --    &  7.3      &  2.7   &  3.0  &  2.0   &  3.2  \\ %
Heavy Flavor Tagging Efficiency  &  0.5  &   --    &   --      &  1.3   &  4.8  &   --   &  0.8  \\ %
Light Flavor Tagging Efficiency  &   --  &   --    &   73      &   --   &   --  &  4.1   &   --  \\ %
Trigger                & 1--4   &   --    &  1--4      &  0.7--4 &  0.7--4&  1--8   &  1--8  \\ %
$Z$ boson $p_T$ Model          &   --  &   --    &  3.3      &  1.5   &  1.4  &   --   &   --  \\ %
$Z$+jets Jet Angles    &   --  &   --    &  1.7      &  2.3   &  2.7  &   --   &   --  \\ %
\alpgen\ MLM             &   --  &   --    &  0.4      &   --   &   --  &   --   &   --  \\ %
\alpgen\ Scale           &   --  &   --    &  0.7      &  0.7   &  0.7  &   --   &   --  \\ %
Multijet Shape for $ee$ channel    &   --  &   59    &   --      &   --   &   --  &   --   &   --  \\ %
Underlying Event       &   --  &   --    &  0.9      &  1.1   &  1.1  &   --   &   --  \\
\end{tabular}
\end{ruledtabular}

\end{center}
\end{table*}

\begin{table*}[phtb]
\begin{center}
\caption{\label{tab:dt_systs}Systematic uncertainties on DT events in the
$\ttbar$-depleted and enriched regions.
Systematic uncertainties for $ZH$ production shown in this Table are obtained for $M_H=125$~GeV.
Relative uncertainties are given in percent.
As these uncertainties change the shape of the global RF distributions,
the numbers refer to average per-bin changes. 
When a range is given, the uncertainty varies by $Z$ boson decay channel.}
\vskip 0.5cm
{\centerline{$ZH \rightarrow \ell\ell \bbbar$ Relative uncertainties (\%) in the $t\bar t$-depleted region for DT events}}
\vskip 0.099cm
\begin{ruledtabular}
\begin{tabular}{  l  c c  c  c  c  c  c  c }  \\
Contribution             & $ZH$ & Multijet& $Z$+LF  &  \zbb & \zcc & Dibosons & \ttbar\\  \hline
Jet Energy Scale      &  0.5  &   --     &  4.6   &  3.0   &  1.3   &  4.5   &  1.4  \\ %
Jet Energy Resolution &  0.4  &   --     &  7.0   &  1.8   &  2.9   &  0.9   &  0.9  \\ %
Jet Identification                &  0.6  &   --     &  7.9   &  0.3   &  0.5   &  0.5   &  0.5  \\ %
Jet Taggability           &  1.7  &   --     &  7.0   &  1.5   &  1.5   &  3.0   &  1.7  \\ %
Heavy Flavor Tagging Efficiency &  4.4  &   --     &   --   &  5.0   &  5.6   &   --   &  3.8  \\ %
Light Flavor Tagging Efficiency &   --  &   --     &  75    &   --   &   --   &  4.7   &   --  \\ %
Trigger               &  0.4--2&   --     &  0.6--6 &  0.3--2 &  0.3--3 &  0.4--2 &  0.6--5\\ %
$Z_{p_T}$ Model       &   --  &   --     &  2.9   &  1.4   &  1.9   &   --   &   --  \\ %
$Z$+jets Jet Angles   &   --  &   --     &  1.9   &  3.5   &  3.8   &   --   &   --  \\ %
\alpgen\ MLM            &   --  &   --     &  0.2   &   --   &   --   &   --   &   --  \\ %
\alpgen\ Scale          &   --  &   --     &  0.4   &  0.5   &  0.5   &   --   &   --  \\ %
Multijet Shape for $ee$ channel   &   --  &    66    &   --   &   --   &   --   &   --   &   --  \\ %
Underlying Event      &   --  &   --     &  0.5   &  0.4   &  0.4   &   --   &   --  \\ %
\end{tabular}
\end{ruledtabular}

\vskip 0.5cm
{\centerline{$ZH \rightarrow \ell\ell \bbbar$ Relative uncertainties (\%) in the $t\bar t$-enriched region for DT events}}
\vskip 0.099cm
\begin{ruledtabular}
\begin{tabular}{  l  c c  c  c  c  c  c  c }  \\
Contribution             & $ZH$ & Multijet& $Z$+LF  &  \zbb & \zcc & Dibosons & \ttbar\\  \hline
Jet Energy Scale      &  6.6  &   --    &  0.8     &  1.6   &  2.2   &  5.9   &  1.5  \\ %
Jet Energy Resolution  &  1.4  &   --    &  267     &  1.4   &  2.1   &  4.0   &  0.4  \\ %
Jet Identification                &  0.9  &   --    &  0.6     &  0.5   &  3.6   &  2.8   &  0.6  \\ %
Jet Taggability           &  2.0  &   --    &  0.9     &  1.6   &  1.9   &  3.1   &  2.1  \\ %
Heavy Flavor Tagging Efficiency &  4.0  &   --    &   --     &  5.1   &  6.6   &   --   &  4.2  \\ %
Light Flavor Tagging Efficiency &   --  &   --    &  72      &   --   &   --   &   --   &   --  \\ %
Trigger               &  1--3  &   --    &  1--3     &  0.6--3 &  0.7--4 &  0.7--4 &  1--3  \\ %
$Z$ boson $p_T$ Model          &   --  &   --    &  1.8     &  1.4   &  1.5   &   --   &   --  \\ %
$Z$+jets Jet Angles   &   --  &   --    &  1.4     &  3.7   &  2.3   &   --   &   --  \\ %
\alpgen\ MLM            &   --  &   --    &  0.5     &   --   &   --   &   --   &   --  \\ %
\alpgen\ Scale          &   --  &   --    &  0.8     &  0.5   &  0.4   &   --   &   --  \\ %
Multijet Shape for $ee$ channel   &   --  &    91   &   --     &   --   &   --   &   --   &   --  \\ %
Underlying Event       &   --  &   --    &  0.9     &  0.7   &  0.5   &   --   &   --  \\
\end{tabular}
\end{ruledtabular}

\end{center}
\end{table*}

\section{Results}\label{sec:results}

We use the global RF output distributions of the four sub-samples (ST and
DT in the $\ttbar$-depleted and $\ttbar$-enriched regions) in each
channel along with the corresponding systematic uncertainties to
extract results for both Higgs boson production
and diboson production. The use of separate channels and sub-samples
takes advantage of the sensitivity from the signal-rich
sub-samples and allows for a better background assessment based on
the signal-poor sub-samples. The binning of each distribution
is chosen such that the statistical uncertainty for each bin
is less than 20\% for the signal-plus-background prediction
and 25\% for the background-only prediction.

We evaluate the consistency of the data with the
background-only ($B$) and signal-plus-background ($S+B$) hypotheses using
a modified frequentist (CL$_S$) method~\cite{cls}. This method uses the negative
log likelihood ratio $LLR=-2\ln(L_{S+B}/L_B)$, where $L_{S+B}$ and 
$L_B$ are the Poisson likelihoods for the $S+B$ 
and the $B$ hypotheses, respectively.

We combine our results by summing the $LLR$ over all bins of all contributing
channels and sub-samples.  The signal and background predictions are functions of
nuisance parameters that account for the presence of systematic
uncertainties.  
We maximize $L_{S+B}$ with respect to the $S+B$ hypothesis and $L_B$ 
with respect to the $B$ hypothesis with independent fits that allow the
sources of nuisance parameters to vary within Gaussian priors~\cite{wade}.
The maximized values of $L_{B}$ and $L_{S+B}$ are then used in the calculation of
the $LLR$.

We integrate the $LLR$ distributions obtained from $B$ and $S+B$
pseudo-experiments to obtain the $p$-values $CL_{B}$ and $CL_{S+B}$ for the two
hypotheses.  If the data are consistent with the $B$
hypothesis, we exclude values of the product of the $ZH$ production cross
section and branching ratios for which $CL_{S}~=~CL_{S+B} / CL_{B} < 0.05$
at the 95\% C.L.

\subsection{Results for Diboson Production}\label{sec:diboson_results}

To validate the search procedure, we search for $ZZ$ production in the
$\ell^+\ell^- \bbbar$  final state.  We use
the same event selection, corrections to our background
models, normalization fit parameters, RF training procedure, and 
statistical analysis methods as for
the $ZH$ search.  Our search also includes contributions from 
$ZZ \rightarrow \ell^+\ell^- \ccbar$ and $WZ$ production in the
$\csbar\ell^+\ell^-$ final state 
where the $c$ jet passes the $b$-tagging requirement. 
We collectively refer to them as $VZ$ production.  
The $WW$ process is considered to be background.

\begin{figure}[htbp]\centering
\includegraphics[height=0.24\textheight]{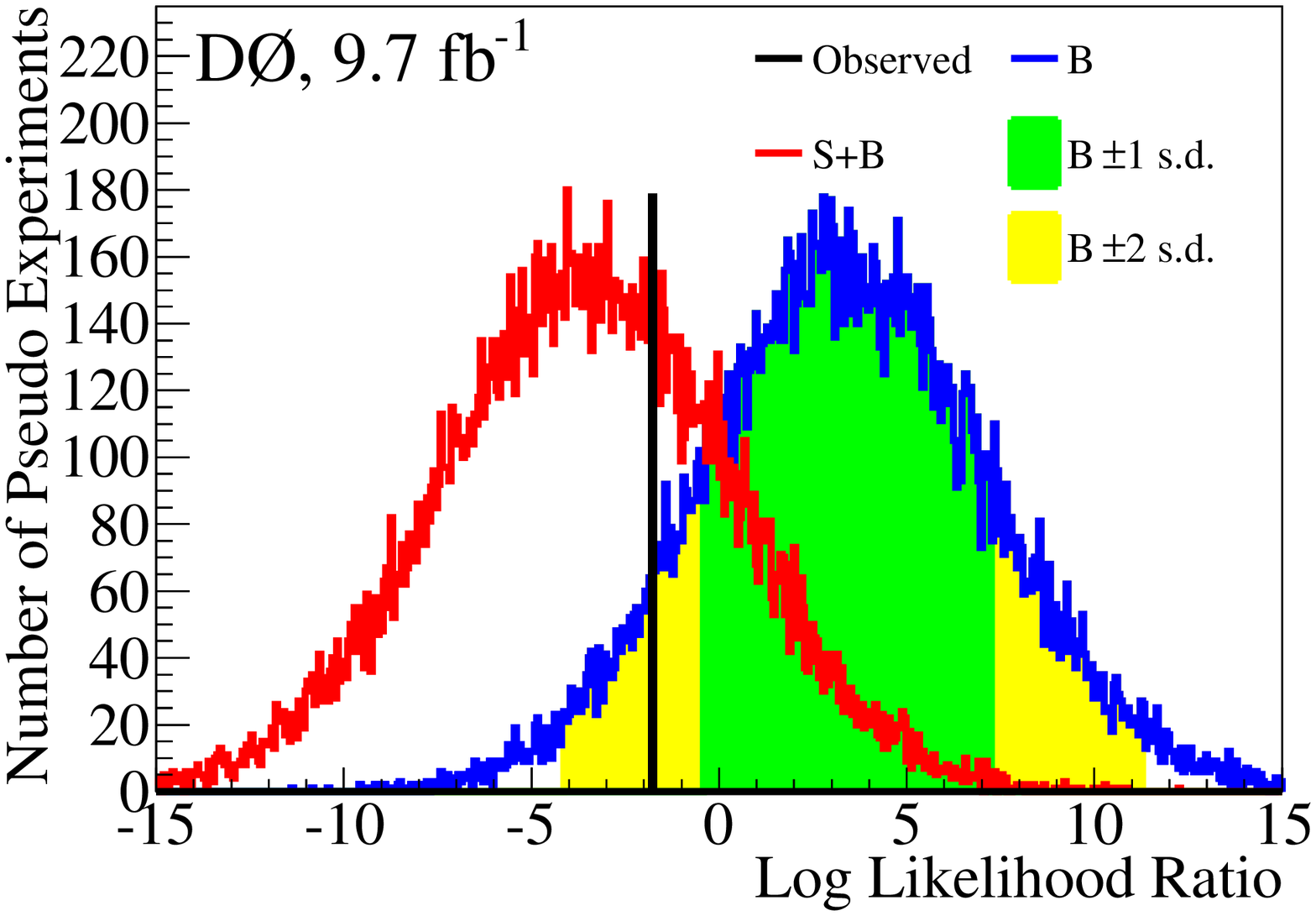}  
\caption{\label{fig:vz_llr} (color online). $LLR$ distributions obtained from
  $B$ and $S+B$ pseudo-experiments, using the global RF output
  as the final variable, for the $VZ$ search. The
  vertical line indicates the $LLR$ obtained from the data.}
\end{figure}

\begin{figure}[ht]\centering
\includegraphics[height=0.24\textheight]{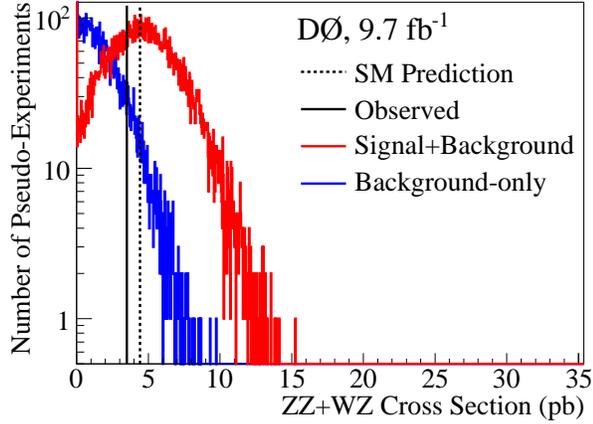}
\caption{\label{fig:vz_xsec_pe} (color online). Distribution of $VZ$
  cross sections obtained from $B$ and $S+B$ 
   pseudo-experiments.  The observed cross section from the 
  data and the SM cross section are also shown.}
\end{figure}

\begin{figure*}[htbp]
\begin{tabular}{cc}
\includegraphics[height=0.24\textheight]{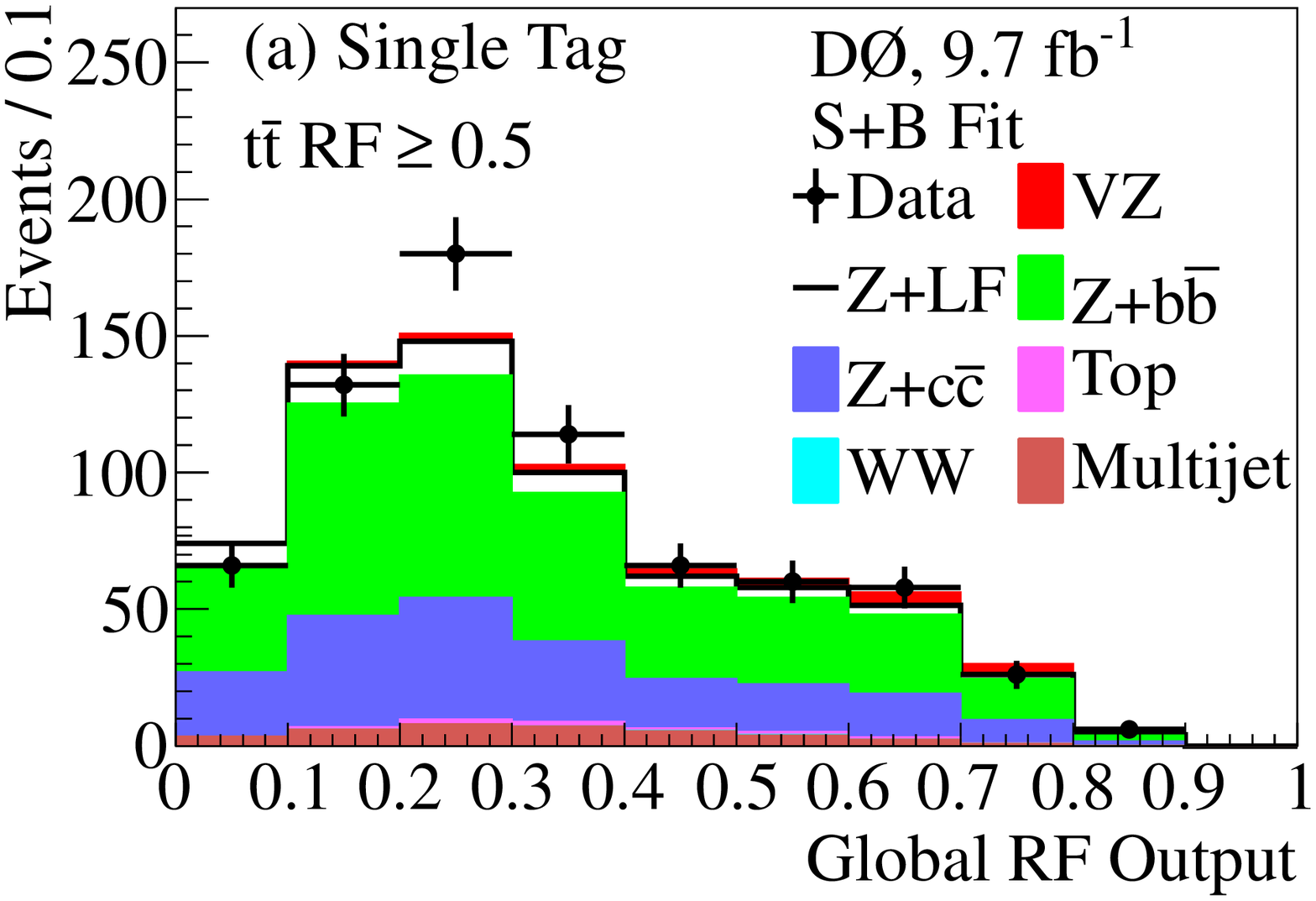}  &
\includegraphics[height=0.24\textheight]{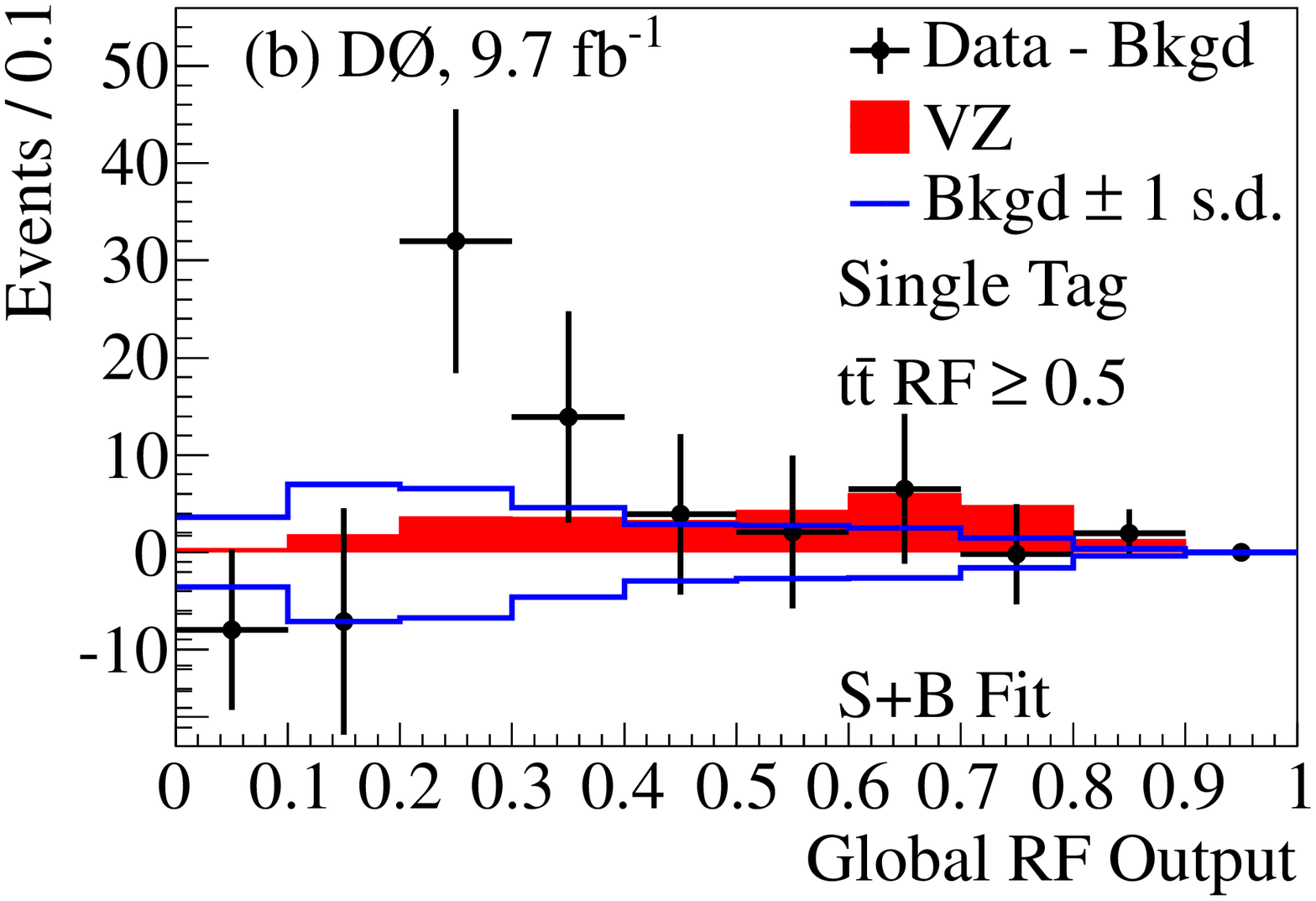}  \\
\includegraphics[height=0.24\textheight]{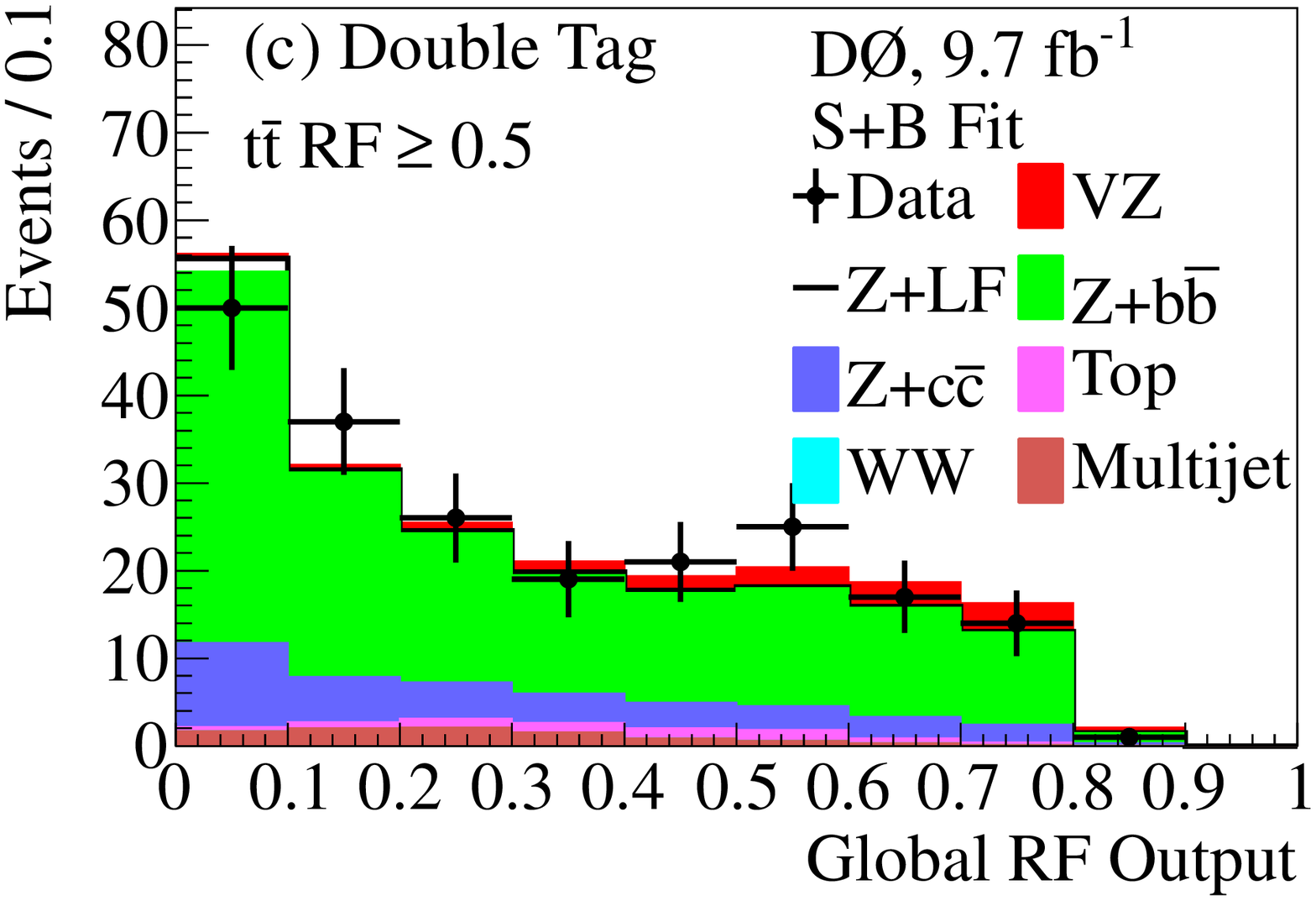} &
\includegraphics[height=0.24\textheight]{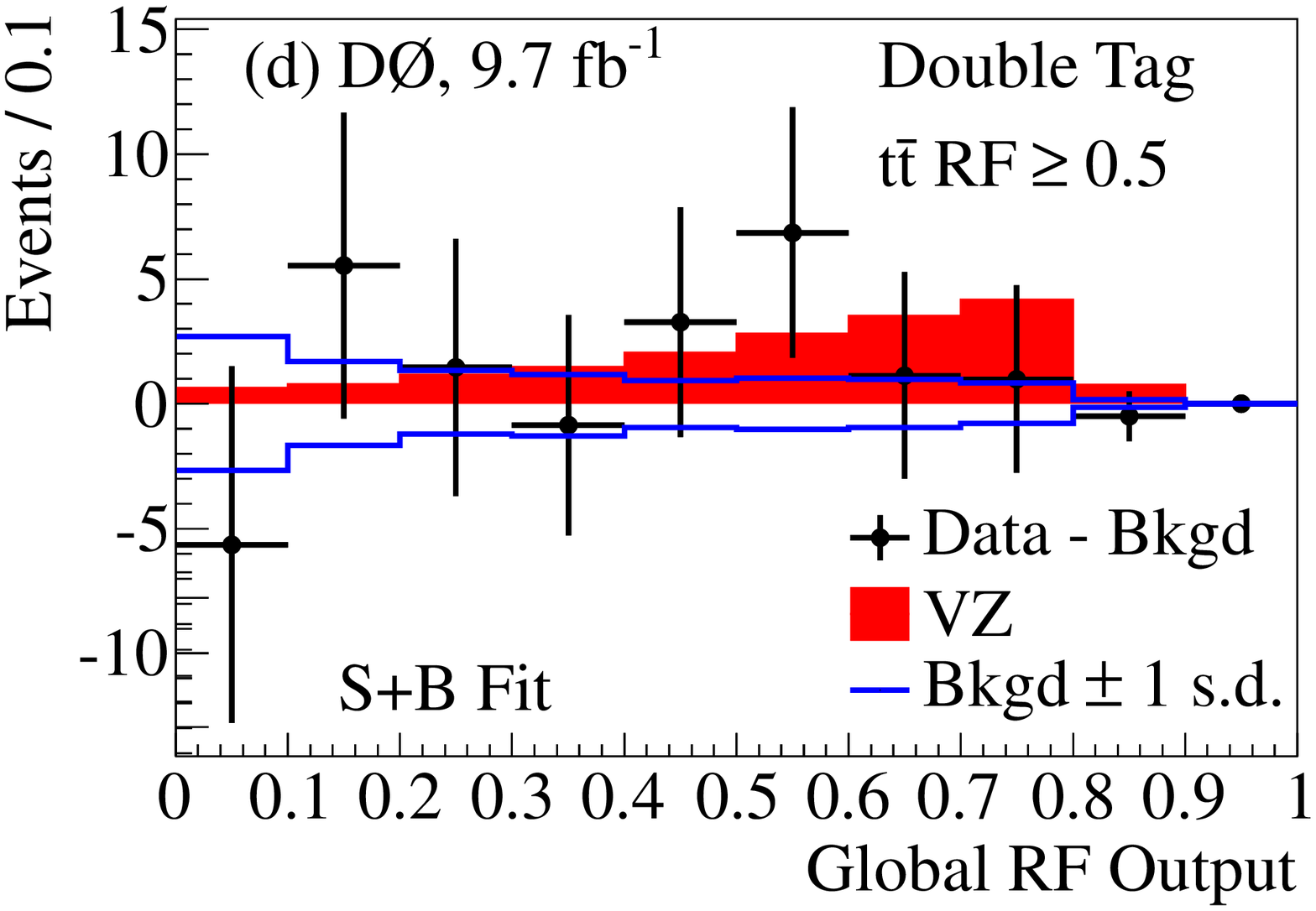} \\
\includegraphics[height=0.24\textheight]{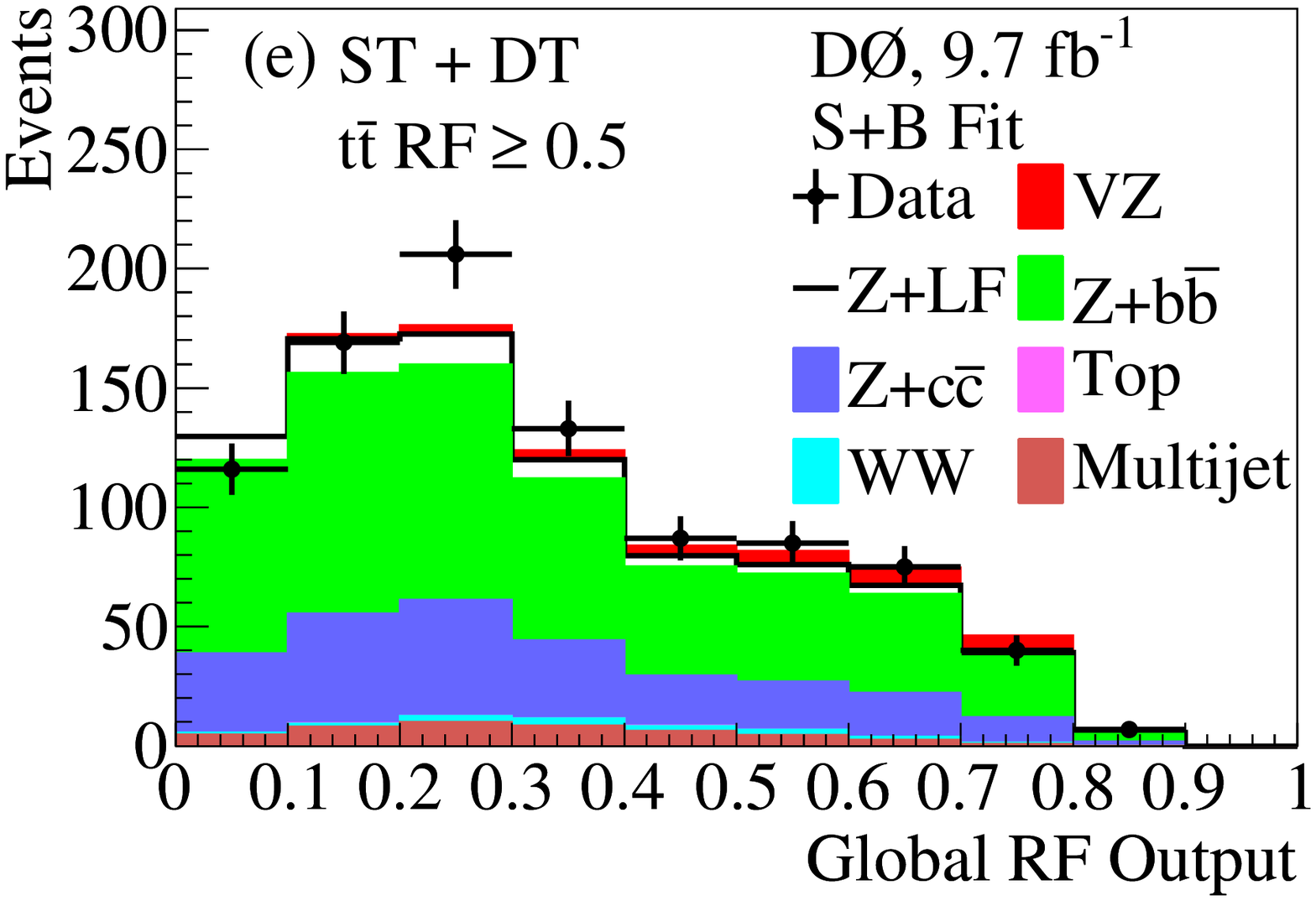} &
\includegraphics[height=0.24\textheight]{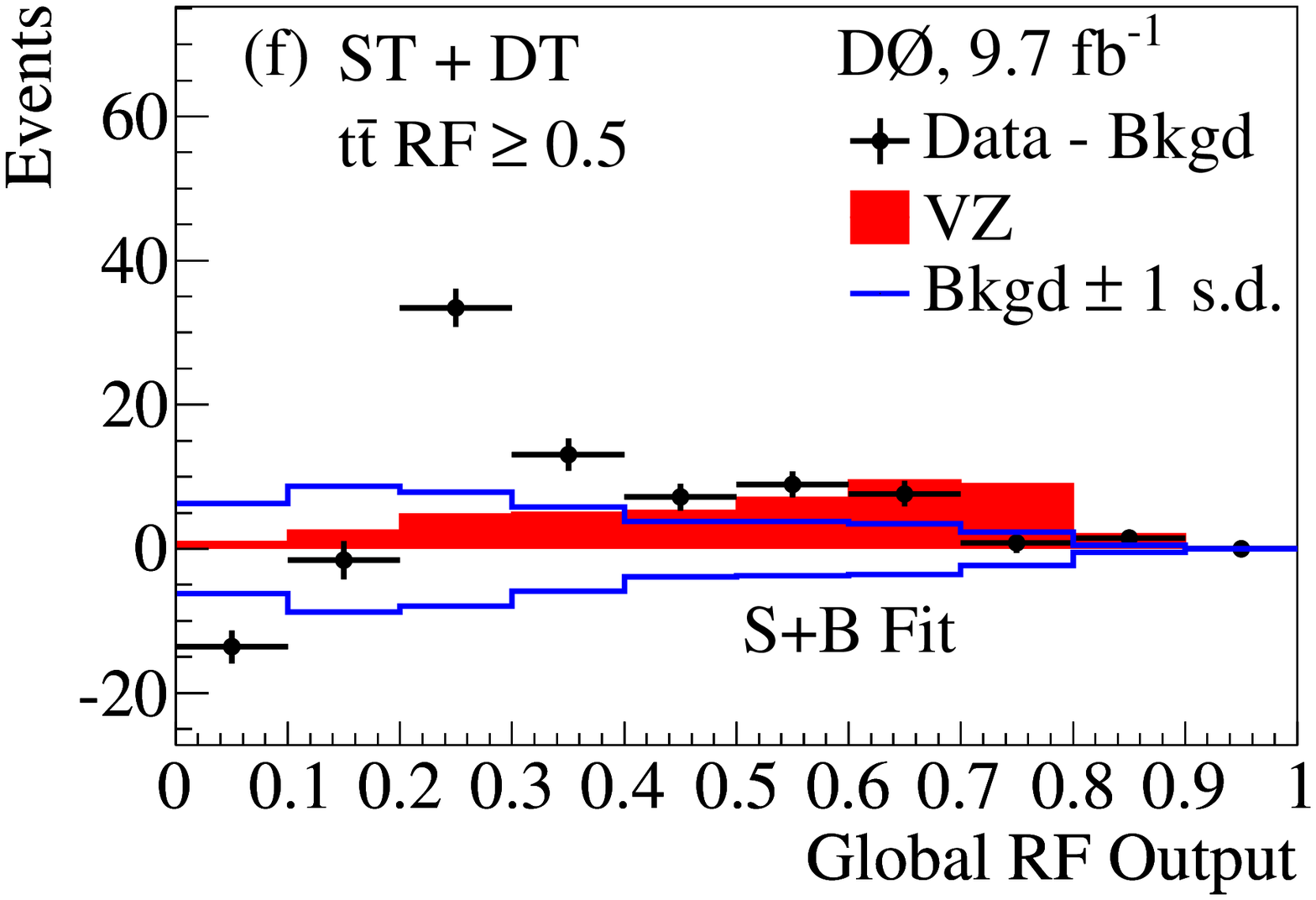} \\
\end{tabular} 
\caption{\label{fig:postfit_vzrf} (color online). Global RF output distributions for the $VZ$ search after
the fit to data in the $S+B$ hypothesis in (a) ST events, (c) DT events, 
and (e) ST and DT events combined.
Distributions are summed over all $Z\to\ell\ell$ channels.  The $VZ$ signal
distribution, scaled to the measured $\sigma_{VZ}$,
is compared to the data after subtracting the fitted background in (b) ST events,
(d) DT events, and (e) ST and DT events combined.  
Data points are shown with Poisson statistical errors.
Also shown is the uncertainty on the background after the
fit.}
\end{figure*}

\begin{figure*}[htbp]
\begin{tabular}{cc}
\includegraphics[height=0.24\textheight]{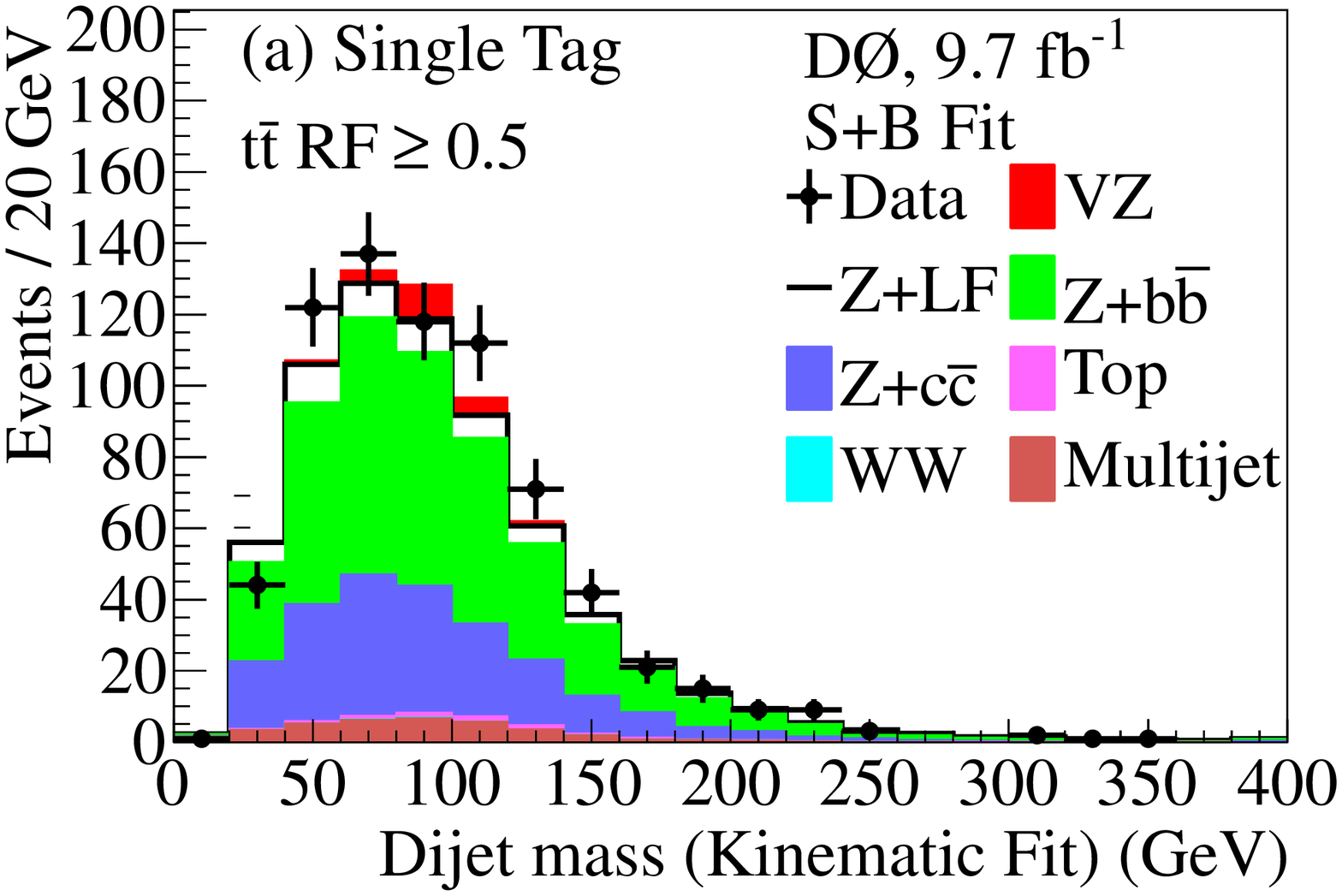}  &
\includegraphics[height=0.24\textheight]{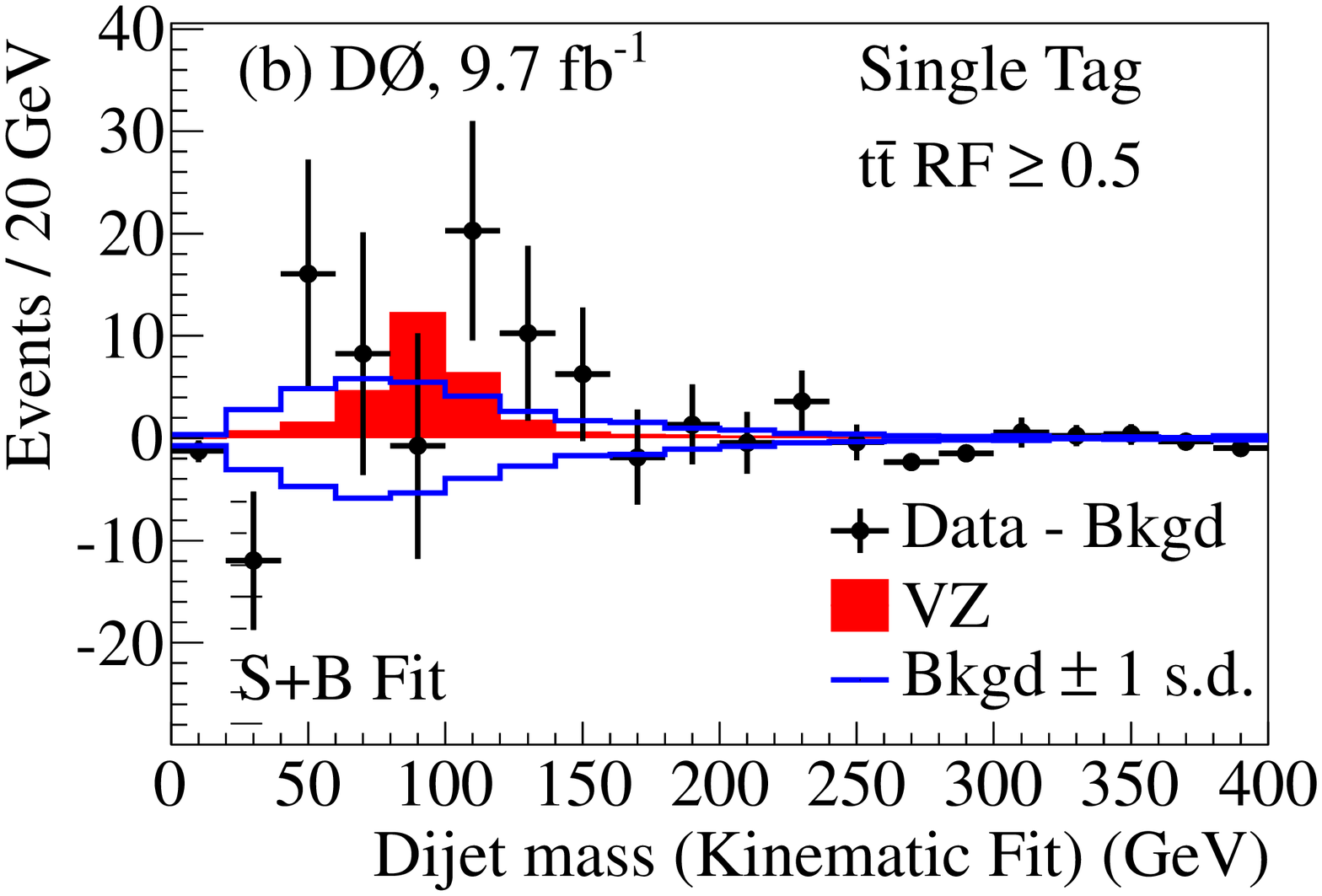}  \\
\includegraphics[height=0.24\textheight]{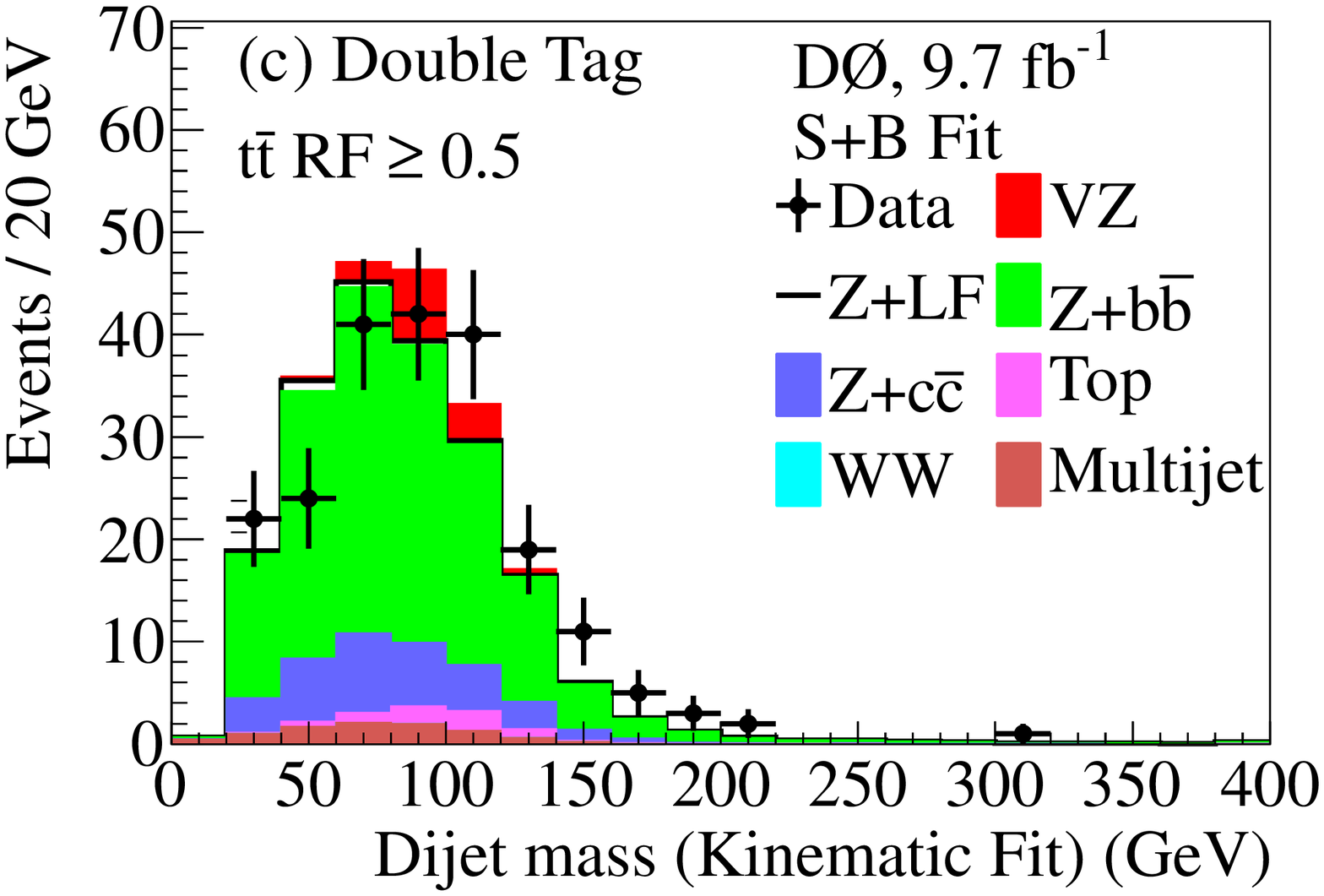} &
\includegraphics[height=0.24\textheight]{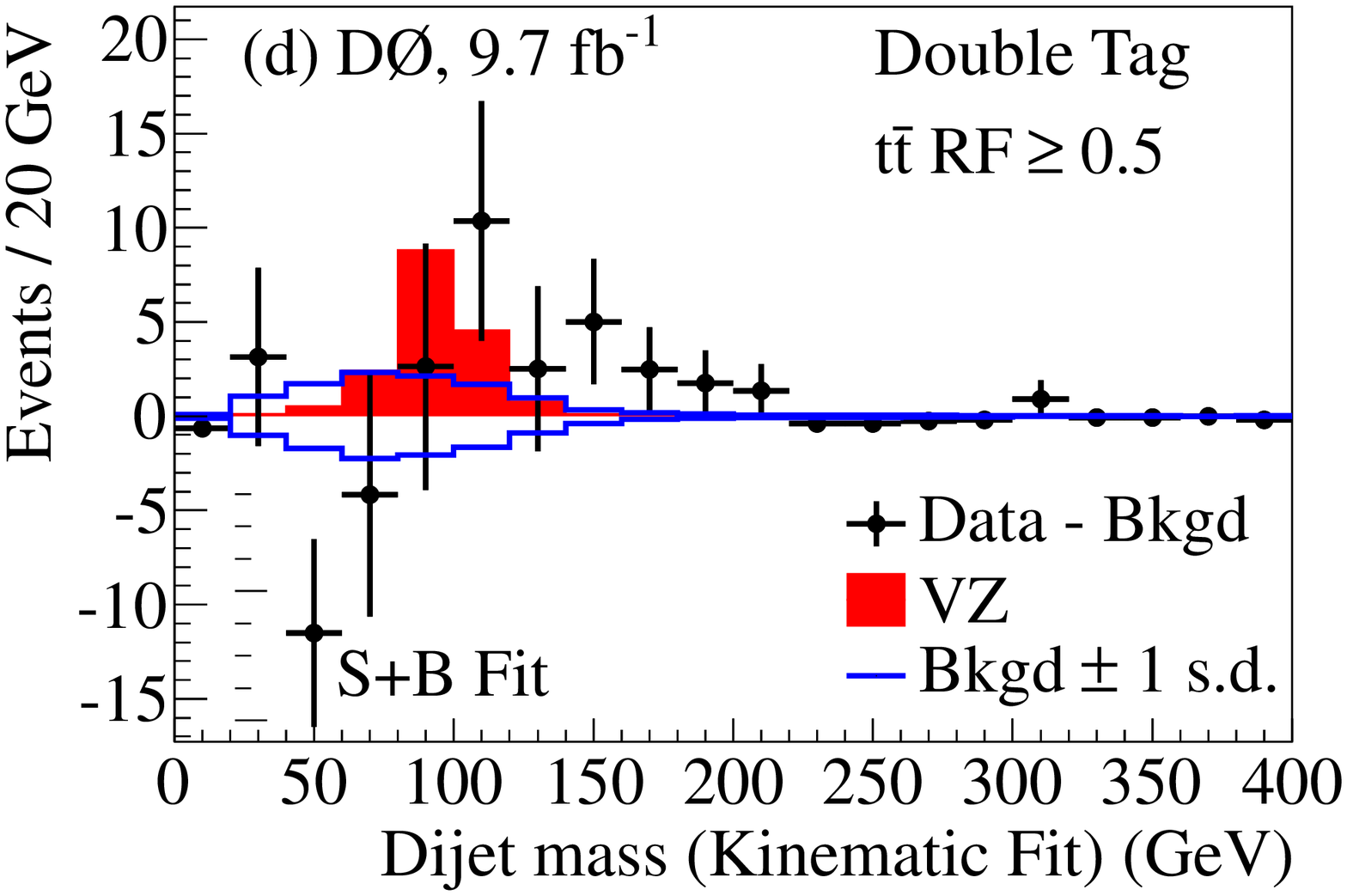} \\
\includegraphics[height=0.24\textheight]{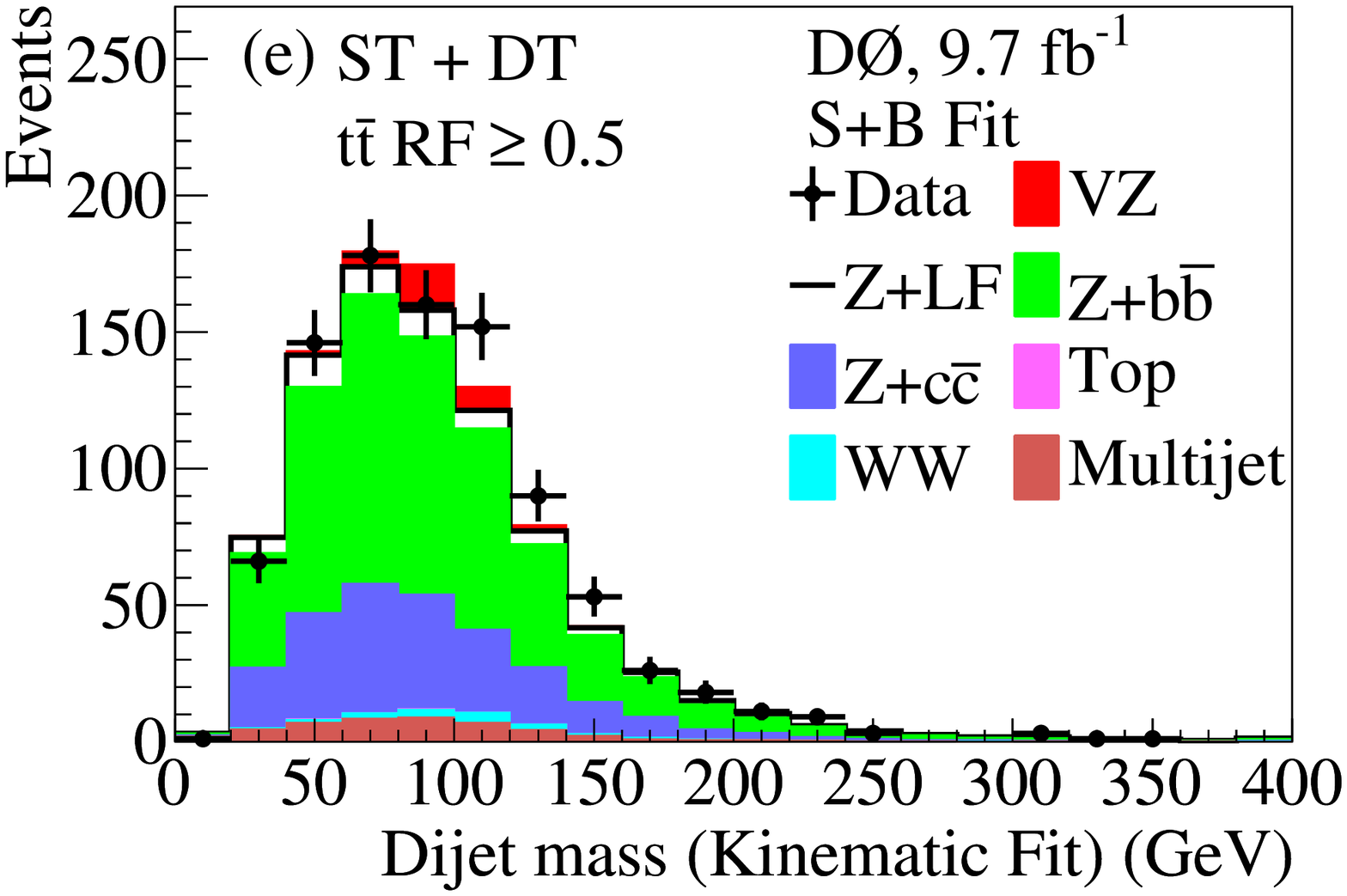} &
\includegraphics[height=0.24\textheight]{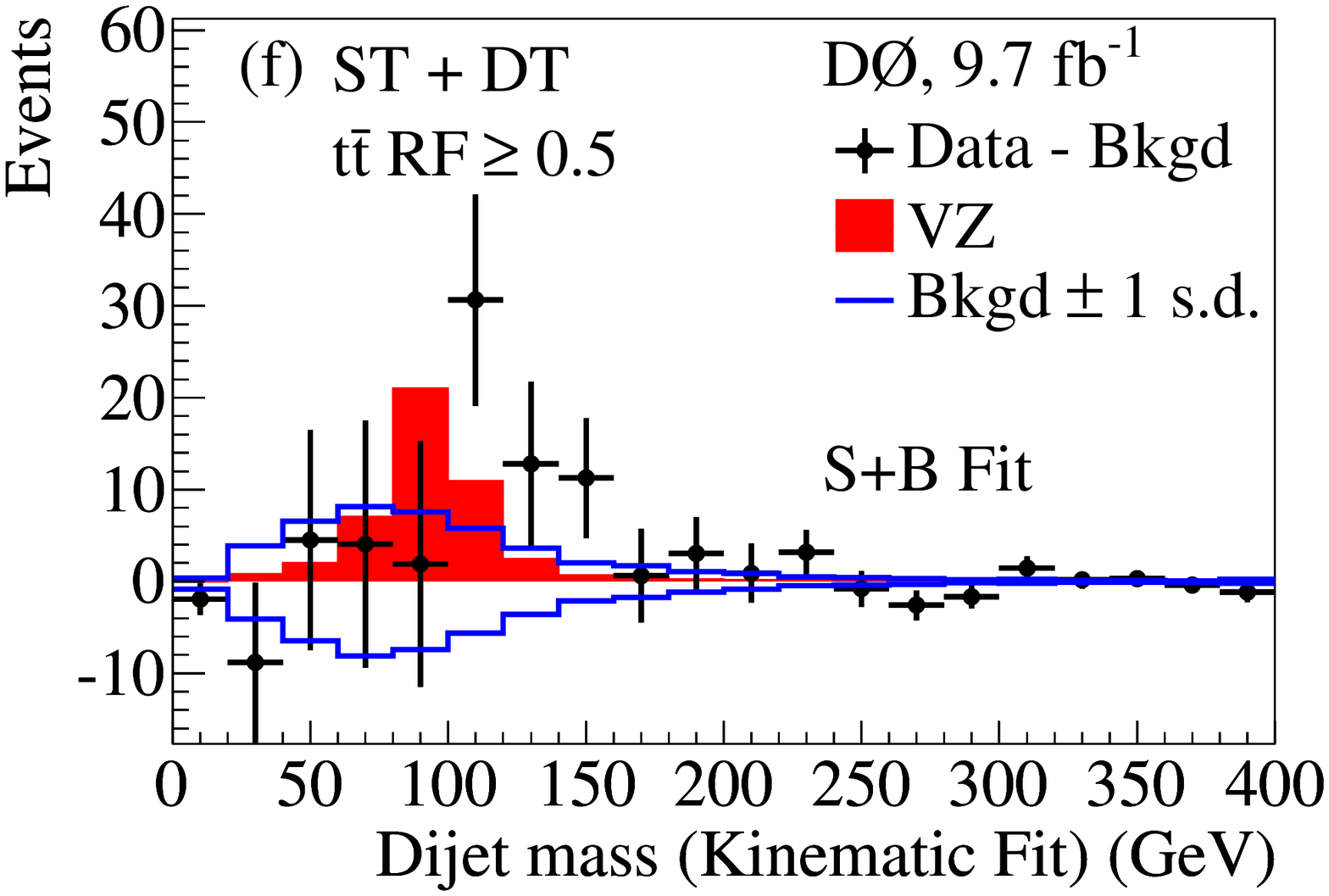} \\
\end{tabular} 
\caption{\label{fig:postfit_vzmjjfit} (color online). Dijet invariant mass distributions
for the $VZ$ search
after the kinematic fit and after the fit to the data in 
the $S+B$ hypothesis in (a) ST events, (c) DT events,
and (e) ST and DT events combined.  Distributions are summed over all 
$Z\to\ell\ell$ channels.  The $VZ$ signal
distribution, scaled to the measured $\sigma_{VZ}$,
is compared to the data after subtracting the fitted background in (b) ST events,
(d) DT events, and (e) ST and DT events combined.  
Also shown is the uncertainty on the background after the $S+B$
fit.}
\end{figure*}

Figure~\ref{fig:vz_llr} compares the $LLR$ value observed in the data to
distributions obtained from $B$ and $S+B$
pseudo-experiments. To obtain $\sigma_{VZ}$
in units of the SM value, we maximize $L_{S+B}$ with respect to the
nuisance parameters and a signal scale factor $f$, keeping the ratio
of the $ZZ$ and $WZ$ cross sections fixed to the SM prediction.  We
find $f=\vzRFobsSF\pm\vzRFerrSF$, which translates to
$\sigma_{VZ}=\vzRFobs\pm\vzRFerr$~pb given the
predicted total SM cross section of 
$\sigma_{VZ}=4.4\pm0.3~\text{pb}$~\cite{mcfm}. Figure~\ref{fig:vz_xsec_pe}
compares this result to the SM cross section and to the
distribution of results obtained from $B$ and $S+B$
pseudo-experiments.  The probability ($p$-value) that the
$B$ hypothesis results in a cross section greater than that
determined from the data is 0.071, equivalent to 1.5 standard
deviations (s.d.).  The expected $p$-value is 0.032, corresponding to
1.9 s.d.  In Figs.~\ref{fig:postfit_vzrf} and
\ref{fig:postfit_vzmjjfit} we show the global RF and 
post-kinematic fit dijet mass distributions after the likelihood fit, separately 
for ST and DT events in the $\ttbar$-depleted region. 
The diboson signal consists of 66\% (93\%) $ZZ$ production
and 34\% (7\%) $WZ$ production in the ST (DT) sample.

\subsection{Higgs Boson Search Results}\label{sec:higgs_results}

\begin{figure*}[htbp]
\begin{tabular}{cc}
\includegraphics[height=0.24\textheight]{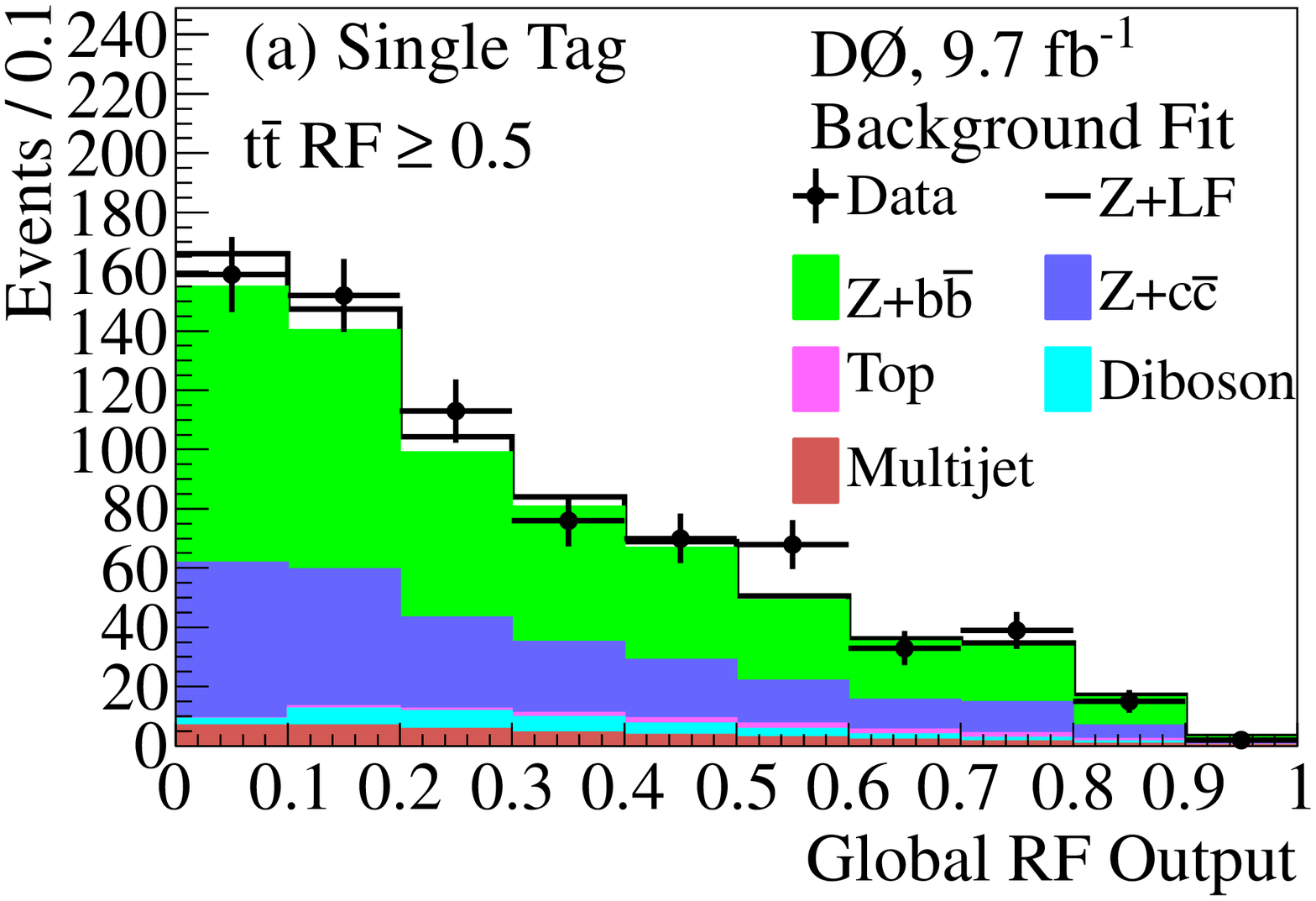}  &
\includegraphics[height=0.24\textheight]{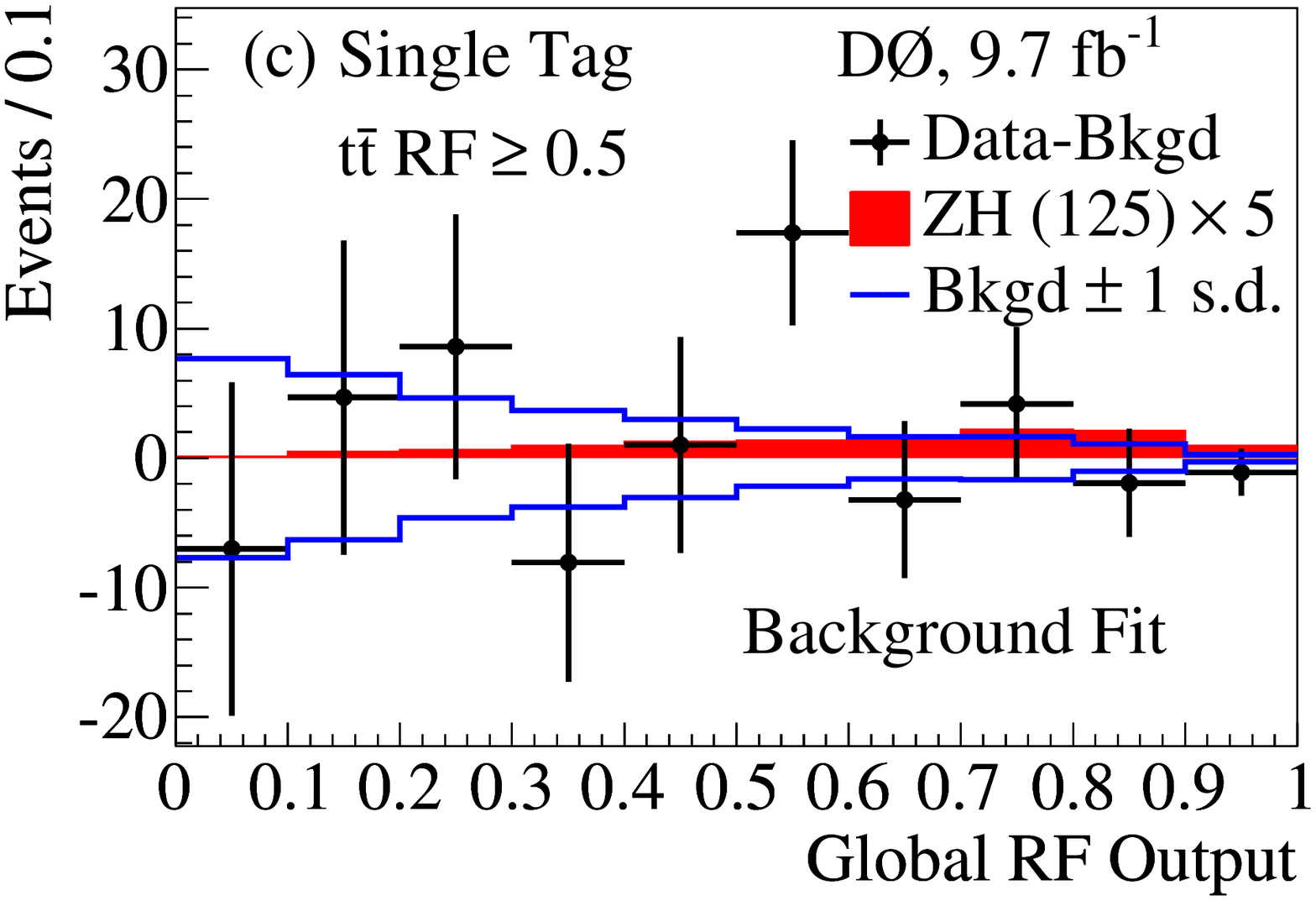}  \\
\includegraphics[height=0.24\textheight]{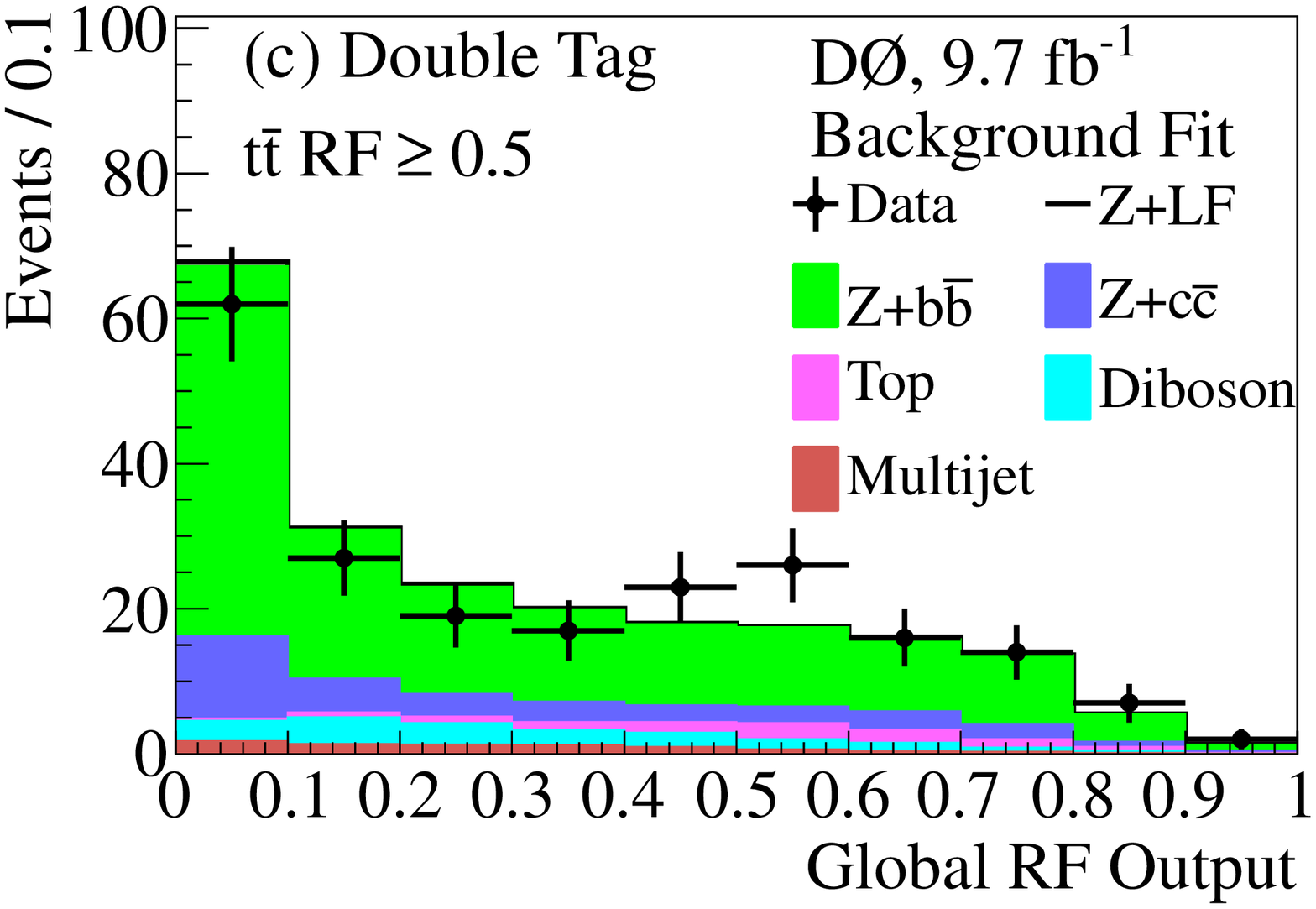} &
\includegraphics[height=0.24\textheight]{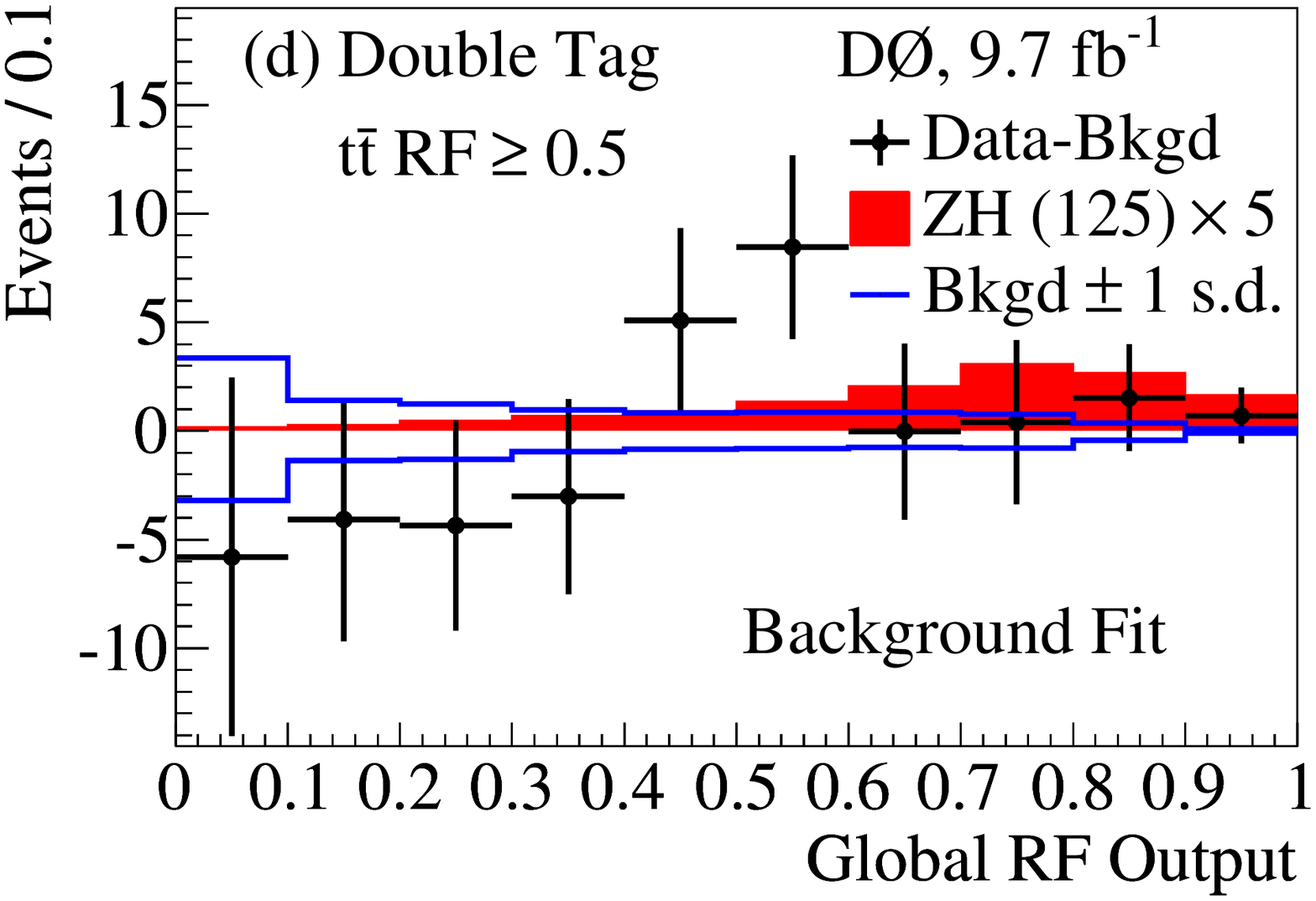} \\
\includegraphics[height=0.24\textheight]{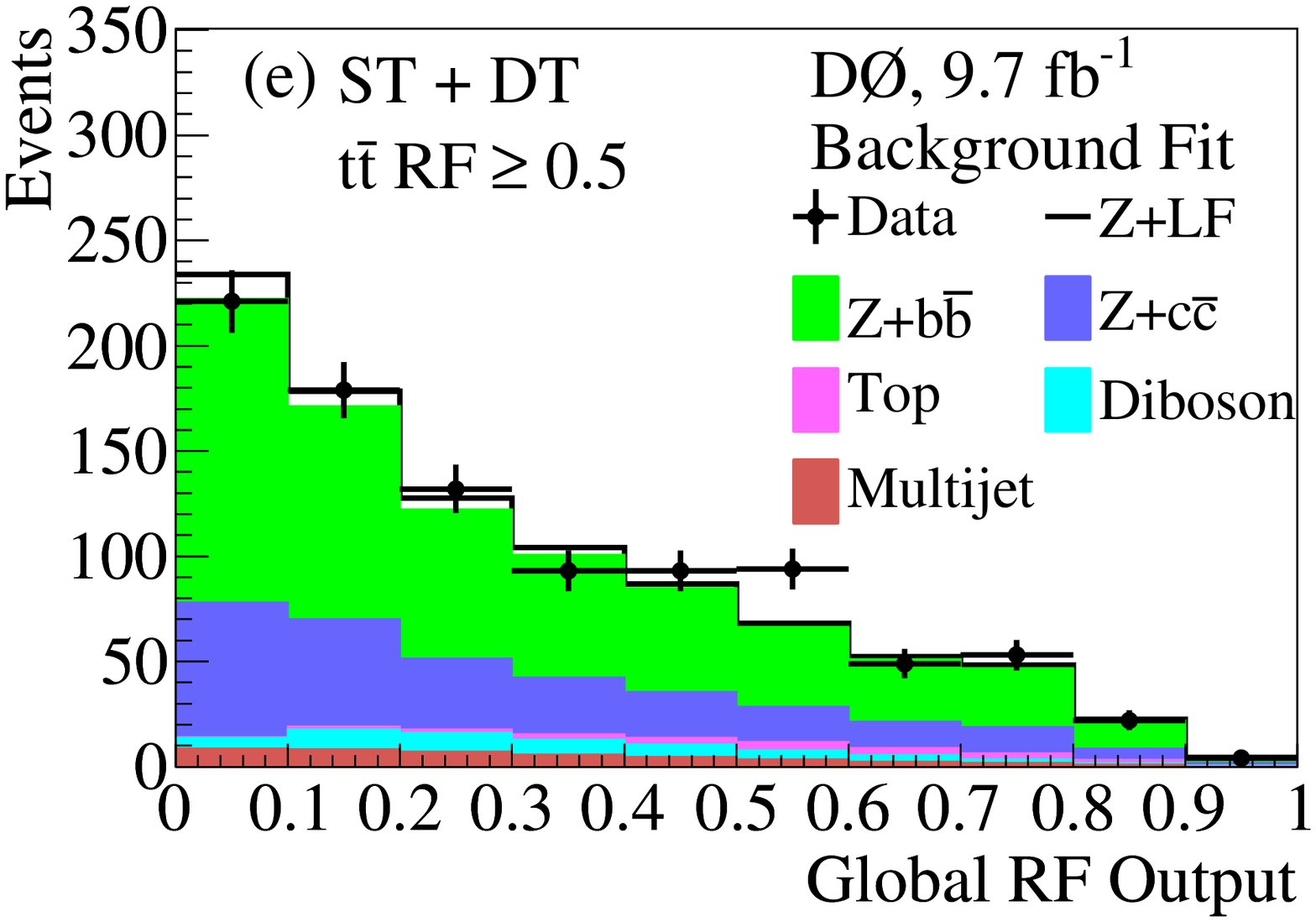} &
\includegraphics[height=0.24\textheight]{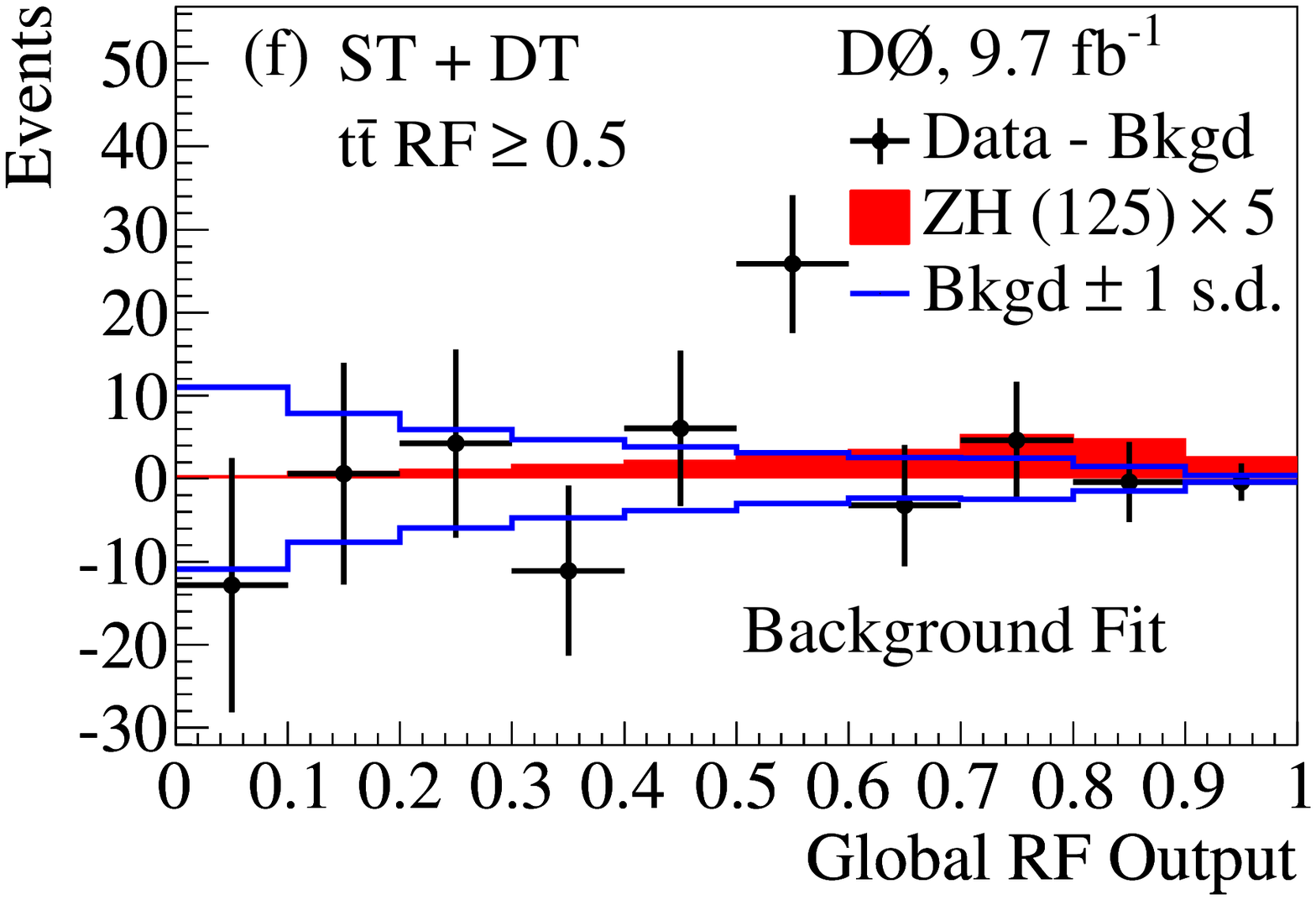} \\
\end{tabular} 
\caption{\label{fig:rf_poor_postfit} (color online). Global RF output distributions in
  the $\ttbar$-depleted region, assuming $M_H=$ 125 GeV, after the 
  fit to the data in the $B$ hypothesis 
  for (a) ST events, (c) DT events, and (e) ST and DT events combined.
  Background-subtracted distributions for (a), (c), and (e) are shown in (b), (d),
  and (f), respectively.  Signal distributions for $M_H=$ 125 GeV are shown with the SM
  cross section scaled by a factor of $5$ in (b), (d), and (f).}
 \end{figure*}

\begin{figure*}[htbp]
\begin{tabular}{cc}
\includegraphics[height=0.24\textheight]{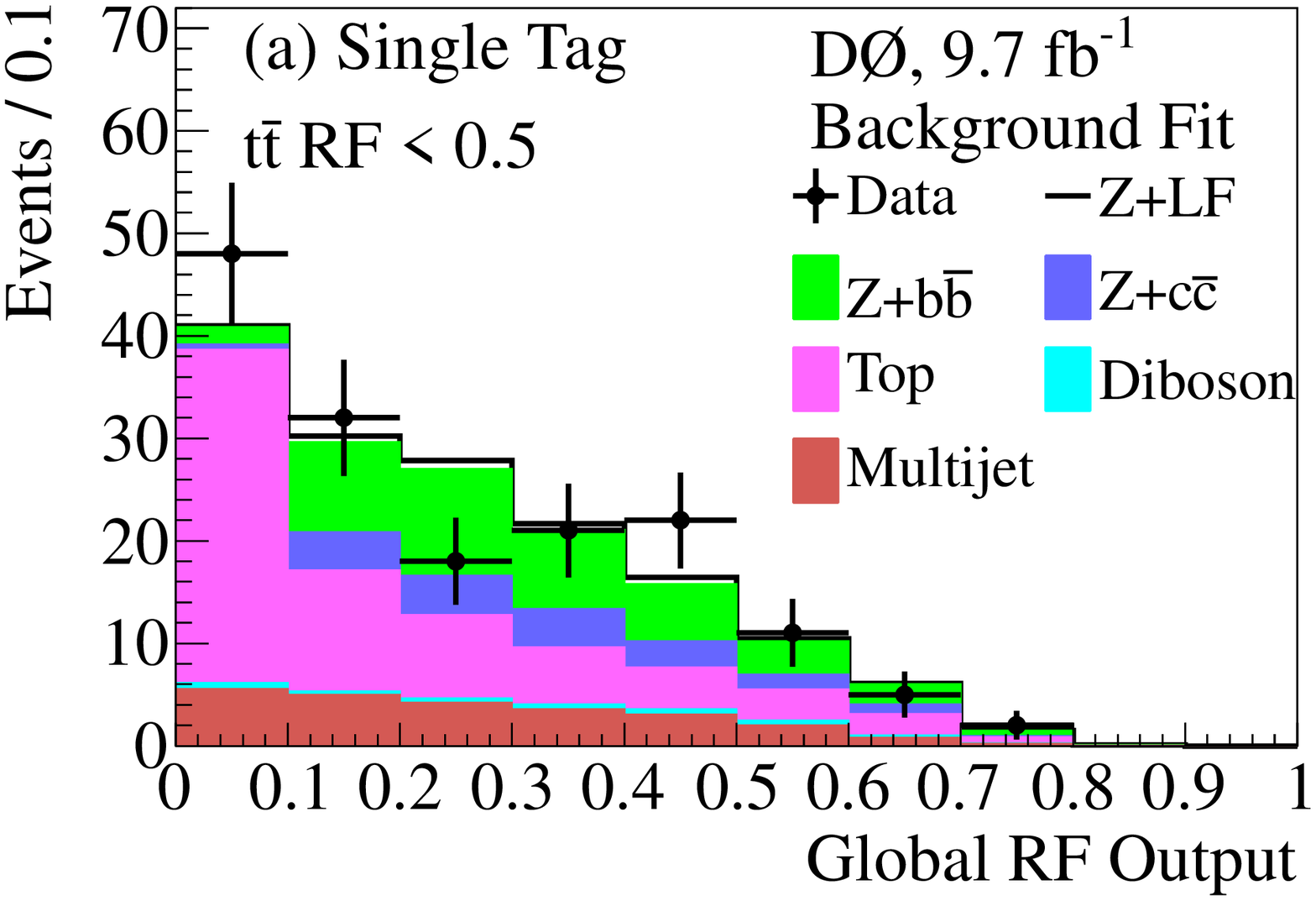}  &
\includegraphics[height=0.24\textheight]{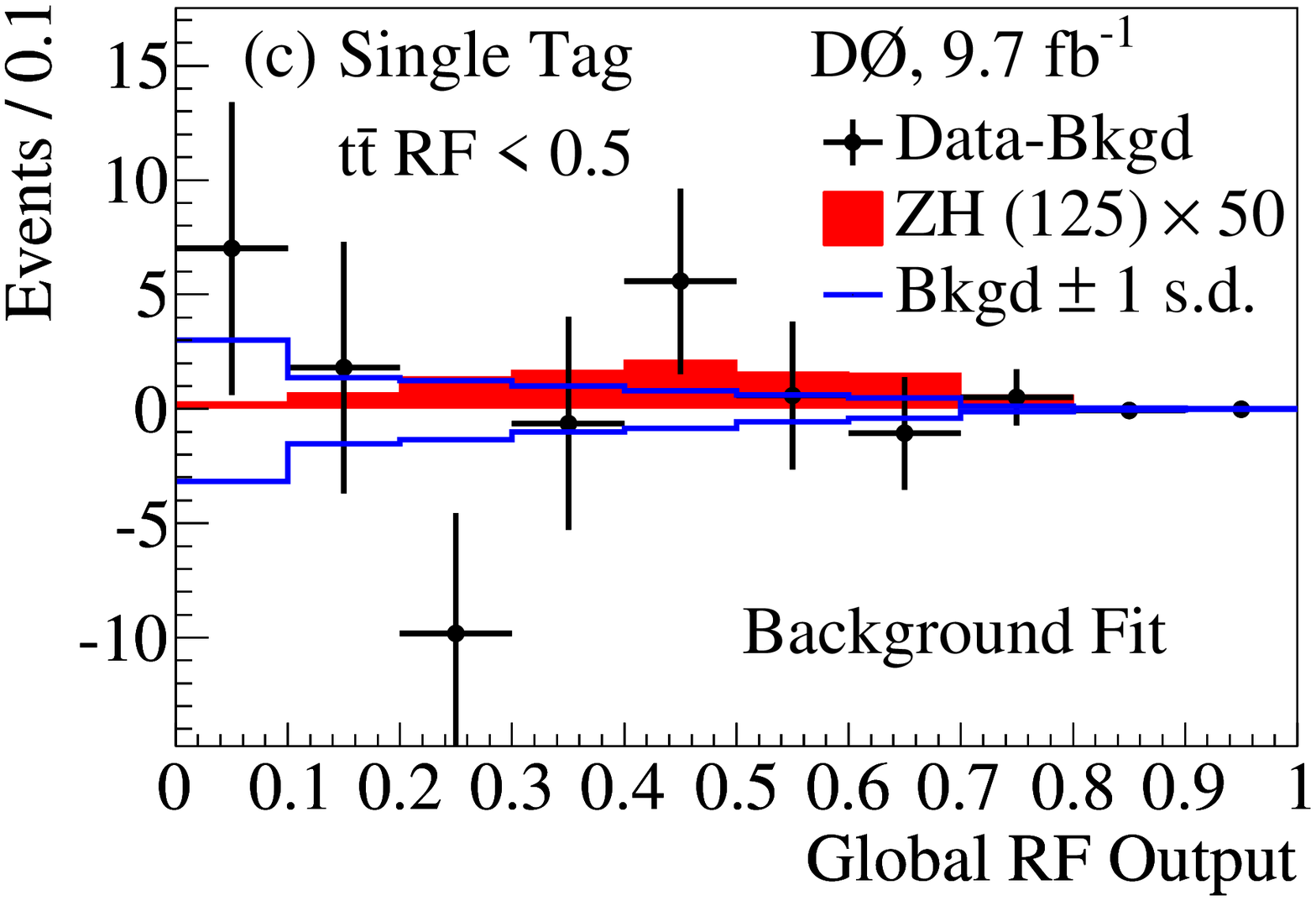}  \\
\includegraphics[height=0.24\textheight]{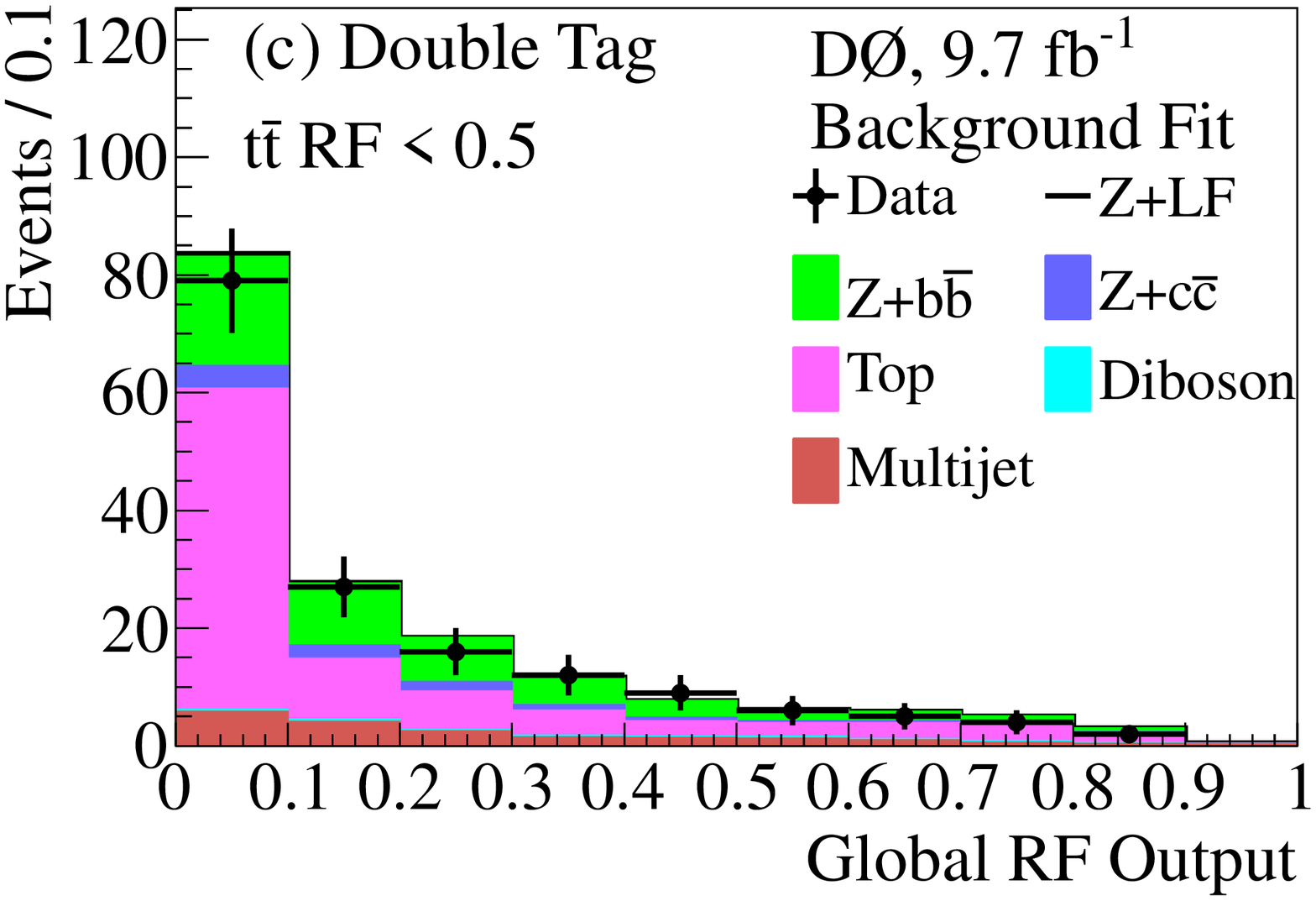} &
\includegraphics[height=0.24\textheight]{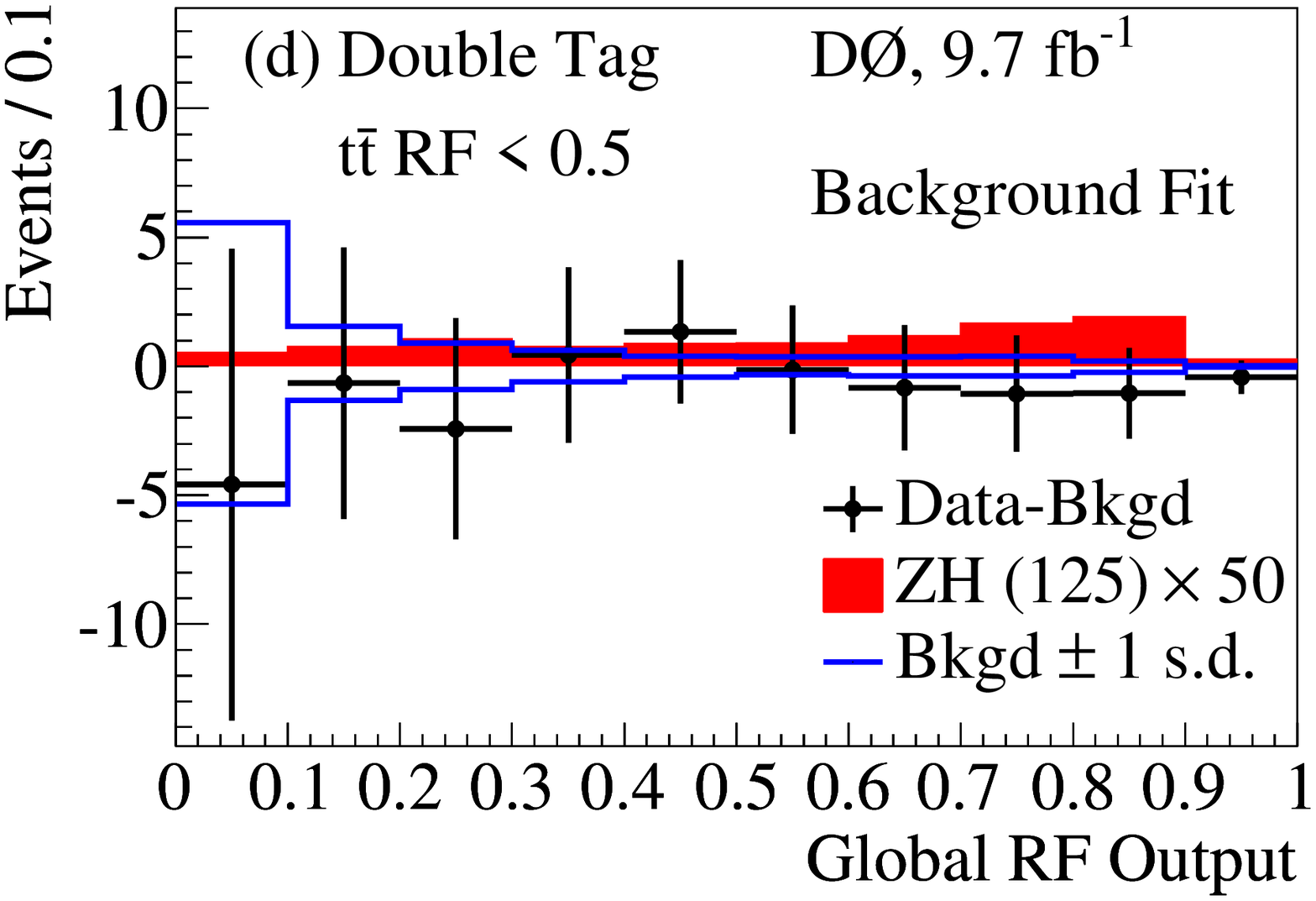} \\
\includegraphics[height=0.24\textheight]{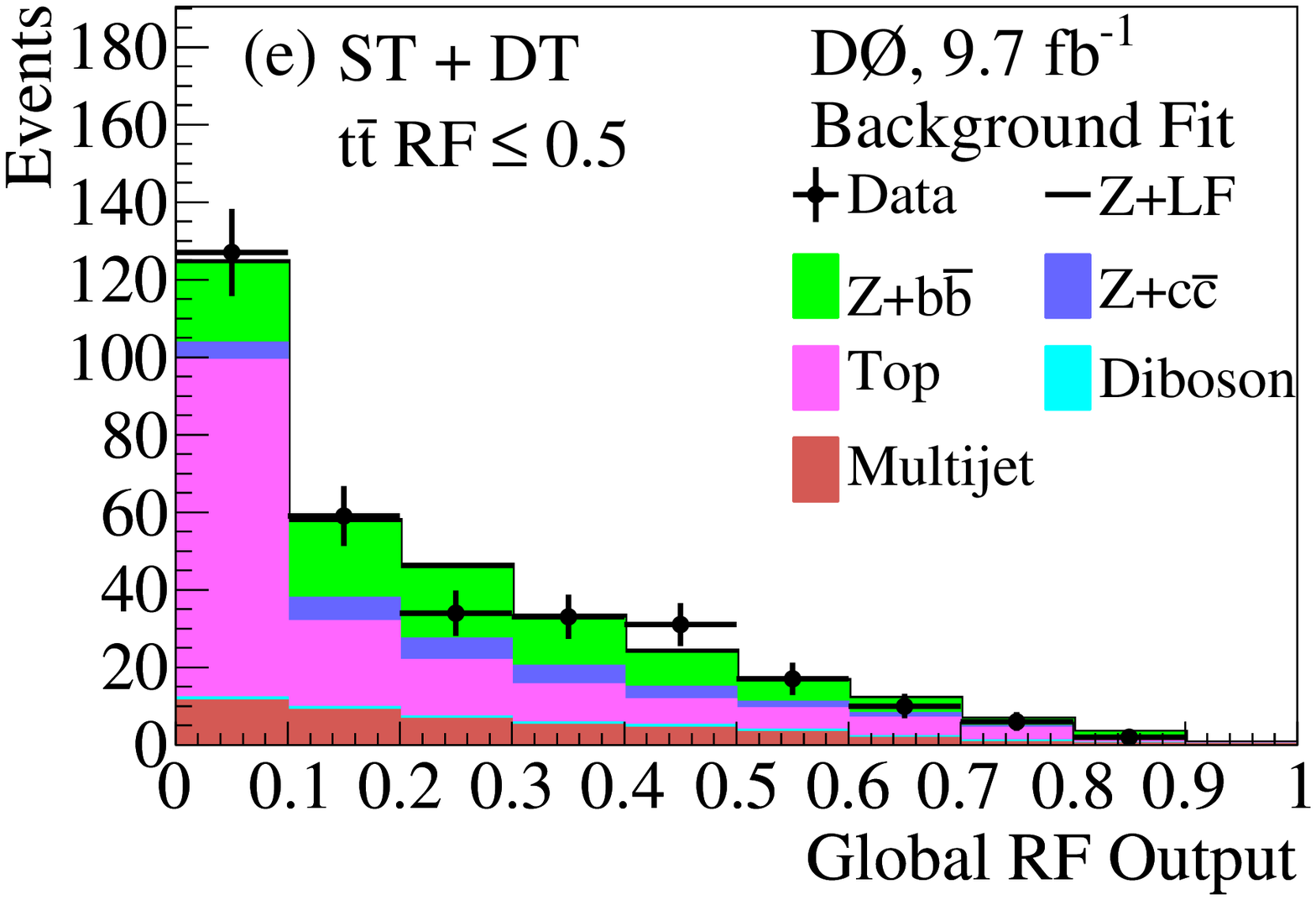} &
\includegraphics[height=0.24\textheight]{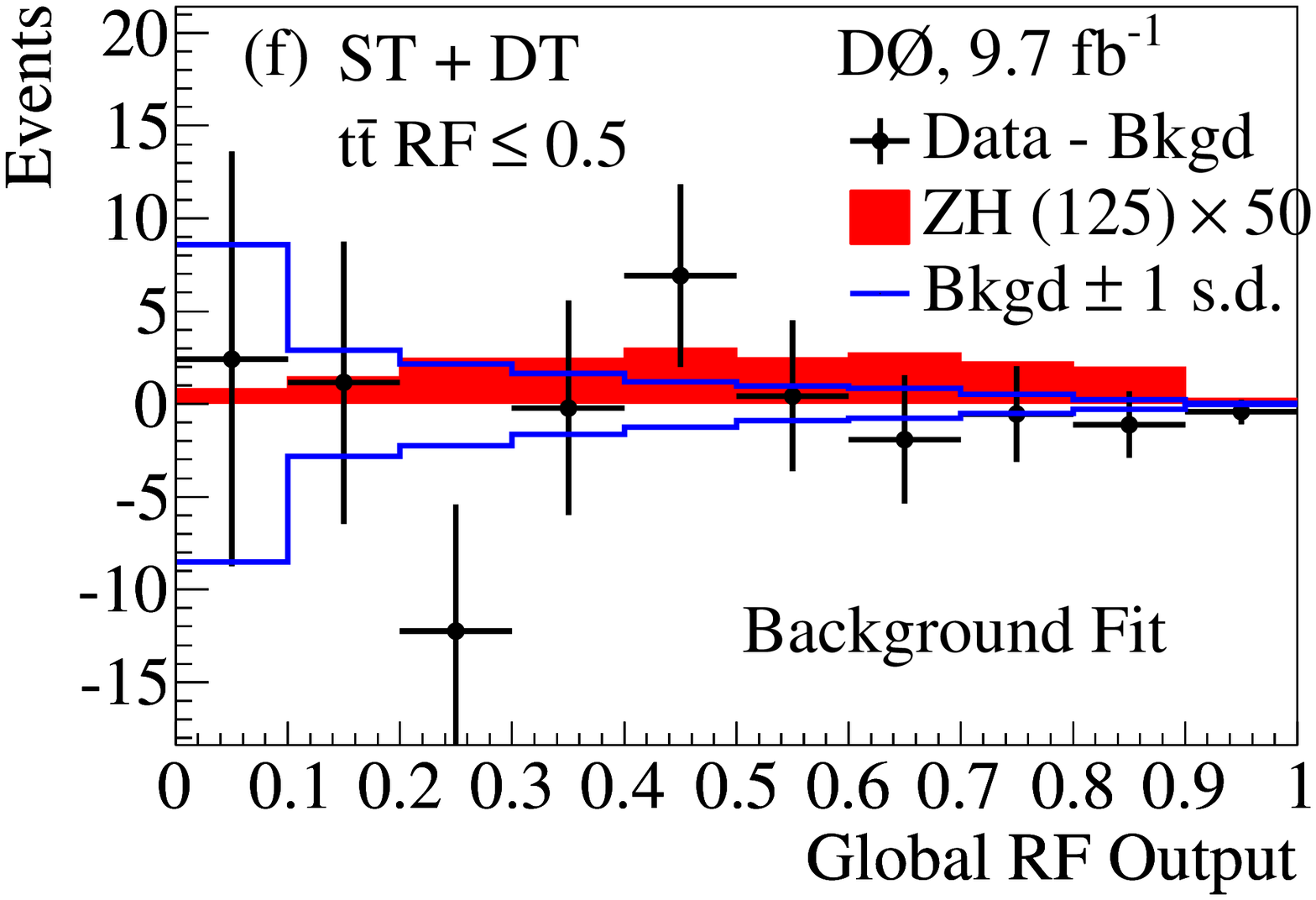} \\
\end{tabular} 
\caption{\label{fig:rf_rich_postfit} (color online). Global RF output distributions in
  the $\ttbar$-enriched region, assuming $M_H=$ 125 GeV, after the fit
  to the data in the $B$ hypothesis for 
  (a) ST events, (c) DT events, and (e) ST and DT events combined.
  Background-subtracted distributions for (a), (c), and (e) are shown in (b), (d),
  and (f), respectively.  Signal distributions for $M_H=$ 125 GeV are shown with the SM
  cross section scaled by a factor of $50$ in (b), (d) and (f).}
\end{figure*}

\begin{figure*}[!hdtp]\centering
\begin{tabular}{ll}
\includegraphics[height=0.24\textheight]{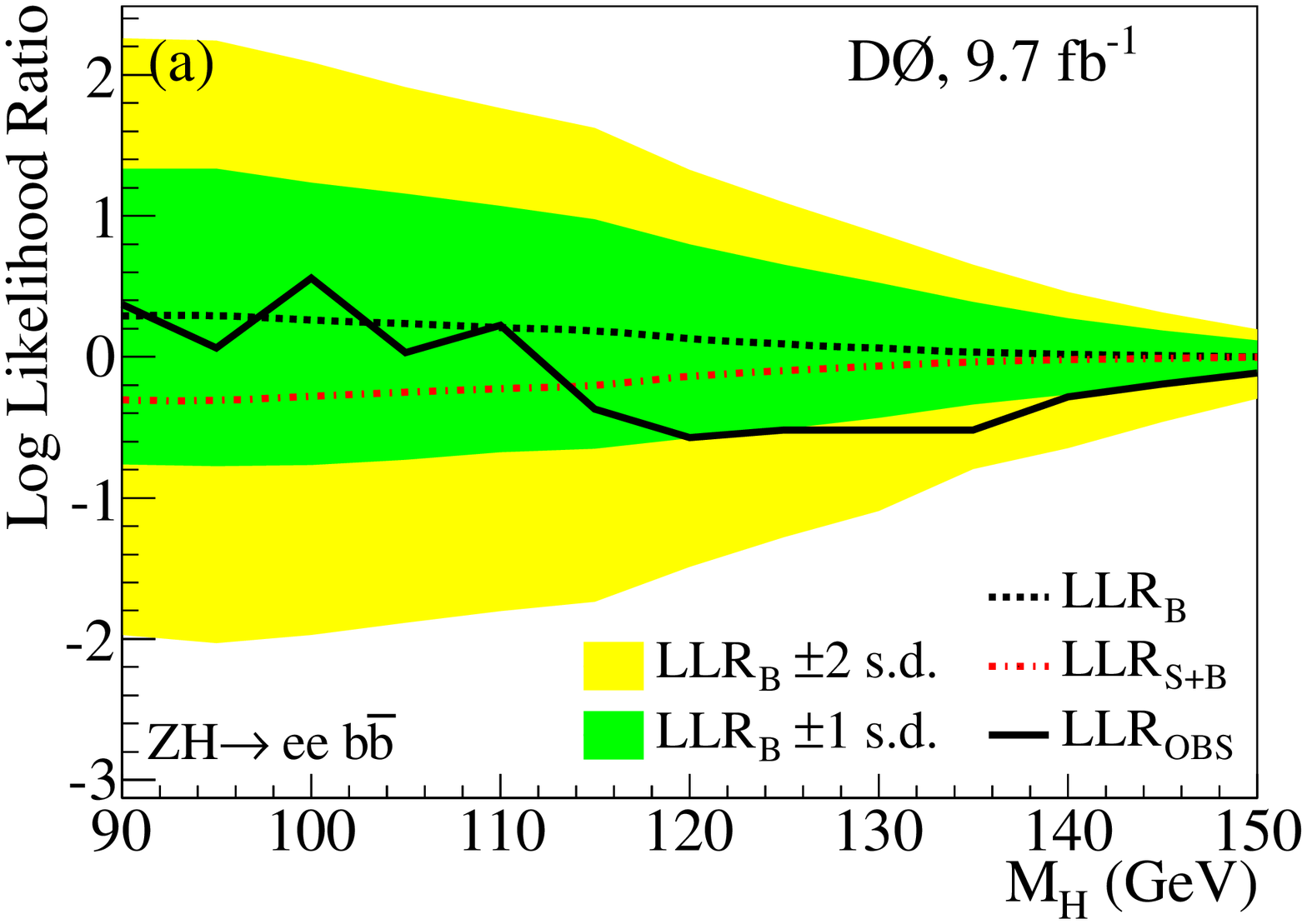}&
\includegraphics[height=0.24\textheight]{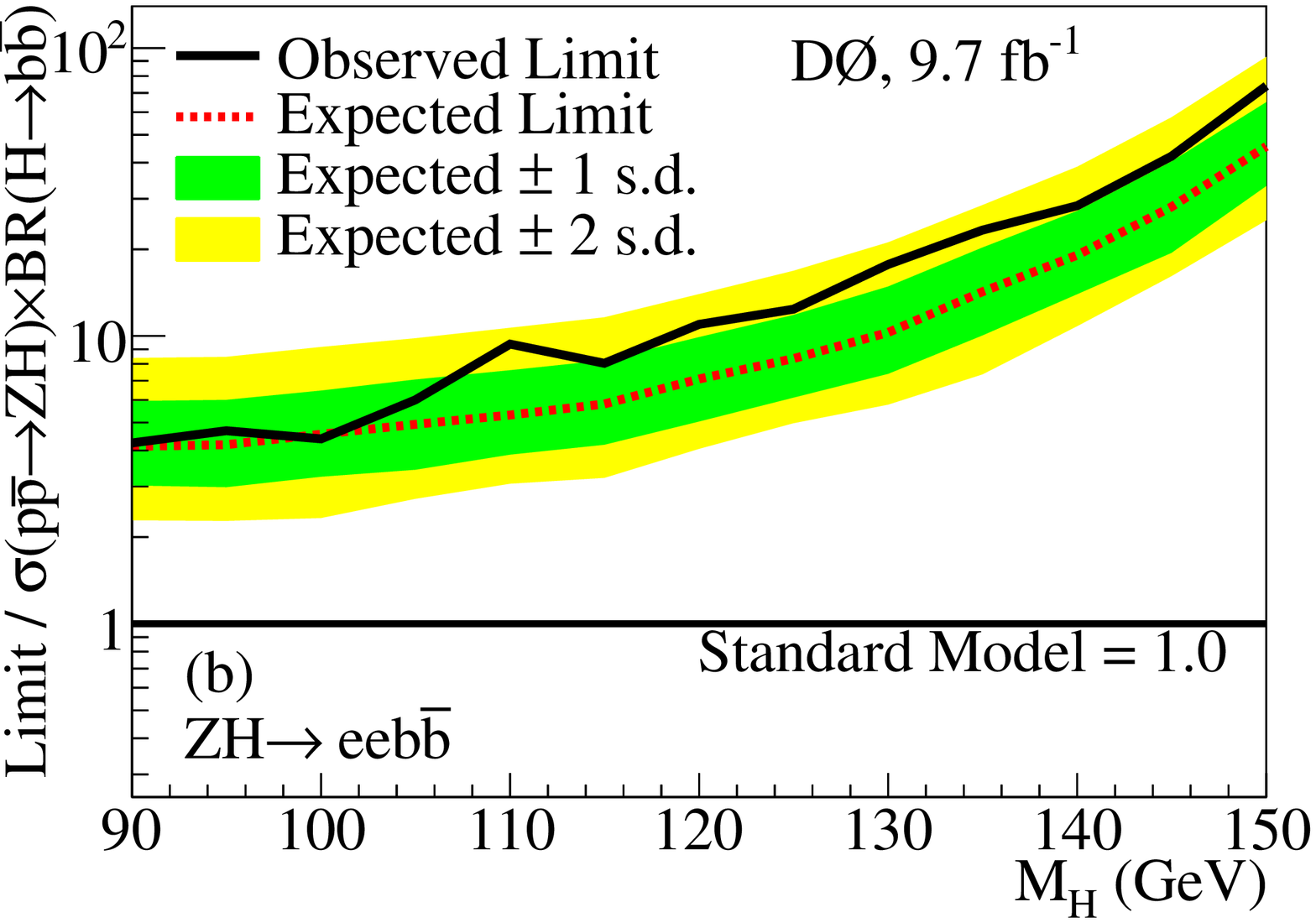}\\
\includegraphics[height=0.24\textheight]{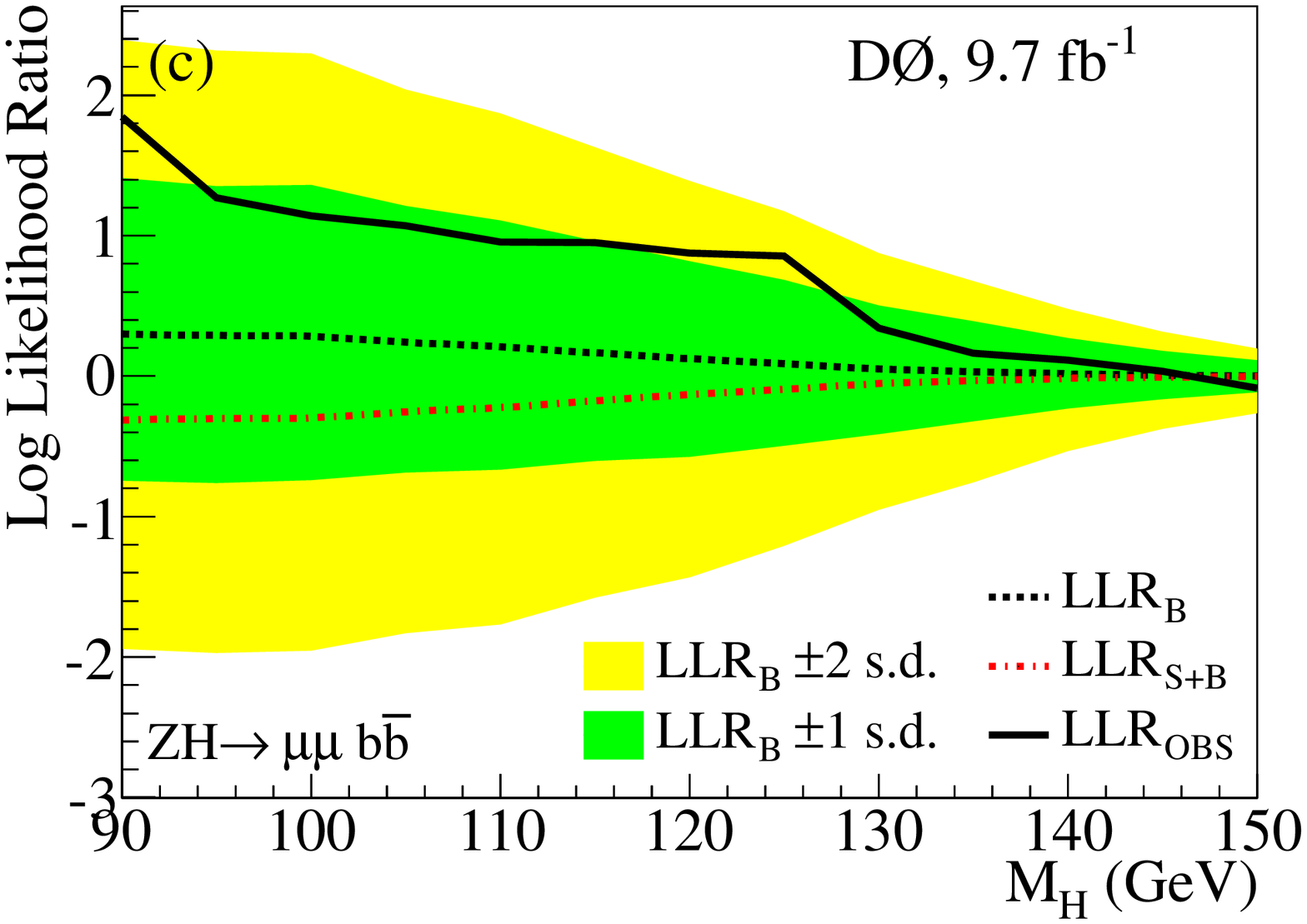}&
\includegraphics[height=0.24\textheight]{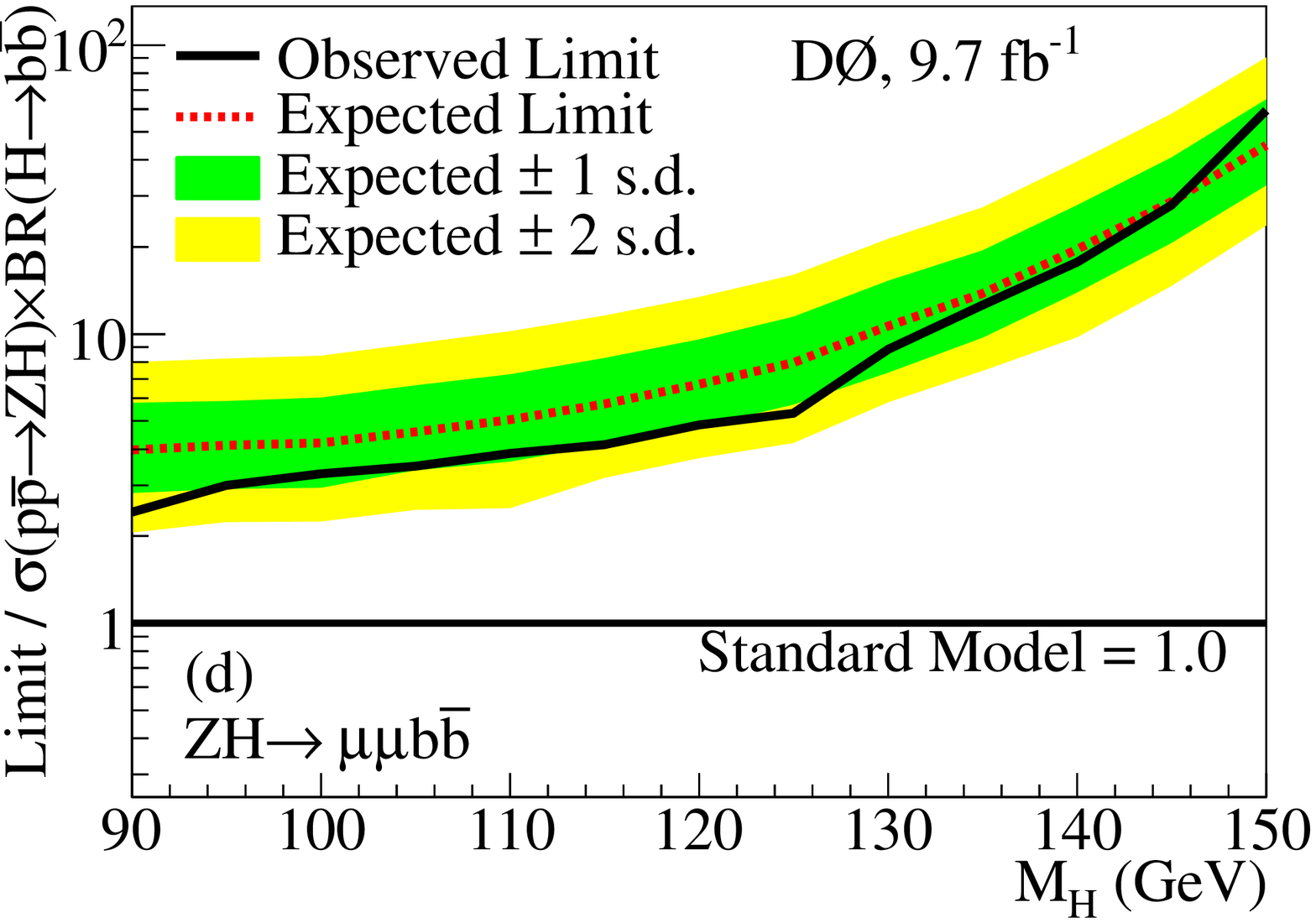}\\
\includegraphics[height=0.24\textheight]{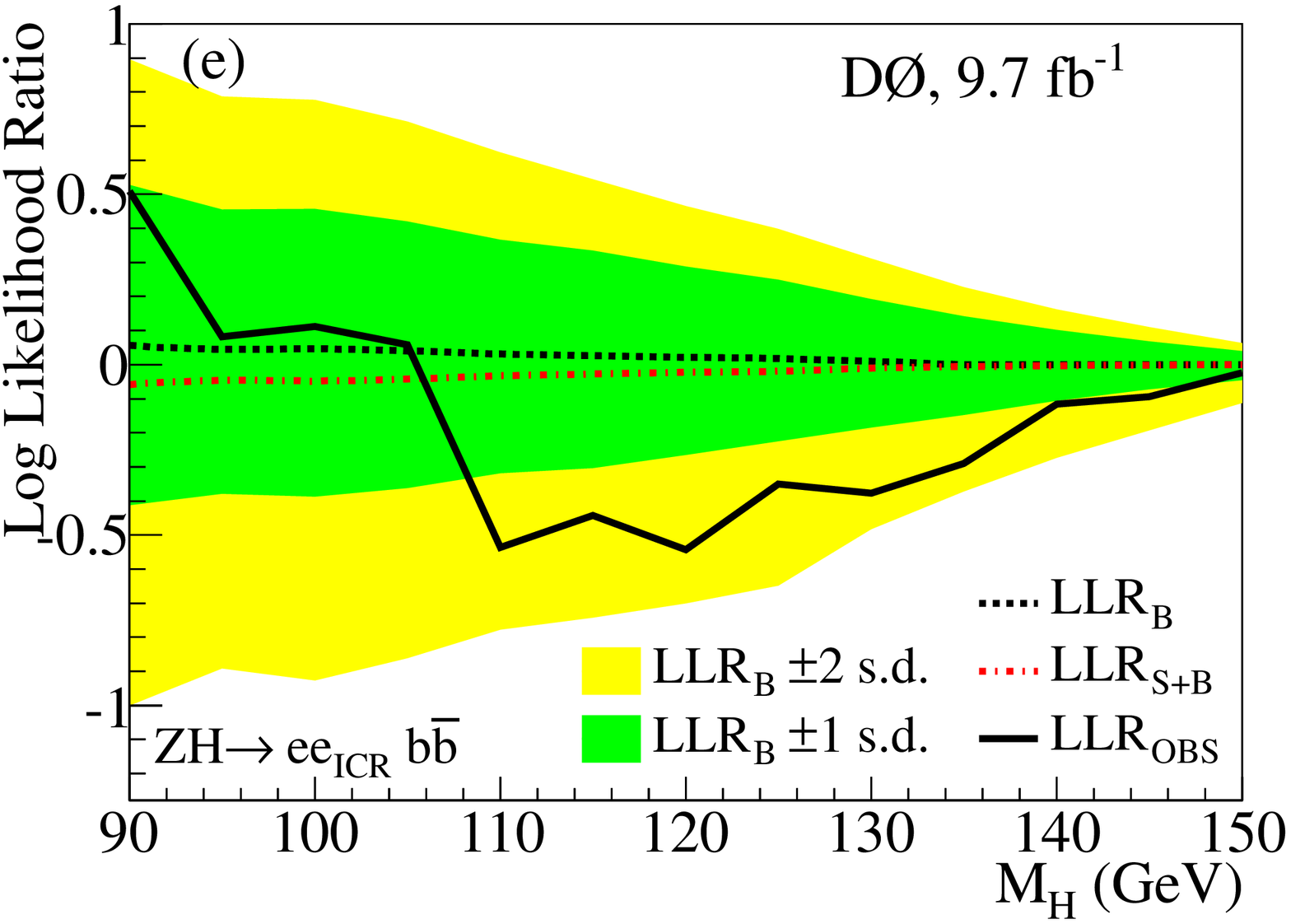}&
\includegraphics[height=0.24\textheight]{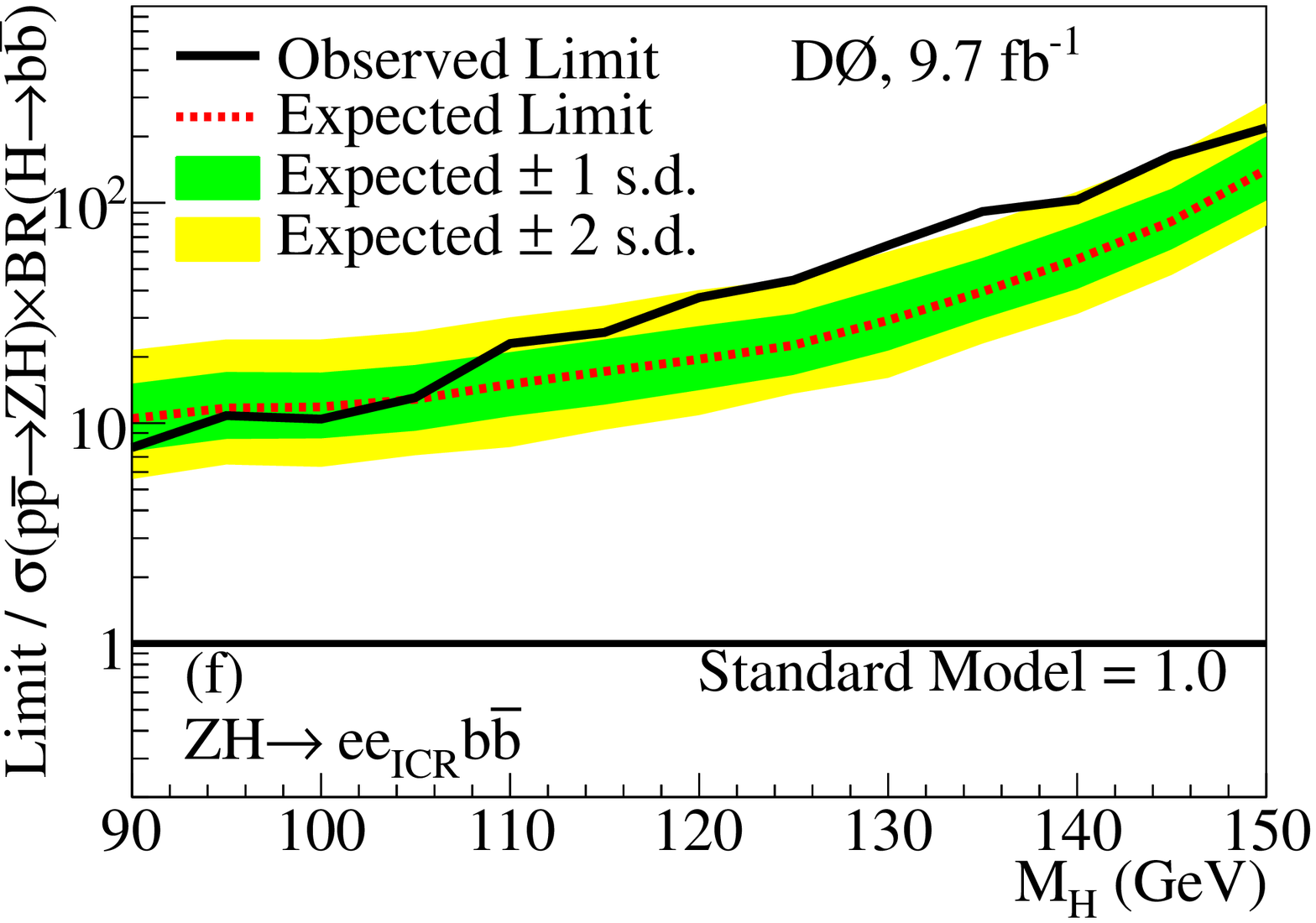}\\
\includegraphics[height=0.24\textheight]{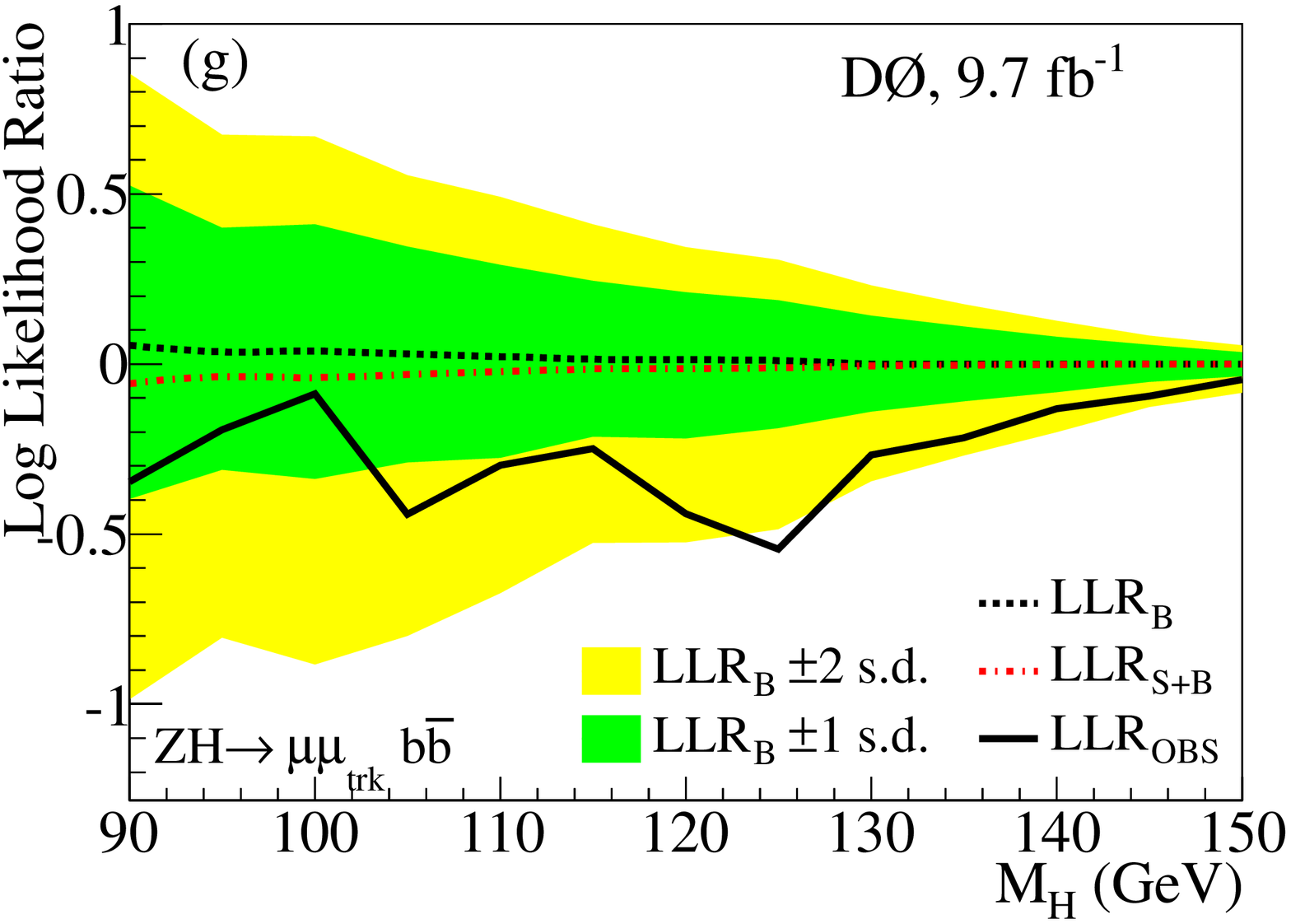}&
\includegraphics[height=0.24\textheight]{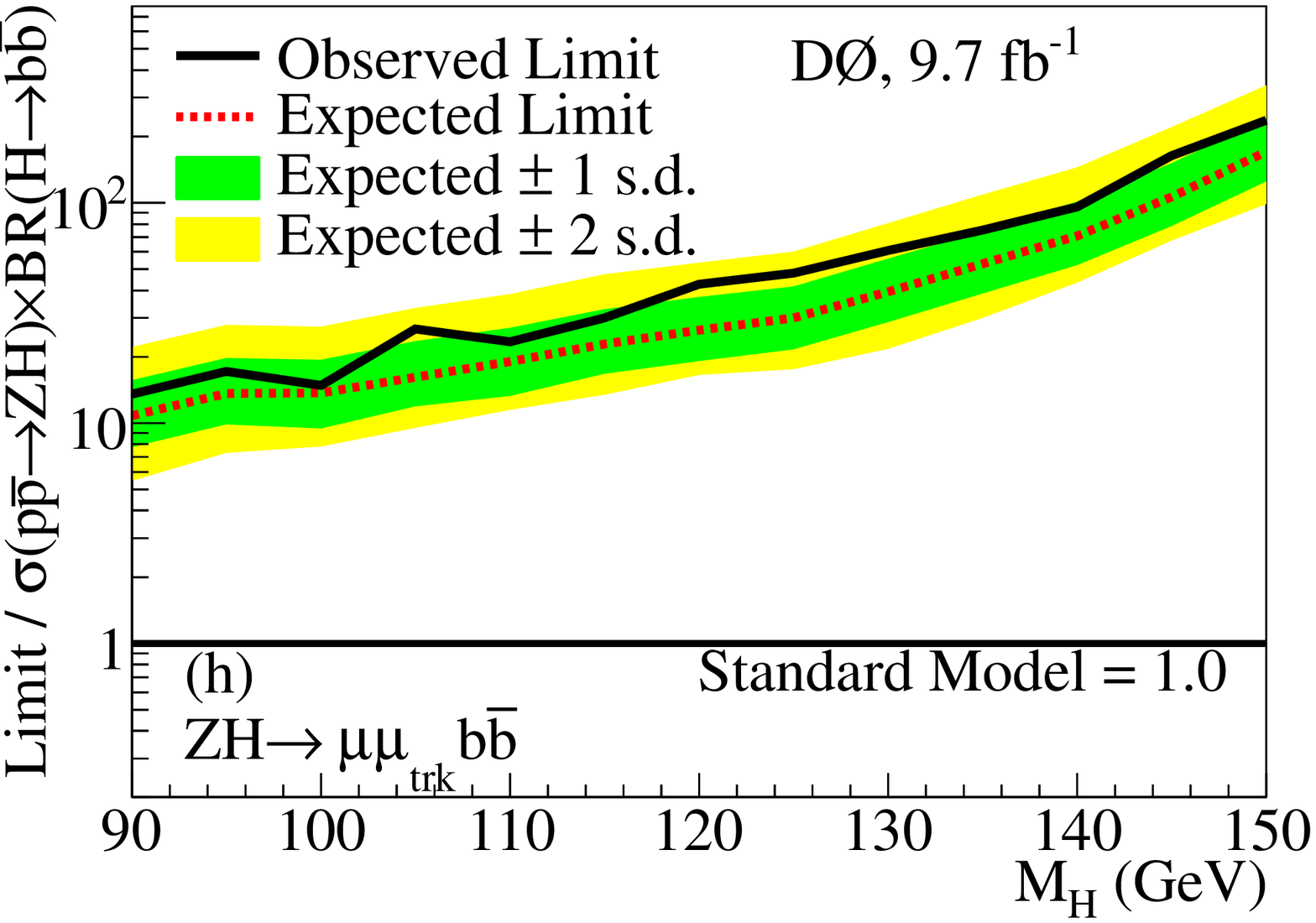}
\end{tabular}
 \caption{(color online). Observed and expected $LLR$ values for the $S+B$ and $B$ hypotheses,
along with the $\pm1$ and $\pm2$ s.d. bands for the $B$ hypotheses, as well
as observed and expected cross section upper limits 
(along with the $\pm1$ and $\pm2$ s.d. bands for the expected limit) relative to the SM 
cross section,  
(a, b) for the \ee~channel, 
(c, d) for the \mumu~channel, 
(e, f) for the \eeicr~channel, and 
(g, h) for the \mumutrk~channel.} 
 \label{fig:per_channel_results}
\end{figure*}

\begin{figure*}[tbp]\centering
\begin{tabular}{ll}
\includegraphics[height=0.24\textheight]{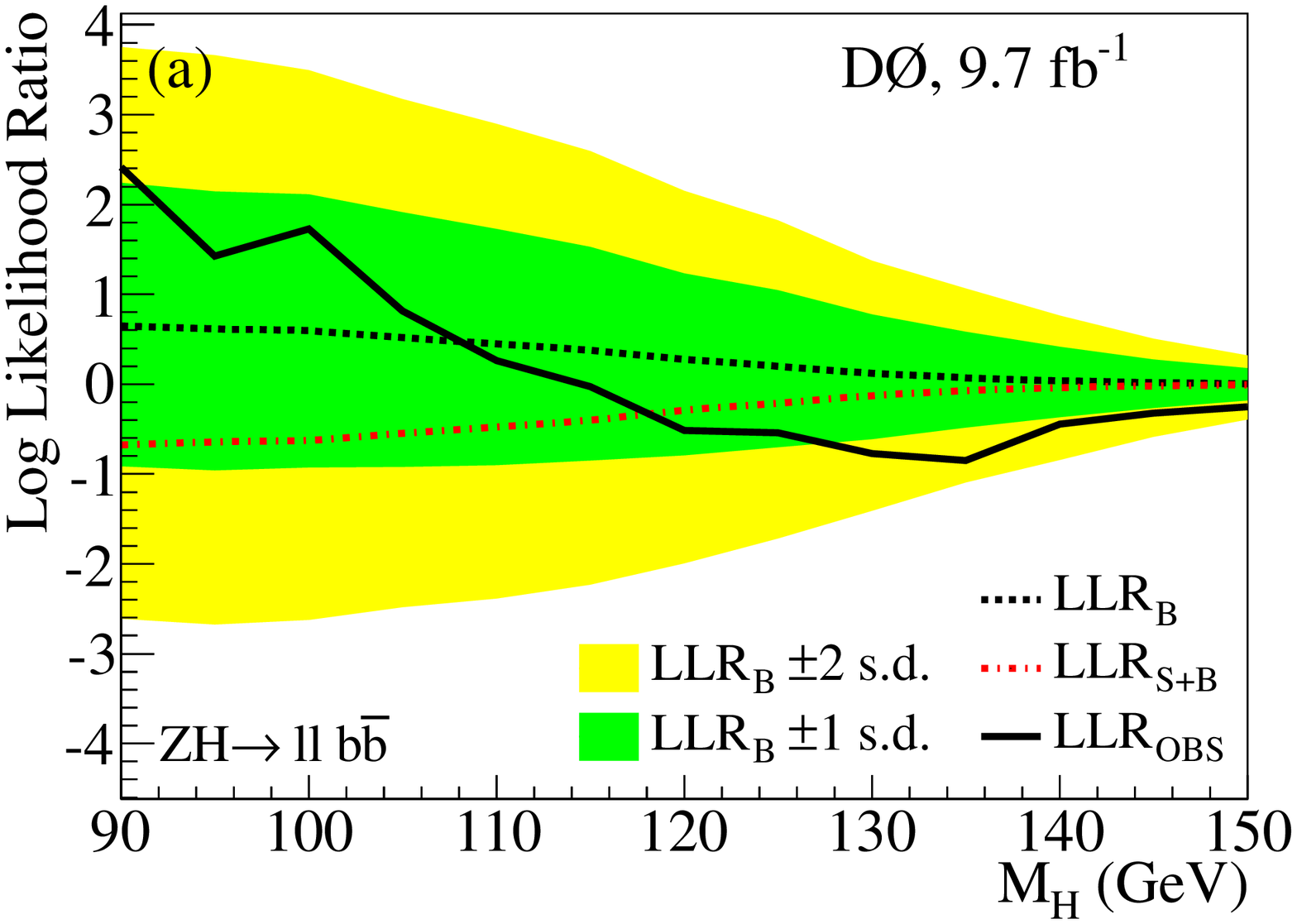} &
\includegraphics[height=0.24\textheight]{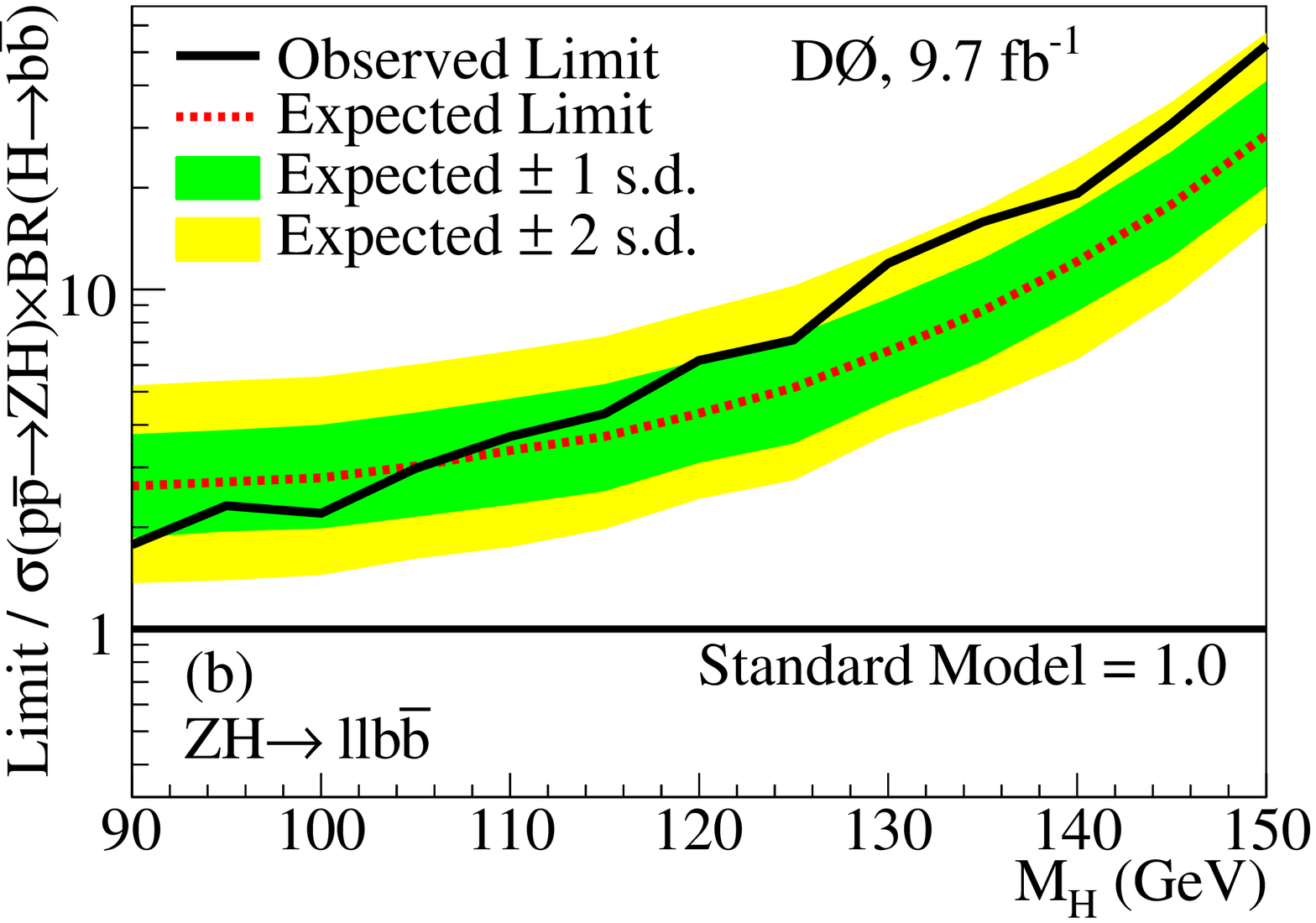} \\
\end{tabular}
\caption{(color online). (a) Observed and expected $LLR$ values as a function of $M_H$
for the $S+B$ and $B$ hypotheses, along with 
the $\pm1$ and $\pm2$ s.d. bands for the $B$ hypotheses,
for all lepton channels combined.
(b) Expected and observed cross section
upper limits at the 95\% C.L. 
for $ZH \rightarrow \ell^+\ell^- b\bar{b}$ production, 
relative to the SM cross section.}
\label{fig:llr_limits}
\end{figure*}

In Figs.~\ref{fig:rf_poor_postfit} and \ref{fig:rf_rich_postfit} we 
show the global RF distributions for $M_H=125~\gev$ after the fit of the nuisance 
parameters to the data in the $B$ hypothesis.
Figure~\ref{fig:per_channel_results} shows the observed and 
expected (median) $LLR$ values for the individual analysis channels.
Also shown are the upper
limits at the 95\% C.L. on the product of the $ZH$ production cross section and branching
ratio for $H\to\bbbar$. The $LLR$ values 
for all lepton channels combined are shown in Fig.~\ref{fig:llr_limits}(a),
and limits are shown in Fig.~\ref{fig:llr_limits}(b)
and Table~\ref{tbl:limits}. The limits 
are expressed as a ratio to the
SM prediction.
At $M_H=125~\gev$ the observed (expected) limit on this ratio is
\obslim~(\explim).

\begin{table*}[htp!]
\caption{The expected and observed upper limits at the 95\% C.L. on the SM Higgs 
boson production cross section for $ZH \rightarrow \ell^+\ell^- b\bar{b}$,
expressed as a ratio to the SM cross section.}
\begin{center}
\begin{tabular}{lccccccccccccc}
\hline\hline
 $M_H~(\gev)$ & 90 & 95 & 100 & 105 & 110 & 115 & 120 & 125 & 130 & 135 & 140 & 145 & 150 \\
\hline
Expected & 2.6  & 2.7  & 2.8 & 3.0 & 3.4 & 3.7 & 4.3 & 5.1 & 6.6 & 8.7 & 12 & 18 & 29\\
Observed & 1.8  & 2.3  & 2.2 & 3.0 & 3.7 & 4.3 & 6.2 & 7.1 &  12 &  16 & 19 & 31 & 53\\
\hline\hline
\end{tabular}
\label{tbl:limits}
\end{center}
\end{table*}

%\clearpage

\section{Summary}\label{sec:summary}

In summary, we have searched for SM Higgs boson production in association
with a $Z$ boson in the final state of two charged leptons (electrons
or muons) and two $b$-quark jets using 9.7~fb$^{-1}$ of
$p\bar{p}$ collisions at $\sqrt{s}$ = 1.96~TeV.  To validate the
methods used in this analysis, we have determined the cross section for $ZZ$
production in the same final state and found it to be a 
factor of $\vzRFobsSF\pm\vzRFerrSF$
relative to the SM prediction, with a significance of 1.5 s.d. 
We have set an upper limit on the product of the $ZH$ production cross section and
branching ratio for $H\to\bbbar$ as a function of $M_H$.  
The observed (expected) limit at the 95\% C.L. for 
$M_H = 125$ GeV is \obslim~(\explim) times the SM 
expectation.

\begin{acknowledgments}
\input{acknowledgement}
\end{acknowledgments}

\end{document}

%% file: author_list.tex
\affiliation{LAFEX, Centro Brasileiro de Pesquisas F\'{i}sicas, Rio de Janeiro, Brazil}
\affiliation{Universidade do Estado do Rio de Janeiro, Rio de Janeiro, Brazil}
\affiliation{Universidade Federal do ABC, Santo Andr\'e, Brazil}
\affiliation{University of Science and Technology of China, Hefei, People's Republic of China}
\affiliation{Universidad de los Andes, Bogot\'a, Colombia}
\affiliation{Charles University, Faculty of Mathematics and Physics, Center for Particle Physics, Prague, Czech Republic}
\affiliation{Czech Technical University in Prague, Prague, Czech Republic}
\affiliation{Center for Particle Physics, Institute of Physics, Academy of Sciences of the Czech Republic, Prague, Czech Republic}
\affiliation{Universidad San Francisco de Quito, Quito, Ecuador}
\affiliation{LPC, Universit\'e Blaise Pascal, CNRS/IN2P3, Clermont, France}
\affiliation{LPSC, Universit\'e Joseph Fourier Grenoble 1, CNRS/IN2P3, Institut National Polytechnique de Grenoble, Grenoble, France}
\affiliation{CPPM, Aix-Marseille Universit\'e, CNRS/IN2P3, Marseille, France}
\affiliation{LAL, Universit\'e Paris-Sud, CNRS/IN2P3, Orsay, France}
\affiliation{LPNHE, Universit\'es Paris VI and VII, CNRS/IN2P3, Paris, France}
\affiliation{CEA, Irfu, SPP, Saclay, France}
\affiliation{IPHC, Universit\'e de Strasbourg, CNRS/IN2P3, Strasbourg, France}
\affiliation{IPNL, Universit\'e Lyon 1, CNRS/IN2P3, Villeurbanne, France and Universit\'e de Lyon, Lyon, France}
\affiliation{III. Physikalisches Institut A, RWTH Aachen University, Aachen, Germany}
\affiliation{Physikalisches Institut, Universit\"at Freiburg, Freiburg, Germany}
\affiliation{II. Physikalisches Institut, Georg-August-Universit\"at G\"ottingen, G\"ottingen, Germany}
\affiliation{Institut f\"ur Physik, Universit\"at Mainz, Mainz, Germany}
\affiliation{Ludwig-Maximilians-Universit\"at M\"unchen, M\"unchen, Germany}
\affiliation{Fachbereich Physik, Bergische Universit\"at Wuppertal, Wuppertal, Germany}
\affiliation{Panjab University, Chandigarh, India}
\affiliation{Delhi University, Delhi, India}
\affiliation{Tata Institute of Fundamental Research, Mumbai, India}
\affiliation{University College Dublin, Dublin, Ireland}
\affiliation{Korea Detector Laboratory, Korea University, Seoul, Korea}
\affiliation{CINVESTAV, Mexico City, Mexico}
\affiliation{Nikhef, Science Park, Amsterdam, the Netherlands}
\affiliation{Radboud University Nijmegen, Nijmegen, the Netherlands}
\affiliation{Joint Institute for Nuclear Research, Dubna, Russia}
\affiliation{Institute for Theoretical and Experimental Physics, Moscow, Russia}
\affiliation{Moscow State University, Moscow, Russia}
\affiliation{Institute for High Energy Physics, Protvino, Russia}
\affiliation{Petersburg Nuclear Physics Institute, St. Petersburg, Russia}
\affiliation{Instituci\'{o} Catalana de Recerca i Estudis Avan\c{c}ats (ICREA) and Institut de F\'{i}sica d'Altes Energies (IFAE), Barcelona, Spain}
\affiliation{Uppsala University, Uppsala, Sweden}
\affiliation{Lancaster University, Lancaster LA1 4YB, United Kingdom}
\affiliation{Imperial College London, London SW7 2AZ, United Kingdom}
\affiliation{The University of Manchester, Manchester M13 9PL, United Kingdom}
\affiliation{University of Arizona, Tucson, Arizona 85721, USA}
\affiliation{University of California Riverside, Riverside, California 92521, USA}
\affiliation{Florida State University, Tallahassee, Florida 32306, USA}
\affiliation{Fermi National Accelerator Laboratory, Batavia, Illinois 60510, USA}
\affiliation{University of Illinois at Chicago, Chicago, Illinois 60607, USA}
\affiliation{Northern Illinois University, DeKalb, Illinois 60115, USA}
\affiliation{Northwestern University, Evanston, Illinois 60208, USA}
\affiliation{Indiana University, Bloomington, Indiana 47405, USA}
\affiliation{Purdue University Calumet, Hammond, Indiana 46323, USA}
\affiliation{University of Notre Dame, Notre Dame, Indiana 46556, USA}
\affiliation{Iowa State University, Ames, Iowa 50011, USA}
\affiliation{University of Kansas, Lawrence, Kansas 66045, USA}
\affiliation{Louisiana Tech University, Ruston, Louisiana 71272, USA}
\affiliation{Northeastern University, Boston, Massachusetts 02115, USA}
\affiliation{University of Michigan, Ann Arbor, Michigan 48109, USA}
\affiliation{Michigan State University, East Lansing, Michigan 48824, USA}
\affiliation{University of Mississippi, University, Mississippi 38677, USA}
\affiliation{University of Nebraska, Lincoln, Nebraska 68588, USA}
\affiliation{Rutgers University, Piscataway, New Jersey 08855, USA}
\affiliation{Princeton University, Princeton, New Jersey 08544, USA}
\affiliation{State University of New York, Buffalo, New York 14260, USA}
\affiliation{University of Rochester, Rochester, New York 14627, USA}
\affiliation{State University of New York, Stony Brook, New York 11794, USA}
\affiliation{Brookhaven National Laboratory, Upton, New York 11973, USA}
\affiliation{Langston University, Langston, Oklahoma 73050, USA}
\affiliation{University of Oklahoma, Norman, Oklahoma 73019, USA}
\affiliation{Oklahoma State University, Stillwater, Oklahoma 74078, USA}
\affiliation{Brown University, Providence, Rhode Island 02912, USA}
\affiliation{University of Texas, Arlington, Texas 76019, USA}
\affiliation{Southern Methodist University, Dallas, Texas 75275, USA}
\affiliation{Rice University, Houston, Texas 77005, USA}
\affiliation{University of Virginia, Charlottesville, Virginia 22904, USA}
\affiliation{University of Washington, Seattle, Washington 98195, USA}
\author{V.M.~Abazov} \affiliation{Joint Institute for Nuclear Research, Dubna, Russia}
\author{B.~Abbott} \affiliation{University of Oklahoma, Norman, Oklahoma 73019, USA}
\author{B.S.~Acharya} \affiliation{Tata Institute of Fundamental Research, Mumbai, India}
\author{M.~Adams} \affiliation{University of Illinois at Chicago, Chicago, Illinois 60607, USA}
\author{T.~Adams} \affiliation{Florida State University, Tallahassee, Florida 32306, USA}
\author{G.D.~Alexeev} \affiliation{Joint Institute for Nuclear Research, Dubna, Russia}
\author{G.~Alkhazov} \affiliation{Petersburg Nuclear Physics Institute, St. Petersburg, Russia}
\author{A.~Alton$^{a}$} \affiliation{University of Michigan, Ann Arbor, Michigan 48109, USA}
\author{A.~Askew} \affiliation{Florida State University, Tallahassee, Florida 32306, USA}
\author{S.~Atkins} \affiliation{Louisiana Tech University, Ruston, Louisiana 71272, USA}
% ...
%
% visitor_addresses.tex                       11 January 2013 
%  available symbols are:
%  $\ast, \dag, \ddag, \S, \P, $\|$, $\ast\ast$, \dag\dag, \ddag\ddag ,\#
%
\collaboration{The D0 Collaboration\footnote{with visitors from
%{alton}
$^{a}$Augustana College, Sioux Falls, SD, USA,
%{burdin}
$^{b}$The University of Liverpool, Liverpool, UK,
%{garcia-guerra}
$^{c}$UPIITA-IPN, Mexico City, Mexico,
%{grohsjean}
$^{d}$DESY, Hamburg, Germany,
%{partridge}
$^{e}$SLAC, Menlo Park, CA, USA,
%{hesketh}
$^{f}$University College London, London, UK,
%{luna-garcia}
$^{g}$Centro de Investigacion en Computacion - IPN, Mexico City, Mexico,
%{podesta-lerma}
$^{h}$ECFM, Universidad Autonoma de Sinaloa, Culiac\'an, Mexico,
%{santos}
$^{i}$Universidade Estadual Paulista, S\~ao Paulo, Brazil,
%{meyer}
$^{j}$Karlsruher Institut f\"ur Technologie (KIT) - Steinbuch Centre for Computing (SCC)
and
%{patwa}
$^{k}$Office of Science, U.S. Department of Energy, Washington, D.C. 20585, USA.
%{falkowski}
%$^{?}$Laboratoire de Physique Theorique, Orsay, FR,
%{hooper}
%$^{?}$Visitor from Bradley University, Peoria, IL, USA.
%{kozminski}
%$^{?}$}Visitor from Lewis University, Romeoville, IL, USA.
%{weber}
%$^{?}$Universit{\"a}t Bern, Bern, Switzerland.
%{deceased}
%$^{\ddag}$Deceased.
}} \noaffiliation
\vskip 0.25cm

%% file: acknowledgement.tex
% acknowledgement.tex                             17 May 2010
%
We thank the staffs at Fermilab and collaborating institutions,
and acknowledge support from the
DOE and NSF (USA);
CEA and CNRS/IN2P3 (France);
FASI, Rosatom and RFBR (Russia);
CNPq, FAPERJ, FAPESP and FUNDUNESP (Brazil);
DAE and DST (India);
Colciencias (Colombia);
CONACyT (Mexico);
KRF and KOSEF (Korea);
CONICET and UBACyT (Argentina);
FOM (The Netherlands);
STFC and the Royal Society (United Kingdom);
MSMT and GACR (Czech Republic);
CRC Program and NSERC (Canada);
BMBF and DFG (Germany);
SFI (Ireland);
The Swedish Research Council (Sweden);
and
CAS and CNSF (China).

%% file: zh_llbb_prd.bbl
\begin{thebibliography}{99}
\bibitem{d0coords}
The D0 detector utilizes a right-handed coordinate system with the $z$~axis
pointing in the direction of the proton beam, the $y$~axis pointing upwards,
and the $x$~axis pointing away from the center of the collider ring.
The azimuthal angle $\phi$~is defined in the $x$-$y$~plane measured from the $x$~axis.
The pseudorapidity is defined as $\eta=-\textup{ln}[\textup{tan}(\theta/2)]$, where
$\theta=\textup{arctan}(\sqrt{x^2+y^2}/z)$. Pseudorapidity calculated from 
the center of the detector at $z=0$, rather than from the measured $\ppbar$ interaction
vertex position,
is denoted $\etadet$.
Transverse variables are defined as projections of the variables onto the $x$-$y$~plane. 
Each category of reconstructed objects is ordered by decreasing $\pt$ or $\et$,
with the highest-$\pt$ or highest-$\et$ object called ``leading'' and the second-highest
called ``sub-leading.''
\bibitem{d0tunea}
D0 Tune A is identical to Tune A~\cite{tunea}, but uses the {\sc CTEQ6L1} PDF set 
and sets $\Lambda_{\mathrm{QCD}} = 0.165$~GeV.
\end{thebibliography}
